\documentclass[trackchanges,twocolumn]{aastex701}

\newcommand\rap{RA$^\prime$}
\newcommand\decp{Dec$^\prime$}
\newcommand\mrap{\mathrm{RA}^\prime}
\newcommand\mdecp{\mathrm{Dec}^\prime}

\newcommand\ldg{\log (\delta_\mathrm{gal}+1)}

\newcommand\logmstar{\log (M_*/\mathrm{M_\odot})}
\newcommand\logmpeak{\log (M_\mathrm{Peak}/\mathrm{M_\odot})}
\newcommand\logmstruct{\log (M_\mathrm{Struct}/\mathrm{M_\odot})}

\begin{document}

\title{Where Galaxies Go to Die: The Environments of Massive Quiescent Galaxies at $3<z<5$}

\author[0000-0002-2446-8770]{Ian McConachie}
\affiliation{Department of Astronomy, University of Wisconsin-Madison, 475 N. Charter St., Madison, WI 53706 USA}
\email[show]{ian.mcconachie@wisc.edu}  

\author[0000-0002-2380-9801]{Anna de Graaff}
\affiliation{Max-Planck-Institut f\"ur Astronomie, K\"onigstuhl 17, D-69117 Heidelberg, Germany}
\email{placeholder@placeholder.edu}  

\author[0000-0003-0695-4414]{Michael V.\ Maseda}
\affiliation{Department of Astronomy, University of Wisconsin-Madison, 475 N. Charter St., Madison, WI 53706 USA}
\email{placeholder@placeholder.edu}  

\author[0000-0001-6755-1315]{Joel Leja}
\affiliation{Department of Astronomy \& Astrophysics, The Pennsylvania State University, University Park, PA 16802, USA} 
\affiliation{Institute for Computational \& Data Sciences, The Pennsylvania State University, University Park, PA 16802, USA}
\affiliation{Institute for Gravitation and the Cosmos, The Pennsylvania State University, University Park, PA 16802, USA}
\email{placeholder@placeholder.edu}  

\author[0000-0001-6454-1699]{Yunchong Zhang} 
\affiliation{Department of Physics and Astronomy and PITT PACC, University of Pittsburgh, Pittsburgh, PA 15260, USA}
\email{placeholder@placeholder.edu}  

\author[0000-0003-4075-7393]{David J. Setton} 
\thanks{Brinson Prize Fellow} \affiliation{Department of Astrophysical Sciences, Princeton University, 4 Ivy Lane, Princeton, NJ 08544, USA}
\email{placeholder@placeholder.edu}  

\author[0000-0001-5063-8254]{Rachel Bezanson}
\affiliation{Department of Physics and Astronomy and PITT PACC, University of Pittsburgh, Pittsburgh, PA 15260, USA}
\email{placeholder@placeholder.edu}

\author[0000-0002-3952-8588]{Leindert A. Boogaard}\affiliation{Leiden Observatory, Leiden University, PO Box 9513, NL-2300 RA Leiden, The Netherlands}\email{placeholder@placeholder.edu}

\author[0000-0003-2680-005X]{Gabriel Brammer}
\affiliation{Cosmic Dawn Center (DAWN), Denmark}
\affiliation{Niels Bohr Institute, University of Copenhagen, Jagtvej 128,
K{\o}benhavn N, DK-2200, Denmark}
\email{placeholder@placeholder.edu}  

\author[0000-0001-7151-009X]{Nikko J. Cleri}
\affiliation{Department of Astronomy \& Astrophysics, The Pennsylvania State University, University Park, PA 16802, USA} 
\affiliation{Institute for Computational \& Data Sciences, The Pennsylvania State University, University Park, PA 16802, USA}
\affiliation{Institute for Gravitation and the Cosmos, The Pennsylvania State University, University Park, PA 16802, USA}
\email{placeholder@placeholder.edu}  

\author[0000-0003-3881-1397]{Olivia R. Cooper}\altaffiliation{NSF Astronomy and Astrophysics Postdoctoral Fellow}
\affiliation{Department for Astrophysical \& Planetary Science, University of Colorado, Boulder, CO 80309, USA}
\email{oliviaraecooper@gmail.com}

\author[0000-0002-3254-9044]{Karl Glazebrook}
\affiliation{Centre for Astrophysics and Supercomputing, Swinburne University of Technology, PO Box 218, Hawthorn, VIC 3122, Australia}
\email{placeholder@placeholder.edu}  

\author[0000-0003-0205-9826]{Rashmi Gottumukkala}
\affiliation{Cosmic Dawn Center (DAWN), Copenhagen, Denmark}
\affiliation{Niels Bohr Institute, University of Copenhagen, Jagtvej 128, Copenhagen, Denmark}
\email{placeholder@placeholder.edu}  

\author[0000-0002-5612-3427]{Jenny E. Greene}
\affiliation{Department of Astrophysical Sciences, Princeton University, 4 Ivy Lane, Princeton, NJ 08544, USA}
\email{placeholder@placeholder.edu}

\author[]{Andy D. Goulding}
\affiliation{Department of Astrophysical Sciences, Princeton University, 4 Ivy Lane, Princeton, NJ 08544, USA}
\email{placeholder@placeholder.edu} 

\author[0000-0002-3301-3321]{Michaela Hirschmann}
\affiliation{Institute of Physics, Laboratory for Galaxy Evolution, EPFL, Observatory of Sauverny, Chemin Pegasi 51, CH-1290 Versoix, Switzerland}
\email{placeholder@placeholder.edu}

\author[0000-0002-2057-5376]{Ivo Labbe}
\affiliation{Centre for Astrophysics and Supercomputing, Swinburne University of Technology, Melbourne, VIC 3122, Australia}
\email{placeholder@placeholder.edu}  

\author[]{Zach Lewis}
\affiliation{Department of Astronomy, University of Wisconsin-Madison, 475 N. Charter St., Madison, WI 53706 USA}
\email{placeholder@placeholder.edu} 

\author[0000-0003-2871-127X]{Jorryt Matthee}
\affiliation{Institute of Science and Technology Austria (ISTA), Am Campus 1, 3400 Klosterneuburg, Austria}
\email{placeholder@placeholder.edu}  

\author[0000-0001-8367-6265]{Tim B. Miller}
\affiliation{Center for Interdisciplinary Exploration and Research in Astrophysics (CIERA), Northwestern University, 1800 Sherman Ave, Evanston, IL 60201, USA}
\email{placeholder@placeholder.edu}  

\author[0000-0003-3729-1684]{Rohan P. Naidu}
\thanks{NHFP Hubble Fellow}
\affiliation{MIT Kavli Institute for Astrophysics and Space Research, Cambridge, MA 02139, USA}
\email{rnaidu@mit.edu}

\author[0000-0003-2804-0648]{Themiya Nanayakkara}
\affiliation{Centre for Astrophysics and Supercomputing, Swinburne University of Technology, Melbourne, VIC 3122, Australia}
\email{placeholder@placeholder.edu}  

\author[0000-0001-5851-6649]{Pascal A.\ Oesch}
\affiliation{Department of Astronomy, University of Geneva, Chemin Pegasi 51, 1290 Versoix, Switzerland}
\affiliation{Cosmic Dawn Center (DAWN), Denmark}
\affiliation{Niels Bohr Institute, University of Copenhagen, Jagtvej 128, K{\o}benhavn N, DK-2200, Denmark}
\email{placeholder@placeholder.edu}  

\author[0000-0002-0108-4176]{Sedona H. Price}
\affiliation{Space Telescope Science Institute, 3700 San Martin Drive, Baltimore, Maryland 21218, USA}
\affiliation{Department of Physics and Astronomy and PITT PACC, University of Pittsburgh, Pittsburgh, PA 15260, USA}
\email{placeholder@placeholder.edu}  

\author[0000-0002-1714-1905]{Katherine A. Suess}
\affiliation{Department for Astrophysical \& Planetary Science, University of Colorado, Boulder, CO 80309, USA}
\email{placeholder@placeholder.edu}  

\author[0000-0001-9269-5046]{Bingjie Wang}
\thanks{NHFP Hubble Fellow} \affiliation{Department of Astrophysical Sciences, Princeton University, Princeton, NJ 08544, USA}
\email{placeholder@placeholder.edu}  

\author[0000-0001-7160-3632]{Katherine E. Whitaker}
\affiliation{Department of Astronomy, University of Massachusetts, Amherst, MA 01003, USA} 
\affiliation{Cosmic Dawn Center (DAWN), Denmark}
\email{placeholder@placeholder.edu}  

\author[0000-0003-2919-7495]{Christina C. Williams}
\affiliation{NSF National Optical-Infrared Astronomy Research Laboratory, 950 North Cherry Avenue, Tucson, AZ 85719, USA}
\email{placeholder@placeholder.edu}

\begin{abstract}

At low redshift, massive quiescent galaxies (MQGs) are most frequently found in massive, rich galaxy clusters, but at high redshift the trend is less clear. Here, we present spectroscopic evidence of the effects of environment on the formation and assembly of high-redshift MQGs. We identify 25 (5) $\logmstar\geq10.5$ ($10.0\leq\logmstar<10.5$) spectroscopically-confirmed quiescent galaxies in the UDS and EGS fields at $3<z<5$ with NIRSpec PRISM spectroscopy from RUBIES and other public JWST NIRSpec programs. We measure the density contrast in these fields by applying a Monte Carlo Voronoi Tesselation density mapping technique to photometric and spectroscopic redshifts of $m_\mathrm{F444W}<27.5$ sources. We robustly detect 12 massive overdense peaks with $\logmpeak\geq13$ and six extended massive protoclusters ($\logmstruct\geq13.85$). We observe that MQGs are preferentially found in these massive peaks and within these massive structures: $\approx50\%$ of MQGs are found in massive peaks, compared to $\approx20\%$ of massive star forming galaxies (MSFGs) and $\approx15\%$ of the overall spectroscopically-confirmed population. We also find an apparent dependence on both quiescent galaxy mass and environment, with $75\%$ of the most massive ($\logmstar\geq10.75$) residing inside overdense peaks. We compare the star formation histories (SFHs) of the MQGs with the high-redshift galaxy stellar mass function from observations and simulated quiescent galaxies at $z>5$, finding that the masses from the inferred MQG SFHs regularly exceed either observed or simulated high-redshift galaxies, which suggests indicates that mergers and ex-situ star formation play a key role in the mass assembly of MQGs in overdense environments.

\end{abstract}


\section{Introduction}

The formation and evolution of massive quiescent galaxies (MQGs) at high redshift ($z>3$) represent the extremes of the universe. Though rare, their remarkable masses place them at the upper end of the galaxy stellar mass function \citep[SMF; e.g.,][]{Muzzin2013, Weaver2023a}, necessitating strong AGN feedback in order to achieve a sustained period of quiescence \citep[e.g.,][]{Man2018, Lagos2024a}. The sensitivity of this population to the exact mechanism driving the cessation of star formation leads to a frustrating difficulty reproducing the observed number densities of quiescent galaxies at $z>3$ \citep{Nanayakkara2024, Baker2025, Baker2025b, Zhang2025}.

In the era of JWST, NIRSpec's high sensitivity has given access to the previously inaccessible stellar continua of the faintest and reddest of these sources, uncovering MQGs at higher redshifts \citep{Carnall2023c, Carnall2024, DeGraaff2024, Weibel2024} and with even older ages \citep{Glazebrook2024, McConachie2025b}. These subsets of the MQG population come into conflict with theoretical models not only with their early deaths, but also by their even earlier births. Fitting the observed photometry and spectra to synthetic models of stars, gas, and dust assembled by sophisticated modeling tools often suggests that incredible amount of mass formed at extremely early cosmic times which, if formed and contained in a single dark matter halo, possibly conflicts with not only models of galaxy formation, but also potentially with fundamental $\Lambda$CDM cosmology \citep{Behroozi2018, Boylan-Kolchin2023}.

But galaxies are social creatures, not strictly isolated systems; environment has been shown to play a critical role in a galaxy's evolution, leading to galaxy properties like stellar mass and star formation rate to differ between galaxies in overdense environments versus the coeval field. 
It is long established that the properties of galaxies are not solely intrinsic, but often strongly depend on their neighbors and environments \citep{Dressler1980}. There is a clear bimodality in galaxy populations in the universe today: early type galaxies are quiescent, old, red, elliptical, and preferentially found in regions of high density, whereas late-type galaxies tend to be less massive, actively forming stars, younger, blue, with spiral morphologies, and live in lower-density environments. This leads to a clear relationship between a galaxy’s local environment and its properties, e.g., star formation rate \citep[SFR;][]{McGee2011, Wetzel2012}, stellar age \citep{Kauffmann2003}, color \citep{Strateva2001, Blanton2003, Baldry2004, Brammer2009}, and morphology \citep{VanderWel2014} which is well-understood and uncontroversial.

At higher redshifts, this bimodality breaks down, in no small part due to the nature of high-redshift ($z>3$) environments. Insufficient time has passed for protoclusters (i.e., cluster progenitors) to virialize and relax, so they often lack the hot ICM with which clusters quench star formation \citep{Gunn1972, Larson1980, Abadi1999, Balogh2000a, Balogh2000b}. While particularly massive, rich, and mature protoclusters may sport a proto-ICM \citep[][]{DiMascolo2023, Lepore2024},  this hot gas in these structures is relatively low density and contained within a small subvolume within the structure -- in other words, it cannot produce the widespread quenching seen in low-redshift clusters. Indeed, rather than the sites of quiescence, overdensities at high redshift are thought to be regions of vigorous enhanced, not suppressed, star formation \citep[e.g.,][]{Lemaux2022a, Staab2024}, fueled by inflows of pristine gas from the early cosmic web.

While the gas-rich nature of high redshift overdensities allow for more in-situ star formation, their high merger fractions provide an additional pathway for mass assembly \citep{Lotz2013, Hine2016, Watson2019, Liu2023, Shibuya2025}. At low redshifts, galaxies in the deep potential wells of virialized clusters have high velocity dispersions, which suppresses galaxy-galaxy interactions \citep{Jian2012, Alonso2012, Omori2023}. Galaxies in high redshift protoclusters, on the other hand, lack the velocity to overcome their gravity attraction to each other. This turns what would have been low-redshift near misses into high-redshift certain collisions, leading to extreme starbursts and active galactic nuclei (AGN) \citep[][]{Lehmer2009b, Lotz2013, Hine2016, Mei2023}.

 The high incidence of $z\gtrsim3$ MQGs discovered in overdense structures \citep{Kubo2021b, Kalita2021, McConachie2022, Ito2023,  Tanaka2024, Kakimoto2024, DeGraaff2024, UrbanoStawinski2024, McConachie2025b, Ito2025, Jespersen2025} or in close proximity to other massive galaxies \citep{Schreiber2018c, Turner2025, Carnall2024, Kawinwanichakij2025} suggests a link between early, rapid growth and early cosmic structure. Clearly, when considering the question  \emph{how} do high redshift MQGs form and quench, it is therefore necessary to consider the question \emph{where} do high-redshift MQGs form and quench.

In this work, we present a first systematic investigation of the environments of spectroscopically-confirmed MQGs in the UDS and EGS fields. In \S\ref{sec:data} we provide an overview of the photometric and spectroscopic data used in this analysis. We present a set of spectroscopically confirmed massive (quiescent) galaxies in \S\ref{sec:MQGs}. \S\ref{sec:env} details the methods used to quantify environment in these fields and identify prominent proto-structures. In \S\ref{sec:results} and \S\ref{sec:disc} we discuss the link between MQGs and cosmic stricture at $3<z<5$ and we summarize our main conclusions in \S\ref{sec:conc}. Throughout, a flat $\Lambda$CDM cosmology with $H_0 = 70 \ \mathrm{km \ s^{-1} \ Mpc^{-1}}$, $\Omega_m = 0.3$, and $\Omega_\Lambda = 0.7$. We utilize a Chabrier initial mass function \citep{Chabrier2003a}. All magnitudes are on the AB system \citep{Oke1983}.

\section{Data}
\label{sec:data}

First, we provide a brief overview of the datasets used in our analyses. First, we describe the data used in the analyses. In our MQG modeling (\S\ref{sec:MQGs}), we match NIRSpec MSA PRISM spectroscopy to the photometric catalogs and model both the photometry and spectroscopy. In the density mapping (\S\ref{sec:env}), we match all JWST NIRSpec MSA spectroscopy and archival spectroscopic catalogs to the photometric catalogs, using spectroscopic redshifts when available and photometric redshifts otherwise.

\subsection{Photometric Catalogs}
\label{ssec:photcat}

In this work we use PSF-matched photometric catalogs in the EGS \citep{Wright2024} and UDS \citep{Cutler2024} fields. The catalogs were created with HST and NIRCam F115W, F150W, F200W, F277W, F356W, F410M, and F444W imaging from the CEERS \citep{Finkelstein2024} and PRIMER \citep[PRIMER also has F090W imaging;][]{Donnan2024} programs. Mosaic images reduced with \texttt{grizli} \citep{grizli} were downloaded from the Dawn JWST Archive \citep[v7.2][]{Valentino2023}. Sources were detected from a combined F277W-F356W-F444W noise-equalized image and aperture photometry was extracted in circular apertures from the individual mosaics after they were degraded to F444W resolution. The full process is detailed in \citet[][]{Weaver2024}.

\subsubsection{Photometric Redshifts}
\label{sssec:photz}

We obtain photometric redshifts with the code \texttt{eazy} \citep{Brammer2008}. \texttt{eazy} matches synthetic template spectra created through linear combinations of empirical templates and finds the solution which best fits the observed photometry at every point along a redshift grid by minimizing $\chi^2$. Because \texttt{eazy} calculates the $\chi^2$ on this grid, it is then straightforward to recover a redshift probability distribution $p(z)$. We use this $p(z)$ (hereafter, the ``photometric $p(z)$'') when statistically resampling redshifts in \S\ref{sec:env}.

We use the updated template set \texttt{agn$\_$blue$\_$sfhz$\_$13}, which contains 12 galaxy templates derived from the flexible stellar population synthesis (FSPS) models \citep{Conroy2009, Conroy2010a}, an additional FSPS template made to match an extreme emission line galaxy at $z=8.5$ (\texttt{fsps$\_$4590}), and a ``Little Red Dot'' template with extreme emission lines and a red optical spectrum \citep[\texttt{j0647agn+torus}][]{Killi2024}.

\subsection{Spectroscopic Catalogs}
\label{ssec:spec}

The bulk of spectroscopy used in this work is sourced from the spectroscopic Red Unknowns: Bright Extragalactic Survey program \citep[RUBIES; GO $\#4233$][]{DeGraaff2025}. RUBIES observed 12/6 JWST/NIRSpec MSA masks in the UDS/EGS fields, obtaining spectroscopy in the NIRSpec PRISM and G395M grating. The RUBIES selection function highly weighted relatively rare red (in $\mathrm{F150W-F444W}$) and bright (in F444W) high redshift ($z_\mathrm{phot}>3$) sources.

To maximize spectroscopic coverage, we further utilize legacy spectroscopic catalogs and all publicly available NIRSpec spectra on the DJA \citep{msaexp, Heintz2024, DeGraaff2025} in the UDS and EGS fields. Arranged by program ID, we include spectra from JWST programs $\#1213, \#1215$ \citep[GTO-WIDE][]{Maseda2024}, $\#1345$ \citep[CEERS-ERS][]{Finkelstein2023, Finkelstein2024}, $\#2565$ \citep{Nanayakkara2024, Nanayakkara2025}, $\#3543$ \citep[EXCELS][]{Carnall2024}, $\#3567$ \citep[DEEPDIVE][]{Ito2025b}, $\#4106$ (PI: E. Nelson), $\#4233$ \citep[RUBIES][]{DeGraaff2025}, $\#4287$ \citep[][]{Tang2025}, $\#5224$ (Mirage or Miracle, PIs: R. Naidu \& P. Oesch), $\#6368$ (CAPERS, PI: M. Dickinson).

\subsubsection{NIRSpec Spectroscopic Redshifts}
\label{sssec:specz}

We produce two lists of spectra, one for SED modeling (\S~\ref{sec:MQGs}) and one for our density mapping procedure (\S~\ref{sec:env}). In both cases, all spectra taken from the DJA v4.4 \citep{brammer2025}. We require a robust redshift (grade=3), for the spectrum to have a match within $0.5''$ in our photometric catalogs, and for that match to be detected in F444W. Of the 4823 unique $m_\mathrm{F444W}<27.5$ matched sources, we visually examined slit placements on NIRCam mosaics for the 14 with $0.3''<r_\mathrm{match}<0.5''$. 12 of these spectra are correctly paired with sources which are extended and/or significantly off-center in the MSA slitlet. The spectrum of one faint source at $z=2.9$ in EGS is misidentified as a neighboring galaxy and one $z=4.8$ spectrum could be an extended source or a misidentified of a neighbor. These two failures in the catalog matching do not impact our results.

\textbf{Spectra for SED modeling:} For this set, we only consider PRISM spectra due to the generally high S/N and good wavelength coverage.
In the case of sources with PRISM spectra from multiple programs/masks, we rank the spectra by summing the S/N of each wavelength element (i.e., $\sum_\lambda F_\lambda/E_\lambda$, where $F$ and $E$ are the flux and error spectrum) for $3500\ \mathrm{\AA}<\lambda_\mathrm{rest}<7500\ \mathrm{\AA}$, the wavelength range which is used in our spectrophotometric modeling in \S\ref{sec:MQGs}. This effectively selects spectra with low S/N or with significant gaps in wavelength coverage. Due to this weighting, typically (but not always) spectra are given the following preference by program: $\#4106$, CAPERS, interchangeably RUBIES or CEERS depending on pointing, $\#2565$, WIDE. Little Red Dots identified in the literature \citep[e.g.,][]{Hviding2025} are removed from the list and not modeled.

\textbf{Spectra for density mapping:} Here the primary quantity we care about is the redshift and its uncertainty, so we place no restriction on spectrograph grating. For each spectrum, we re-measure the spectroscopic redshift with \texttt{msaexp} \citep{msaexp}, fitting the continuum with a 41-degree spline and the emission lines with Gaussians on a redshift grid. From the $\chi^2$ values on the redshift grid, use the same routine used with \texttt{eazy} in \S~\ref{sssec:photz} to obtain a $p(z)$ distribution from each \texttt{msaexp} spectral fit, which we later use in \S~\ref{sssec:sccat}. In the case of sources with spectra from multiple programs, masks, or gratings, we select the spectrum with the highest redshift precision, which we define as $\delta_{z} = (z_{84}-z_{16})/(2(1+z_{50}))$, where $z_{N}$ is the $N$th percentile of the spectroscopic $p(z)$ distribution. In this way, we select for narrower $p(z)$s (i.e., typically grating spectra with emission lines when available). Overall, NIRSpec redshifts are generally reliable and our results are robust to this choice, even weighting PRISM spectra over grating spectra. Of the 5318/4581 spectra in UDS/EGS, 3180/2895 are unique sources.

\subsubsection{Other Spectroscopy}
\label{sssec:otherspec}

Due to the wealth of legacy programs in these fields, we are able to supplement our list of spectra for density mapping with spectroscopic catalogs from the literature. We include the ground-based spectroscopic catalogs from the Deeper than DEEP survey \citep{Stawinski2024} in EGS and VANDELS \citep{Garilli2021} in UDS. Both of these ground-based catalogs feature abundant robust spectroscopic redshifts in the $3<z<5$ range to help fill in the gaps between NIRSpec pointing footprints. We incorporate the 3DHST catalog \citep{Skelton2014, Momcheva2016a} in our analyses for both fields. In addition to photometric redshifts and grism redshifts, the 3DHST catalog includes ground-based spectroscopic redshifts from DEEP2/3 surveys \citep[][]{Cooper2007a, Newman2013} in EGS and redshifts from a heterogeneous set of small ground-based surveys in the UDS. While vast majority of redshifts in this catalog are at $z<2$ and do not contribute directly to our $3<z<5$ sample, including them allows us to fix these sources to low redshift and prevent them from contaminating our high redshift analyses. 


\section{Massive Quiescent Galaxies}
\label{sec:MQGs}

To assemble a larger sample of massive quiescent and massive star-forming galaxies, from the list of PRISM spectra from \S\ref{ssec:spec} we select all 200 galaxies with $m_\mathrm{F444W}\leq24$ (the F444W magnitude limit used for the $2<z<5$ quiescent galaxy analysis in \citealt{Zhang2025}), and fit them with \texttt{Prospector}

\begin{figure*}[!htb]
\includegraphics[width=\linewidth]{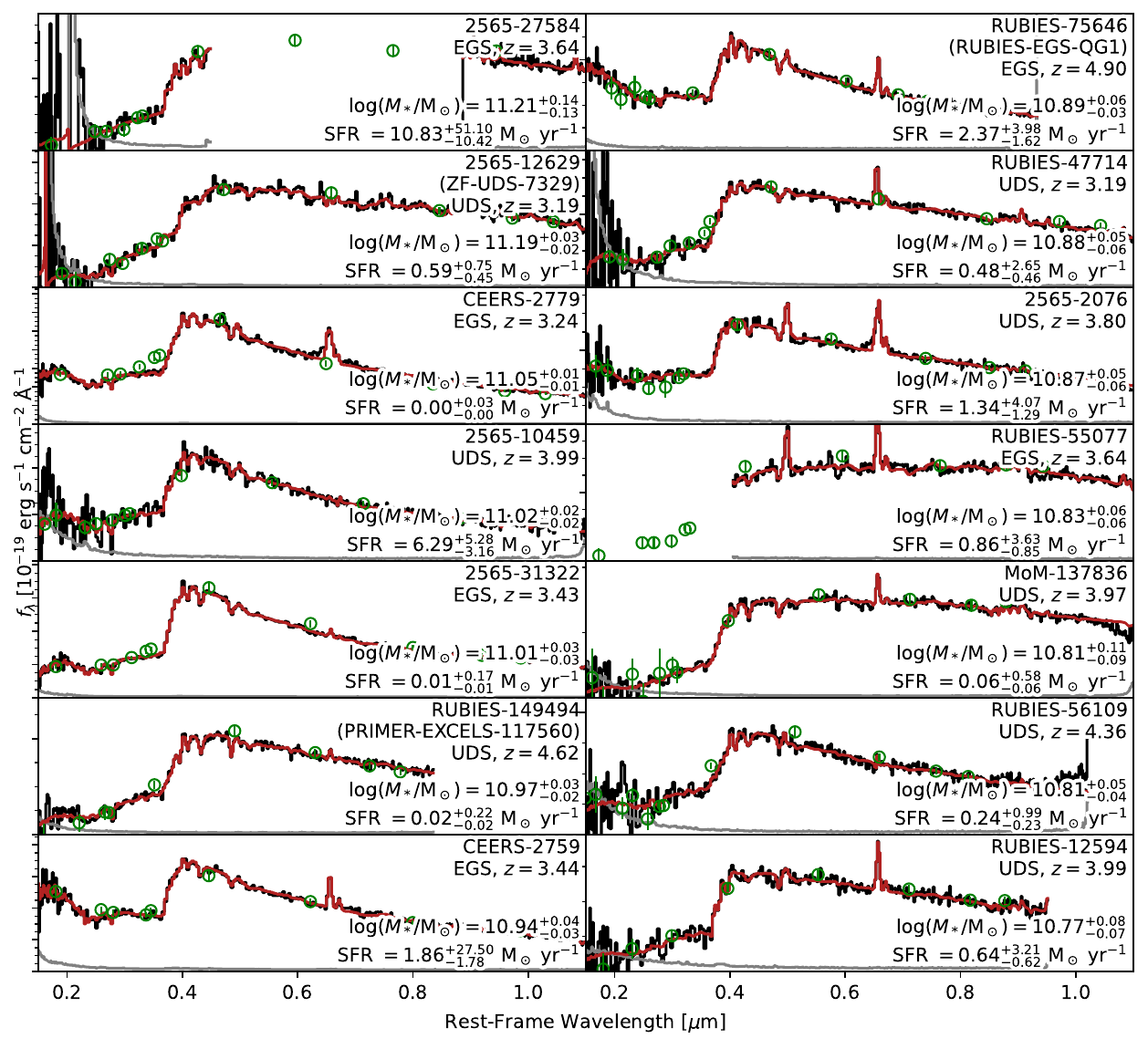}
\caption{14 of the 25 spectroscopically-confirmed $\logmstar\geq10.5$ quiescent galaxies which form our sample. In each panel the PRISM spectrum and uncertainty are shown in black and gray with the MAP \texttt{Prospector} model spectrum overlaid in red (pink for marginally quiescent galaxies). Green open circles with error bars show the photometry. To scale the observed, error, and model spectra to observed photometry, the inverted calibration vector is applied. Each major (minor) tick along the vertical axis is $(0.25\times)10^{-19}\ \mathrm{erg\ s^{-1}\ cm^{-2}\ \AA^{-1}}$. In each panel we give the spectrum ID, field, redshift, stellar mass and SFR (see also Table~\ref{tab:prospector_output}). Sources from \citet{Glazebrook2024, Carnall2024, DeGraaff2024} have their names from the literature given in parentheses; we use these names hereafter.}
\label{fig:prisms1}
\end{figure*}

\begin{figure*}[!htb]
\includegraphics[width=\linewidth]{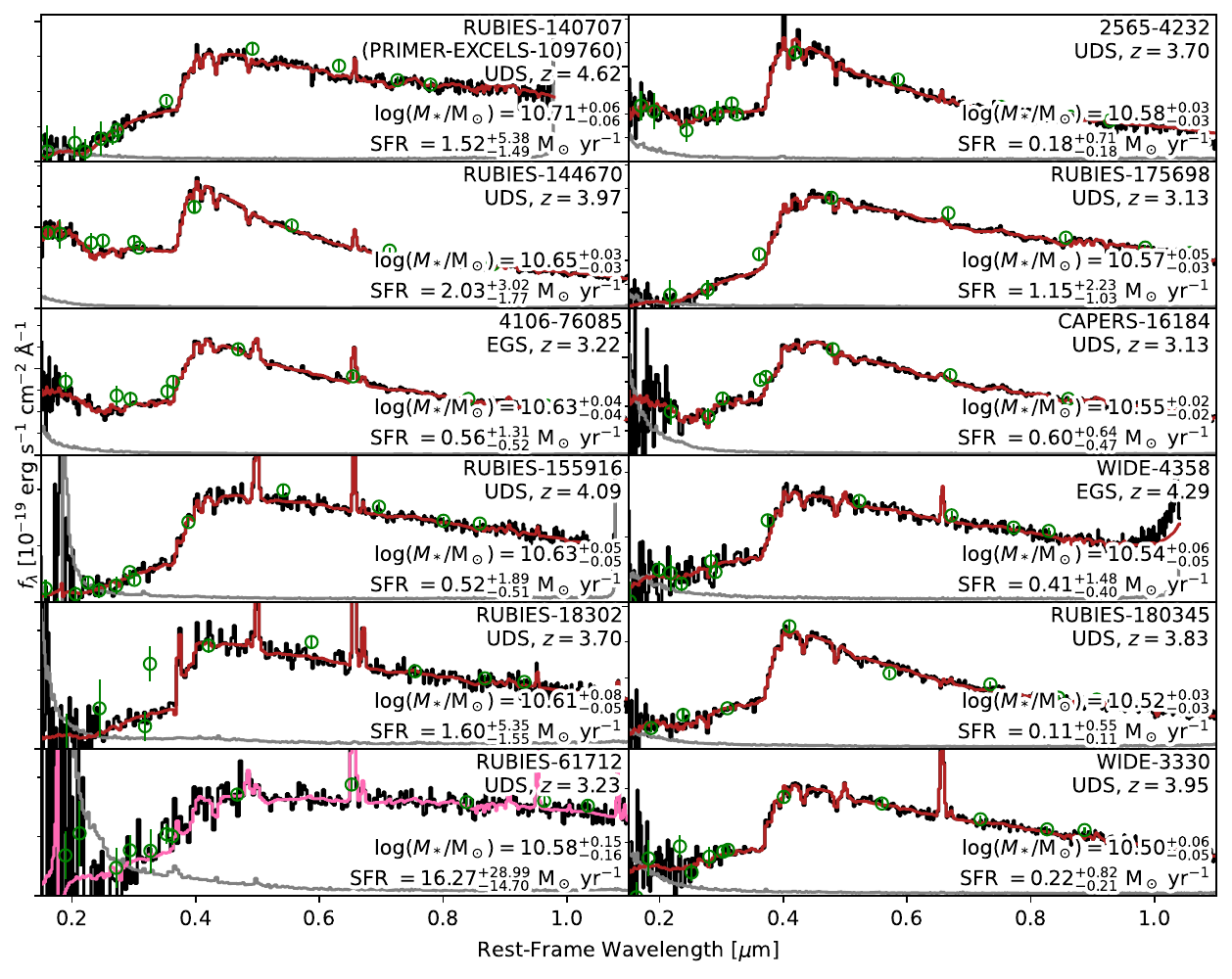}
\caption{Figure~\ref{fig:prisms1}, continued. The remaining 11 massive quiescent galaxies, plus one $\logmstar\geq10.5$ marginally quiescent galaxy (pink model).}
\label{fig:prisms2}
\end{figure*}

\begin{figure*}[!htb]
\includegraphics[width=\linewidth]{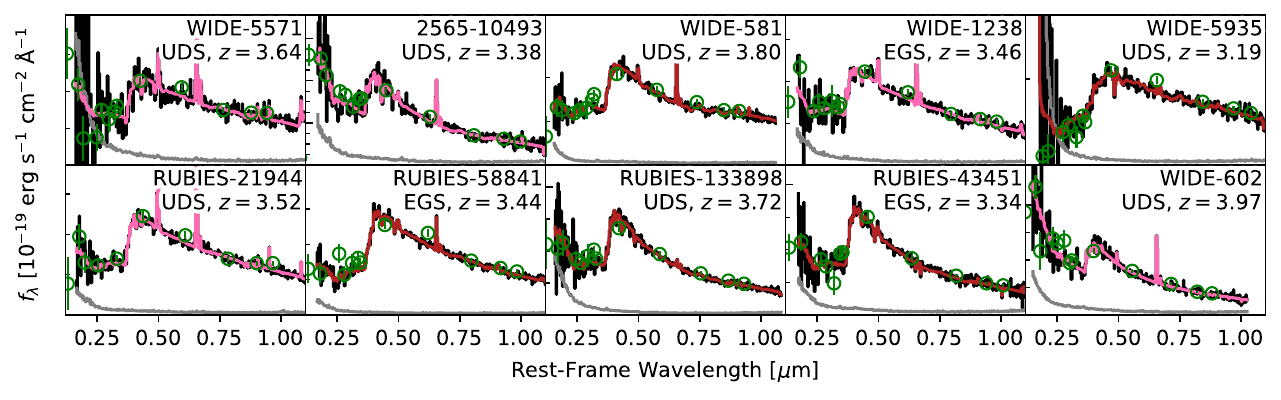}
\caption{As Figures~\ref{fig:prisms1} and \ref{fig:prisms2} but for low-mass QGs ($10\leq\logmstar<10.5$.}
\label{fig:lilprisms}
\end{figure*}

\subsection{Prospector Modeling}

We use the Bayesian inference package \texttt{Prospector} \citep{Johnson2021, Leja2017} with nested sampler \texttt{dynesty}  \citep{Speagle2020} to simultaneously model the PRISM spectra and observed $0.3-5\ \mathrm{\mu m}$ HST and JWST photometry of all 200\footnote{We exclude F814W and F606W fluxes from the fit of RUBIES-EGS-QG1 due to the presence of a diffraction spike from a nearby star in HST ACS imaging.} galaxies with $m_\mathrm{F444W} \leq 24$ in the UDS and EGS fields. We use the MILES spectral stellar library \citep{Sanchez-Blazquez2006, Falcn-Barroso2011} and MIST isochrones \citep{Choi2016, Dotter2016}. Below we detail the parameters and priors of our model. All priors are uniform/flat unless stated otherwise.
Redshift is fit in a narrow range $z_\mathrm{spec}\in [z_\mathrm{DJA} -0.05, z_\mathrm{DJA}+0.05]$, though in all cases we find excellent agreement between $ z_\mathrm{spec}$ and $z_\mathrm{DJA}$. Though F444W bright sources are expected to be preferentially more massive, we apply a wide prior to formed stellar mass $\log(M_\mathrm{*,formed}/\mathrm{M_\odot})\in[7,12]$. Metallicity is fit with a prior $\log(Z/\mathrm{Z_\odot})\in[-1,0.19]$ (i.e., a logarithmic prior $Z\in[0.1-2.5]$). For the parametrization of the model’s star formation history (SFH), we choose to use the ``continuity’’ SFH \citep{Leja2019}, which fits constant SFRs in $n$ fixed user-defined bins. We use 10 bins, spaced thus: three young bins at $[0,10]\ \mathrm{Myr}$, $[10,50]\ \mathrm{Myr}$, $[50,100]\ \mathrm{Myr}$ (i.e., bins with widths of 10, 40, and 50 Myr) to recover recent star formation, followed by seven uniformly logarithmically spaced bins in lookback time until the Big Bang. The base continuity SFH utilizes a Student’s t-distribution prior between neighboring SFH bins centered at $\Delta \log \mathrm{SFR}=0$ with $\sigma=0.3$ and $\nu = 2.0$. 
We parametrize extinction from dust with a \citet{Kriek2013} dust law with free optical depth $\tau_{\mathrm{dust}}\in[0,6]$ and deviation from the \citet{Calzetti2000a} dust law slope $\delta\in[-1, 0.4]$. We set attenuation around stars younger than 10 Myr to be twice that of older populations.

We marginalize over nebular emission lines (i.e., we fit them with simple gaussian profiles rather than tie emission to active star formation). To flux-calibrate our spectra, we scale the spectrum to the model with a seven-degree Chebyshev polynomial at each likelihood call with the \texttt{PolySpecModel} procedure.
To account for broadening from the low resolution of the NIRSpec PRISM, we convolve our models with the JDOX resolution curve, which we multiply by a factor of 1.3 \citep{Slob2024, DeGraaff2023, Nanayakkara2024}. We fit the velocity dispersion of stars and gas as two independent parameters, $\sigma_*\in[0,1000]\ \mathrm{km\ s^{-1}}$ and $\sigma_\mathrm{gas}\in[0,1000]\ \mathrm{km\ s^{-1}}$. We include a spectroscopic outlier model: the uncertainties of a fraction of spectroscopic outlier data points $f_\mathrm{out}\in[10^{-5}, 0.2]$ are inflated by a factor of five to account for bad pixels and mismatch between the model and spectra. We also fit a noise jitter term $j_\mathrm{spec} \in[0.5,3.0]$ as a factor multiplied to all spectroscopic uncertainties.

\begin{figure*}[!htb]
\centering
\includegraphics[width=\linewidth]{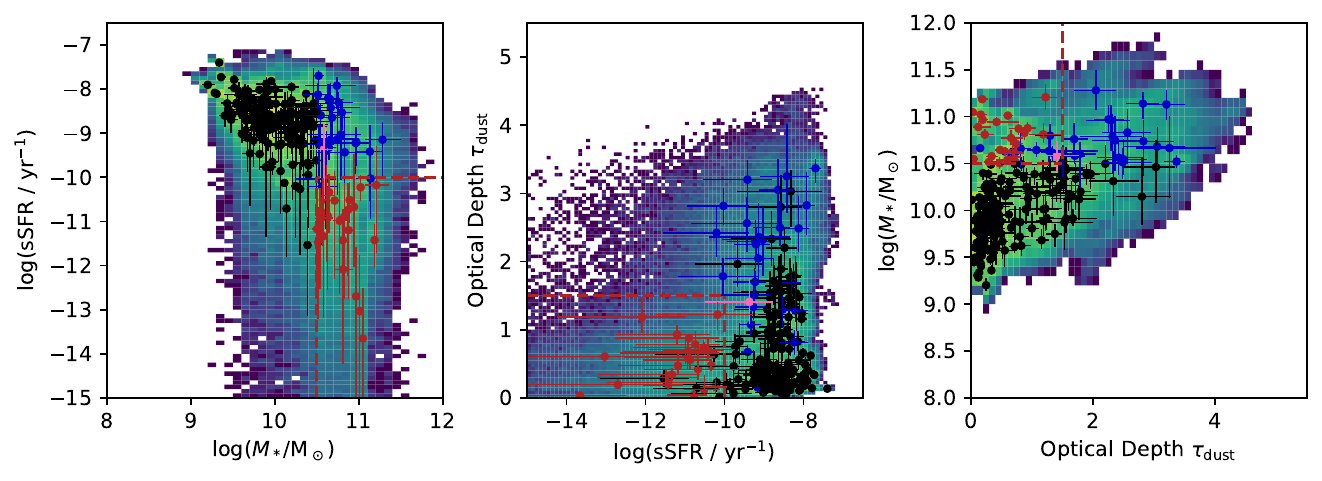}
\caption{\textbf{Left:} A 2D histogram of $\logmstar$ and $\log\mathrm{(sSFR/yr^{-1})}$ from the posterior draws of all $m_\mathrm{F444W}<24$ spectroscopically-confirmed galaxies modeled with \texttt{Prospector}. The individual galaxies and their uncertainties are indicated by black/blue/red for galaxies classified as $\logmstar<10.5$/$\logmstar\geq10.5$ star-forming/$\logmstar\geq10.5$ quiescent based on their inferred posterior medians. The red box denotes the region we use to define MQGs.  \textbf{Middle:} As left, but for  $\log\mathrm{(sSFR/yr^{-1})}$ and optical depth $\tau_{\mathrm{dust}}$. Galaxies with high $\tau_{\mathrm{dust}}$ and low sSFR appear to be dusty star-forming galaxies based on visual inspection of their spectra and SEDs. \textbf{Right:} As left, but for $\logmstar$ and dust optical depth $\tau_\mathrm{dust}$.}%
\label{fig:massdustssfr}
\end{figure*}

\begin{figure*}[!htb]
\centering
\includegraphics[width=\linewidth]{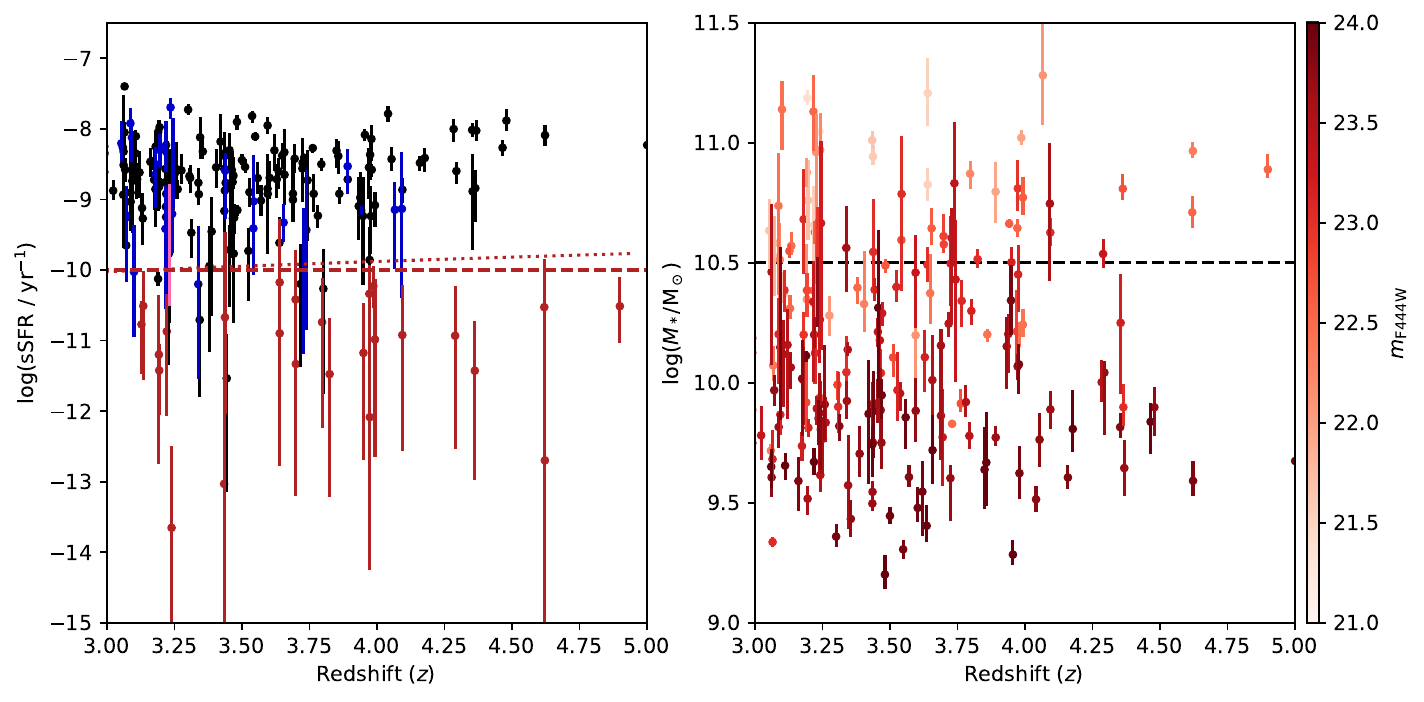}
\caption{\textbf{Left:} Specific star formation rate versus redshift for spectroscopically-confirmed galaxies modeled with \texttt{Prospector}. The dashed red line indicates the $\log (\mathrm{sSFR/yr^{-1}})<-10$ selection we use in this work and the dotted red line shows the alternative $\mathrm{sSFR < 0.2\ yr^{-1}}$ selection. Our results are robust to the choice in sSFR cut. The top histogram shows the non-normalized distribution of redshifts of all galaxies (black), massive star-forming galaxies (blue), and MQGs (red).
\textbf{Right:} Stellar mass versus redshift, with points colored by F444W magnitude to demonstrate the mass completeness limit selected.}%
\label{fig:redshift_evolution}
\end{figure*}

\subsection{Identifying Massive Quiescent Galaxies}

\citet[][]{Zhang2025} showed that $m_\mathrm{F444W}<24$ is mass complete down to $\logmstar=10.5$ for the quiescent galaxies in RUBIES. We adopt this ``massive galaxy'' mass cutoff for the analysis presented here (right panel in Figure~\ref{fig:redshift_evolution}). In Figure~\ref{fig:massdustssfr} we show how we classify galaxies as star forming or quiescent based on their inferred properties. For each galaxy, we sample the posterior inferred with \texttt{Prospector} 2000 times and take the median of the posterior samples of each galaxy's SFR/$M_*$ to be the specific star formation rate (sSFR). We require quiescent galaxies to have $\log \mathrm{(sSFR/yr^{-1})} < -10$ (using an alternative selection, e.g. $\mathrm{sSFR} < 0.2/t_\mathrm{univ}$ does not affect our results; top right of Figure~\ref{fig:redshift_evolution}). Galaxies with $\log\ \mathrm{(sSFR_{16}/yr^{-1})<-10}$ are labeled ``marginally quiescent'' \citep[as in][]{Zhang2025} and are shown in pink in Figures \ref{fig:prisms1}, \ref{fig:prisms2} \ref{fig:lilprisms}, \ref{fig:massdustssfr}, and \ref{fig:redshift_evolution}.

We also perform a cut in dust optical depth. After visually inspecting all of the PRISM spectra and models, we note that three galaxies with inferred $\logmstar\geq10.5$ and $\log (\mathrm{sSFR/yr^{-1}})<-10$ have spectra and SEDs more consistent with dusty star-forming galaxies than quiescent galaxies. The uncertainties on $\log \mathrm{sSFR/yr^{-1}}\approx1$ dex and inferred SFHs all appear flat and relatively constant with high uncertainties, as might be expected for prior-dominated fits. Though there is evidence of high dust obscuration in quiescent galaxies \citep{Setton2024, Siegel2025}, out of an abundance of caution, we assume that all dusty galaxies ($\tau_{\mathrm{dust}}\geq1.5$) are also star-forming. Three galaxies are rejected from the massive quiescent category, five are rejected from the massive marginally quiescent category, and two are rejected from the low-mass marginally quiescent category due to this selection. We show their PRISM spectra and best-fit \texttt{Prospector} models in Appendix~\ref{app:dusty}. Because only a few fits have both low sSFR and high dust optical depths, our results do not change if the $\tau_{\mathrm{dust}}<1.5$ selection is not implemented for quiescent galaxies.

We show the PRISM spectra and maximum a posteriori SED models in Figures~\ref{fig:prisms1} and \ref{fig:prisms2} for the 25 (1) PRISM spectra for sources classified as massive (marginally) quiescent galaxies (see also Table~\ref{tab:prospector_output}). The massive quiescent galaxy SEDs all feature prominent Balmer breaks and low UV flux indicative of evolved stellar populations and minimal ongoing star formation. We show $10\leq\logmstar<10.5$ quiescent and marginally quiescent galaxies along the bottom. In these instances, galaxies classified as quiescent also show strong breaks and little UV flux, but the galaxies selected as ``marginally quiescent'' exhibit notably blue UV slopes.

\begin{deluxetable*}{ccccccc}
\centerwidetable
\tablewidth{0pt}
\tablecaption{Massive Quiescent Galaxies \label{tab:prospector_output}}
\tablehead{\colhead{ID} & RA & Dec & \colhead{$z_\mathrm{spec}$} & \colhead{$\log M_*/\mathrm{M_\odot}$} & \colhead{$\mathrm{SFR_{10\ Myr}}$} & \colhead{$\ldg$}\\
\colhead{} & \colhead{[Degrees]} & \colhead{[Degrees]} & \colhead{} & \colhead{} &\colhead{$[\mathrm{M_\odot\ yr^{-1}}]$} &  \colhead{}}
\startdata
2565-EGS-27584$^{1}$ & 214.853898 & 52.861365 & $3.637^{+0.001}_{-0.001}$ & $11.21^{+0.14}_{-0.13}$  & $10.8^{+51.1}_{-10.4}$ & $0.62^{+0.36}_{-0.14}$\\
ZF-UDS-7329$^{1,2,3}$ & 34.255885 & -5.233871 & $3.196^{+0.001}_{-0.001}$ & $11.19^{+0.03}_{-0.02}$  & $0.6^{+0.7}_{-0.4}$ & $0.23^{+0.30}_{-0.19}$\\
CEERS-EGS-2779 & 214.895621 & 52.856496 & $3.253^{+0.001}_{-0.001}$ & $11.05^{+0.01}_{-0.01}$  & $0.0^{+0.0}_{-0.0}$ & $0.37^{+0.12}_{-0.14}$\\
2565-UDS-10459$^{1,3}$ & 34.340345 & -5.241309 & $3.970^{+0.002}_{-0.002}$ & $11.02^{+0.02}_{-0.02}$  & $6.3^{+5.3}_{-3.2}$ & $0.97^{+0.05}_{-0.73}$\\
2565-EGS-31322$^{1, 6, 7}$ & 214.866054 & 52.884256 & $3.436^{+0.001}_{-0.001}$ & $11.01^{+0.03}_{-0.03}$  & $0.0^{+0.2}_{-0.0}$ & $0.44^{+0.27}_{-0.11}$\\
PRIMER-EXCELS-117560$^{3,4}$ & 34.399676 & -5.136348 & $4.630^{+0.002}_{-0.002}$ & $10.97^{+0.03}_{-0.02}$  & $0.0^{+0.2}_{-0.0}$ & $0.40^{+0.24}_{-0.25}$\\
CEERS-EGS-2759 & 214.871231 & 52.845066 & $3.439^{+0.002}_{-0.002}$ & $10.94^{+0.04}_{-0.03}$  & $1.9^{+27.5}_{-1.8}$ & $0.48^{+0.20}_{-0.21}$\\
RUBIES-EGS-QG1$^{4,5}$ & 214.915546 & 52.949018 & $4.896^{+0.001}_{-0.001}$ & $10.89^{+0.06}_{-0.03}$  & $2.4^{+4.0}_{-1.6}$ & $0.69^{+0.20}_{-0.16}$\\
RUBIES-UDS-47714$^{1,4}$ & 34.258910 & -5.232334 & $3.198^{+0.001}_{-0.001}$ & $10.88^{+0.05}_{-0.06}$  & $0.5^{+2.7}_{-0.5}$ & $0.19^{+0.28}_{-0.16}$\\
2565-UDS-2076$^{1}$ & 34.289492 & -5.269849 & $3.799^{+0.001}_{-0.002}$ & $10.87^{+0.05}_{-0.06}$  & $1.3^{+4.1}_{-1.3}$ & $0.78^{+0.31}_{-0.15}$\\
RUBIES-EGS-55077$^{4}$ & 214.853899 & 52.861365 & $3.641^{+0.001}_{-0.001}$ & $10.83^{+0.06}_{-0.06}$  & $0.9^{+3.6}_{-0.8}$ & $0.62^{+0.36}_{-0.14}$\\
MoM-UDS-137836$^{4}$ & 34.360159 & -5.153092 & $3.982^{+0.001}_{-0.001}$ & $10.81^{+0.11}_{-0.09}$  & $0.1^{+0.6}_{-0.1}$ & $1.38^{+0.21}_{-0.00}$\\
RUBIES-UDS-56109$^{4}$ & 34.280515 & -5.217214 & $4.385^{+0.003}_{-0.005}$ & $10.81^{+0.05}_{-0.04}$  & $0.2^{+1.0}_{-0.2}$ & $0.59^{+0.26}_{-0.36}$\\
RUBIES-UDS-12594$^{4}$ & 34.368535 & -5.299475 & $3.994^{+0.002}_{-0.002}$ & $10.77^{+0.08}_{-0.07}$  & $0.6^{+3.2}_{-0.6}$ & $0.39^{+0.19}_{-0.24}$\\
PRIMER-EXCELS-109760$^{3,4}$ & 34.365084 & -5.148848 & $4.619^{+0.001}_{-0.001}$ & $10.71^{+0.06}_{-0.06}$  & $1.5^{+5.4}_{-1.5}$ & $0.50^{+0.17}_{-0.10}$\\
MoM-UDS-144670 & 34.242593 & -5.143121 & $3.974^{+0.001}_{-0.001}$ & $10.65^{+0.03}_{-0.03}$  & $2.0^{+3.0}_{-1.8}$ & $0.48^{+0.31}_{-0.24}$\\
4106-EGS-76085 & 214.836844 & 52.873457 & $3.221^{+0.001}_{-0.001}$ & $10.63^{+0.04}_{-0.04}$  & $0.6^{+1.3}_{-0.5}$ & $0.19^{+0.23}_{-0.18}$\\
RUBIES-UDS-155916$^{4}$ & 34.317031 & -5.127611 & $4.098^{+0.002}_{-0.002}$ & $10.63^{+0.05}_{-0.05}$  & $0.5^{+1.9}_{-0.5}$ & $0.10^{+0.21}_{-0.22}$\\
RUBIES-UDS-18302$^{4}$ & 34.233628 & -5.283850 & $3.703^{+0.002}_{-0.002}$ & $10.61^{+0.08}_{-0.05}$  & $1.6^{+5.3}_{-1.5}$ & $0.41^{+0.26}_{-0.30}$\\
RUBIES-UDS-61712$^{4}$ & 34.240916 & -5.205573 & $3.236^{+0.001}_{-0.001}$ & $10.58^{+0.15}_{-0.16}$  & $16.3^{+29.0}_{-14.7}$ & $0.14^{+0.20}_{-0.21}$\\
2565-UDS-4232$^{1}$ & 34.290455 & -5.262115 & $3.718^{+0.003}_{-0.003}$ & $10.58^{+0.03}_{-0.03}$  & $0.2^{+0.7}_{-0.2}$ & $0.18^{+0.25}_{-0.23}$\\
RUBIES-UDS-175698$^{4}$ & 34.227642 & -5.099286 & $3.142^{+0.001}_{-0.002}$ & $10.57^{+0.05}_{-0.03}$  & $1.2^{+2.2}_{-1.0}$ & $0.13^{+0.28}_{-0.22}$\\
CAPERS-UDS-16184 & 34.513656 & -5.157829 & $3.123^{+0.002}_{-0.002}$ & $10.55^{+0.02}_{-0.02}$  & $0.6^{+0.6}_{-0.5}$ & $0.09^{+0.33}_{-0.27}$\\
WIDE-EGS-4358 & 215.039079 & 53.002774 & $4.283^{+0.003}_{-0.003}$ & $10.54^{+0.06}_{-0.05}$  & $0.4^{+1.5}_{-0.4}$ & $-0.03^{+0.29}_{-0.27}$\\
RUBIES-UDS-180345$^{4}$ & 34.209839 & -5.091602 & $3.830^{+0.003}_{-0.003}$ & $10.52^{+0.03}_{-0.03}$  & $0.1^{+0.6}_{-0.1}$ & $0.10^{+0.20}_{-0.23}$\\
WIDE-UDS-3330 & 34.338199 & -5.166202 & $3.937^{+0.002}_{-0.002}$ & $10.50^{+0.06}_{-0.05}$  & $0.2^{+0.8}_{-0.2}$ & $0.11^{+0.27}_{-0.16}$\\
\hline
\hline
WIDE-UDS-5571 & 34.437763 & -5.226583 & $3.644^{+0.001}_{-0.001}$ & $10.49^{+0.07}_{-0.07}$  & $7.5^{+8.8}_{-5.6}$ & $0.17^{+0.23}_{-0.18}$\\
RUBIES-UDS-21944 & 34.469218 & -5.283563 & $3.528^{+0.001}_{-0.001}$ & $10.40^{+0.06}_{-0.06}$  & $4.6^{+5.2}_{-3.6}$ & $0.68^{+0.37}_{-0.27}$\\
2565-UDS-10493 & 34.274382 & -5.241174 & $3.390^{+0.002}_{-0.002}$ & $10.40^{+0.04}_{-0.06}$  & $2.9^{+6.5}_{-2.7}$ & $0.21^{+0.29}_{-0.21}$\\
RUBIES-EGS-58841 & 214.879098 & 52.888065 & $3.451^{+0.002}_{-0.002}$ & $10.39^{+0.04}_{-0.04}$  & $0.1^{+0.2}_{-0.1}$ & $1.20^{+0.02}_{-0.00}$\\
WIDE-UDS-581 & 34.282266 & -5.244876 & $3.805^{+0.002}_{-0.003}$ & $10.30^{+0.04}_{-0.05}$  & $1.1^{+2.8}_{-1.0}$ & $0.51^{+0.21}_{-0.22}$\\
RUBIES-UDS-133898 & 34.485164 & -5.157813 & $3.704^{+0.003}_{-0.003}$ & $10.25^{+0.04}_{-0.05}$  & $1.1^{+2.6}_{-1.0}$ & $0.16^{+0.25}_{-0.14}$\\
WIDE-EGS-1238 & 215.011053 & 52.908065 & $3.455^{+0.002}_{-0.003}$ & $10.21^{+0.05}_{-0.05}$  & $3.1^{+3.9}_{-2.6}$ & $0.31^{+0.31}_{-0.18}$\\
RUBIES-EGS-43451 & 214.909553 & 52.875028 & $3.340^{+0.002}_{-0.003}$ & $10.14^{+0.03}_{-0.02}$  & $0.3^{+0.6}_{-0.3}$ & $0.27^{+0.26}_{-0.20}$\\
WIDE-UDS-5935 & 34.254473 & -5.234889 & $3.207^{+0.003}_{-0.009}$ & $10.11^{+0.01}_{-0.02}$  & $1.0^{+0.2}_{-0.2}$ & $0.20^{+0.30}_{-0.18}$\\
WIDE-UDS-602 & 34.467800 & -5.244913 & $3.965^{+0.007}_{-0.008}$ & $10.07^{+0.05}_{-0.05}$  & $1.6^{+3.0}_{-1.5}$ & $0.07^{+0.26}_{-0.23}$\\
\enddata
\tablecomments{Massive quiescent and marginally quiescent galaxies ordered by $\logmstar$. Survey identifiers are assigned based on the spectrum used for SED modeling. For the prominent MQGs featured in \citet{Glazebrook2024, Carnall2024, DeGraaff2024} we use the names given in those works (PRISM spectra for the \citealt{Carnall2024} sources were taken in RUBIES; see Figure~\ref{fig:prisms1}), otherwise MQG IDs are given as [survey]-[source ID] or [program ID]-[source ID] if a survey name is not available.
$^{1}$ \citet{Nanayakkara2024, Nanayakkara2025};
$^{2}$ \citet{Glazebrook2024};
$^{3}$ \citet{Carnall2024};
$^{4}$ \citet{Zhang2025};
$^{5}$ \citet{DeGraaff2024};
$^{6}$ \citet{Ito2025};
$^{7}$ \citet{Jin2024a}.}
\end{deluxetable*}

\section{Quantifying Environment}
\label{sec:env}

In the era of JWST, both the number and maximum redshift of proto-structures and proto-structure candidates have increased dramatically. Many of these were discovered by looking for overdensities using redshifts from photometry \citep{Pan2025},
slit spectroscopy \citep{Morishita2023b, Carnall2024, DeGraaff2024, Napolitano2025},
slitless spectroscopy \citep{Helton2024, Champagne2025, Champagne2025a, Lin2025, Fudamoto2025}, or some combination thereof \citep{Laporte2022, Jin2024a, Helton2024a, UrbanoStawinski2024}. 

\subsection{Overdensity Maps}

To recover the underlying cosmic density field, we employ a Voronoi tesselation method, applied to both spectroscopy and photometry. \citet{Darvish2015} first introduced the ``weighted Voronoi tesselation estimator'' among a collection of other density estimator methods. Of all methods presented, it was shown that the Voronoi method and the weighted adaptive kernel method best recovered the true underlying density. While weighted kernel density estimators have been used to measure environments in other studies \citep[][]{Chartab2020, McConachie2022, UrbanoStawinski2024, McConachie2025a}, they are sensitive to the size and form of the kernel and are typically spatially symmetric, which can be disadvantageous when trying to recover the shapes of often asymmetric and ``lumpy'' proto-structures \citep[e.g.,][]{Muldrew2015, Remus2023}. Conversely, a Voronoi tesselation requires no size calibration and assumes no underlying geometry. The framework for our Voronoi tesselation method was based on and most closely resembles the ``Voronoi Monte Carlo'' (VMC) method first presented in \citet{Lemaux2018}, which has since been subsequently utilized in e.g., \citet{Tomczak2017a, Cucciati2018, Hung2020, Shen2021, Lemaux2022a, Forrest2023, Shah2024, Hung2024}. We provide a brief summary in the following paragraph; those interested in the specifics can find a detailed version in Appendix~\ref{app:mcvt}.

First, we produce cartesian grids of the UDS and EGS fields in a transformed coordinate plane such that the fields are equatorial (RA$^\prime$ and Dec$^\prime$; EGS is also rotated by 49 degrees counter-clockwise in the prime frame). We resample redshifts from $p(z)$ distributions, using spectroscopic data when available and photometric redshifts otherwise. For a given target redshift, we retain galaxies with redshifts lying within a narrow slice with line-of-sight width equal to 10 cMpc and centered on the target redshift. We partition the field grid using a Voronoi tesselation. The pixels in each cell are assigned values equal to the inverse of the entire cell's area (i.e., the cell's galaxy surface density $\Sigma_\mathrm{gal}$). We perform 100 iterations and take the median, 16th, and 84th percentile values of each pixel to be the density and its associated lower and upper uncertainties. Galaxy surface densities are converted into galaxy overdensities $\delta_\mathrm{gal}=\Sigma_\mathrm{gal}/\overline{\Sigma}_\mathrm{gal}$.

\begin{figure*}[!htb]
\includegraphics[width=\linewidth]{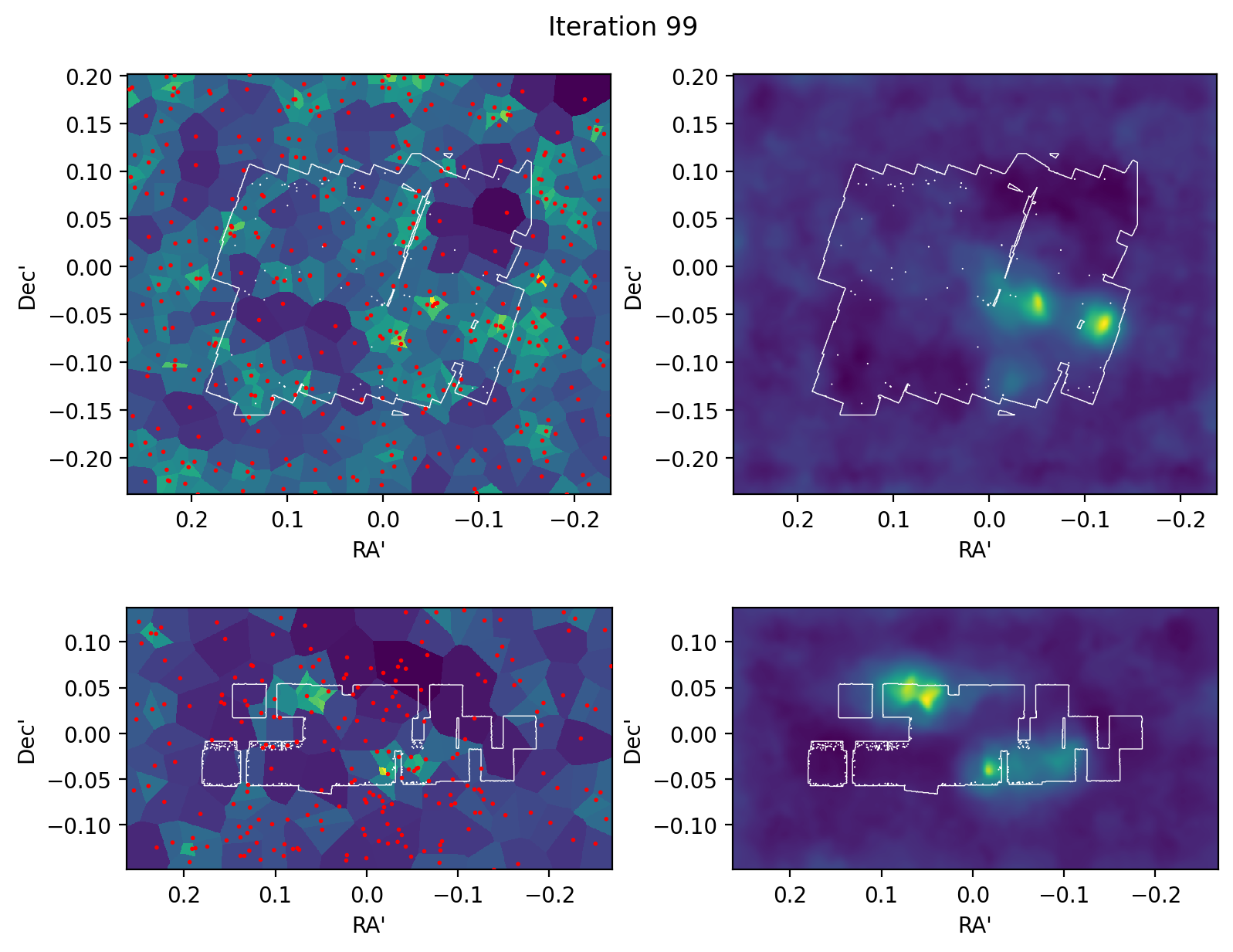}
\caption{The MCVT process at $z=3.193$ in UDS (top) and $z=4.891$ in EGS (bottom). The left panels show a single iteration of the MCVT process. Red points are the coordinates of galaxies which fall into the redshift slice in the given iteration. Each cell is colored according to the inverse of its area. Panels on the right show the stacked medians of all 100 iterations, smoothed by a gaussian filter with a width of 2.5 pixels. In both panels the white contour shows the outline for the F444W footprints and the color density scaling is left arbitrary. The fields are shown in the RA$^\prime$ and Dec$^\prime$ coordinate frame. An animated version showing the construction of the median image for all 100 iterations is available in the online journal. }
\label{fig:MCVT}
\end{figure*}

We refer to this method as Monte Carlo Voronoi Tesselation, or MCVT for short; similar to Voronoi Monte Carlo mapping \citep[VMC; e.g.,][]{Lemaux2018} but distinct in setup and execution. In Figure~\ref{fig:MCVT} we show a single iteration on the left and the stacked median on the right (in the online journal an animation shows all iterations and the development of the median image). Red points denote the location of the redshift slice members for a given iteration and the field grid pixels are colored (with arbitrary scaling) by their $\log \Sigma_{\mathrm{gal}}$. White contours indicate the boundaries of the F444W footprint. With successive iterations, structures which are not detectable in individual iteration maps begin to appear in the median image.

\begin{figure*}[!htb]
\includegraphics[width=\linewidth]{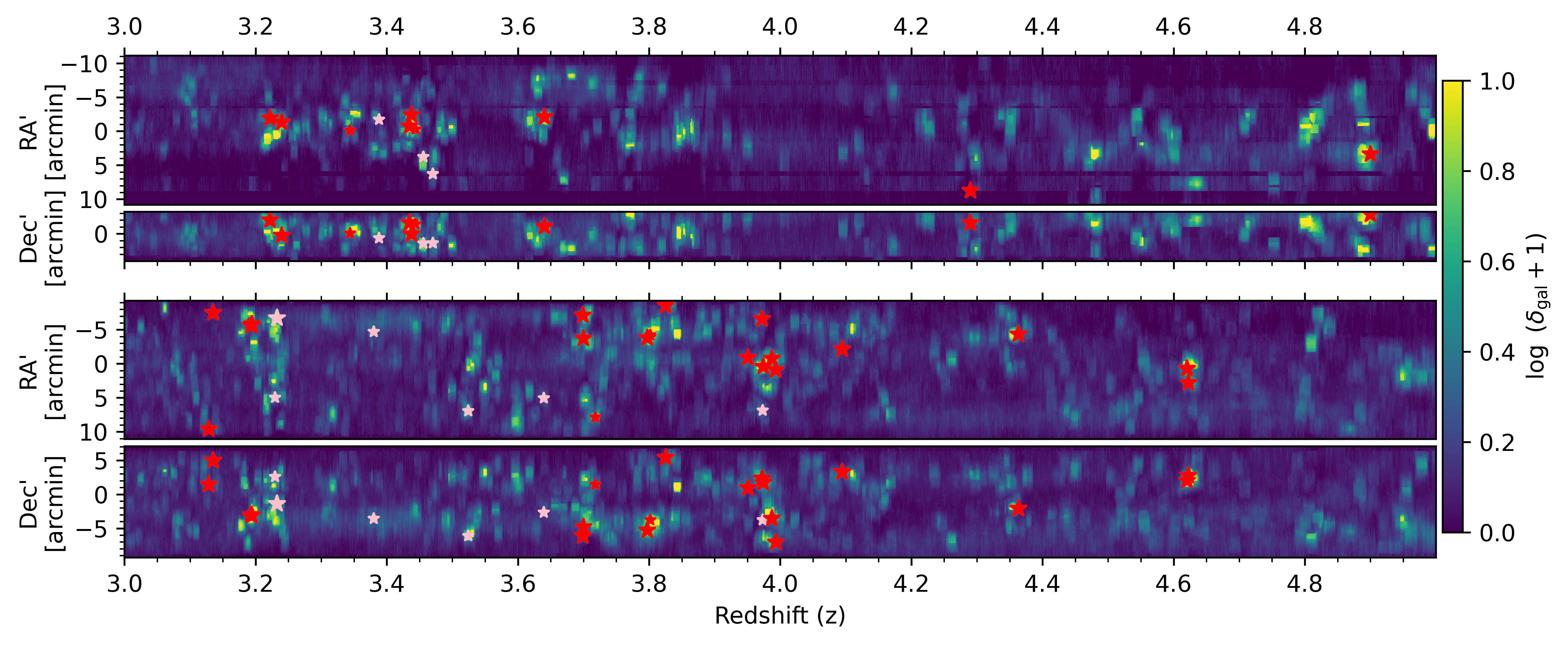}
\caption{Cross-sections of the $3<z<5$ density grids in the EGS (top) and UDS (bottom) with redshift along the x-axis and \decp\ and \rap\ on the y-axes. The locations of the quiescent galaxies from \S\ref{sec:MQGs} are indicated by larger red stars and marginally quiescent galaxies by pink stars. Smaller stars show the positions of $10\leq\logmstar<10.5$ (marginally) quiescent galaxies. Several previously identified overdensities are immediately recognizable: in UDS, the $z=3.2$ overdensity in UDS around ZF-UDS-7329 \citep{Turner2025, Kawinwanichakij2025, Jespersen2025}, the $z=4.6$ overdensity around PRIMER-EXCELS-117560 and PRIMER-EXCELS-109760 \citep{Carnall2024, Jespersen2025}; in EGS, the Cosmic Vine at $z=3.4$ \citep{Jin2024a, Ito2025} and the overdensity around the $z=4.9$ quiescent galaxy RUBIES-EGS-QG1 \citep{DeGraaff2024, UrbanoStawinski2024}. Remarkably, almost all spectroscopically-confirmed quiescent galaxies seem to be found in or near an overdensity.}
\label{fig:crosssec}
\end{figure*}

With the ability to produce galaxy overdensity maps of the UDS and EGS fields at any redshift $z_\mathrm{slice}$, we then walk through redshift space from $3\leq z_\mathrm{slice}\leq5$ in 1 cMpc steps, performing the MCVT at each redshift step. Indexing $z_\mathrm{slice}$ by $k$, for each field we construct a three-dimensional density grid of voxels with values $\delta_\mathrm{gal}^{i,j,k}$, where $i$, $j$, and $k$ are tied to locations in \rap, \decp, and redshift $z$ space. We show cross-sections of the density grids in Figure~\ref{fig:crosssec} with the top pair showing UDS and the bottom pair showing EGS. Of each pair, the upper plot shows overdensity in \rap\ and redshift maximized over \decp\ and the lower shows overdensity in \decp\ and redshift maximized over \rap\ (i.e., $\delta_\mathrm{gal}^{i,j}$ and $\delta_\mathrm{gal}^{i,k}$, respectively). The locations of MQGs are shown by large red stars and massive marginally quiescent galaxies by pink large stars. Galaxies with $10\leq\logmstar<10.5$ classified as quiescent or marginally quiescent are shown by smaller stars with the same coloring scheme. Remarkably, the MQG population appears to coincide strongly with regions of high overdensity in the galaxy maps. We explore this trend in greater detail in \S\ref{sec:results}.

\subsection{Structure Identification}
\label{ssec:struct}

With our three-dimensional RA-Dec-$z$ density grids in hand, we next identify prominent density peaks and surrounding overdense structures. First, we confirm that $\mathrm{median}\ \ldg=0$, and iteratively fit a $3\sigma$ clipped Gaussian centered at zero to $\ldg$. We take the dispersion of this Gaussian to be $\sigma_{\ldg}$. 

\subsubsection{Overdense Peaks}
\label{sssec:peaks}

To identify overdense peaks, we isolate all voxels with $\ldg>5\sigma_{\ldg}$ (hereafter ``peak voxels''). We use the clustering algorithm DBSCAN in the \texttt{scikit-learn} python package to produce groups of contiguous peak voxels. Each group is flagged as an individual overdense peak. For each peak, we convert the galaxy overdensity to a matter density, which we integrate over the volume of the peak to estimate a peak mass:

\begin{equation}
M_\mathrm{peak} =  \sum\limits_{i,j,k}^\mathrm{Peak} M_{i,j,k} =  \sum\limits_{i,j,k}^\mathrm{Peak} V_{i,j,k} \rho_m(1+\frac{\delta_{gal}^{i,j,k}}{b(z_k)})
\end{equation}

where $V_{i,j,k}$ is the comoving volume of each peak voxel with index $i,j,k$ (because each $z$-slice is 10 cMpc thick and we take 1 cMpc steps in $z$, we set the line-of-sight depth of every voxel to be 1 cMpc), $\rho_m$ is the comoving matter density of the Universe, and $b(z_k)$ is the redshift-dependent galaxy bias parameter which relates galaxy density the underlying matter density distribution $b = \delta_\mathrm{gal}/ \delta_\mathrm{m}$ taken from \citep{BaroneNugent2014}.

We rank the robustness of peaks by their mass and galaxy membership. We consider peaks with $10^{12}\leq M_\mathrm{peak}<10^{13}\mathrm{M_\odot}$ to be low confidence overdensity detections (hereafter, minor peaks or subpeaks). Upon visual inspection these minor peaks often correspond with low overdensity over a large volume or moderate overdensity in a small volume with $\lesssim3$ spectroscopic members, and are often located near field edges where density measurements are closer to the median and highly uncertain.
Peaks with $M_\mathrm{peak}\geq10^{13}\mathrm{M_\odot}$ are taken to be moderate confidence overdensity detections, they represent regions with high galaxy density over a large volume. For a peak to be considered a high confidence overdensity detection (hereafter a ``robust overdensity''), we require $M_\mathrm{peak}\geq10^{13}\mathrm{M_\odot}$ and the peak to contain at least one $\logmstar\geq10.5$ galaxy, as an indication that the galaxy number overdensity also traces a large mass. In future works we will calibrate this peak selection method and explore how $M_\mathrm{peak}$ relates to $M_\mathrm{halo}$ in comparisons with simulations.

For each moderate and robust peak we calculate a barycenter position \citep[as in e.g.,][]{Cucciati2018, Shen2021, Forrest2023, Shah2024}. We convert each voxel's coordinates $\mrap_i, \mdecp_j, z_k$ into comoving coordinates $X_i, Y_j, Z_k$. For each comoving coordinate $C$ and associated index $m$, the barycenter comoving coordinate is given by $C_\mathrm{bary}=\sum_m(C_m \delta_\mathrm{gal}^m) / \sum_m \delta_\mathrm{gal}^m$ (i.e., the overdensity-weighted average of comoving coordinate $C$). Robust peaks are assigned identifiers [field]-Peak-z[redshift] and moderate density peaks are assigned [field]-cPeak-z[redshift], where cPeak indicates that these are candidate overdensities that will require further analysis for confirmation. We give peak IDs, barycenter locations in RA, Dec, and redshift, peak mass and volume, and the number of massive and NIRSpec-confirmed members in Table~\ref{tab:peaks} for robust peaks. The same quantities for moderate confidence peaks (i.e., candidate overdensities) are given in Appendix~\ref{app:candidates}, minus the number of massive spectroscopically confirmed members.

\begin{deluxetable*}{cccccccc}
\centerwidetable
\tablewidth{0pt}
\tablecaption{Robust Overdensity Peaks \label{tab:peaks}}
\tablehead{\colhead{ID} & \colhead{$z_\mathrm{Bary}$}& \colhead{RA$_\mathrm{Bary}$} & \colhead{Dec$_\mathrm{Bary}$} & \colhead{$\logmpeak$} & \colhead{Volume} & \colhead{$n_{\mathrm{Massive}}$} & \colhead{$n_\mathrm{NIRSpec}$}\\
& & \colhead{[Degrees]} & \colhead{[Degrees]} & & \colhead{cMpc$^{3}$}}
\startdata
UDS-Peak-z3.2-1 & 3.193 & 34.236749 & -5.242305 & 13.0 &  130 &  1 & 5\\
UDS-Peak-z3.2-2 & 3.228 & 34.267254 & -5.238437 & 13.6 &   480&  1 & 17\\
UDS-Peak-z3.5 & 3.532 & 34.344261 & -5.260252 & 13.4 &351 	&  1 & 8\\  
UDS-Peak-z3.7 & 3.703 & 34.267005 & -5.240434 & 13.7 &644 	&  1 & 8\\  
UDS-Peak-z3.8 & 3.801 & 34.289024 & -5.259379 & 13.9 & 1181&  1 & 14\\  
UDS-Peak-z4.0-1 & 3.978 & 34.351154 & -5.188871 & 13.8 &   874&  2 & 22\\
UDS-Peak-z4.0-2 & 3.983 & 34.399437 & -5.294207 & 13.2 &  253 &  1 & 9\\
UDS-Peak-z4.6 & 4.625 & 34.365338 & -5.143818 & 13.5 & 474 &  2 & 13\\  
EGS-Peak-z3.2 & 3.228 & 214.907479 & 52.883765 & 13.7 &   603&  2 & 30\\
EGS-Peak-z3.4 & 3.438 & 214.875399 & 52.877572 & 13.6 &   431&  3 & 37\\
EGS-Peak-z3.6 & 3.640 & 214.879960 & 52.868047 & 13.2 &  222&  2 & 9\\  
EGS-Peak-z4.9 & 4.894 & 214.925068 & 52.946404 & 13.6 &   448&  1 & 16\\
\hline
\hline
& & & & & & \\
\enddata
\tablecomments{We include a table of candidate overdensities (i.e., moderate confidence overdensity peaks) in Appendix~\ref{app:candidates}}
\end{deluxetable*}

\subsubsection{Protocluster Identification}
\label{sssec:plss}

\begin{figure*}[!htb]
\begin{center}
\includegraphics[width=0.6\linewidth]{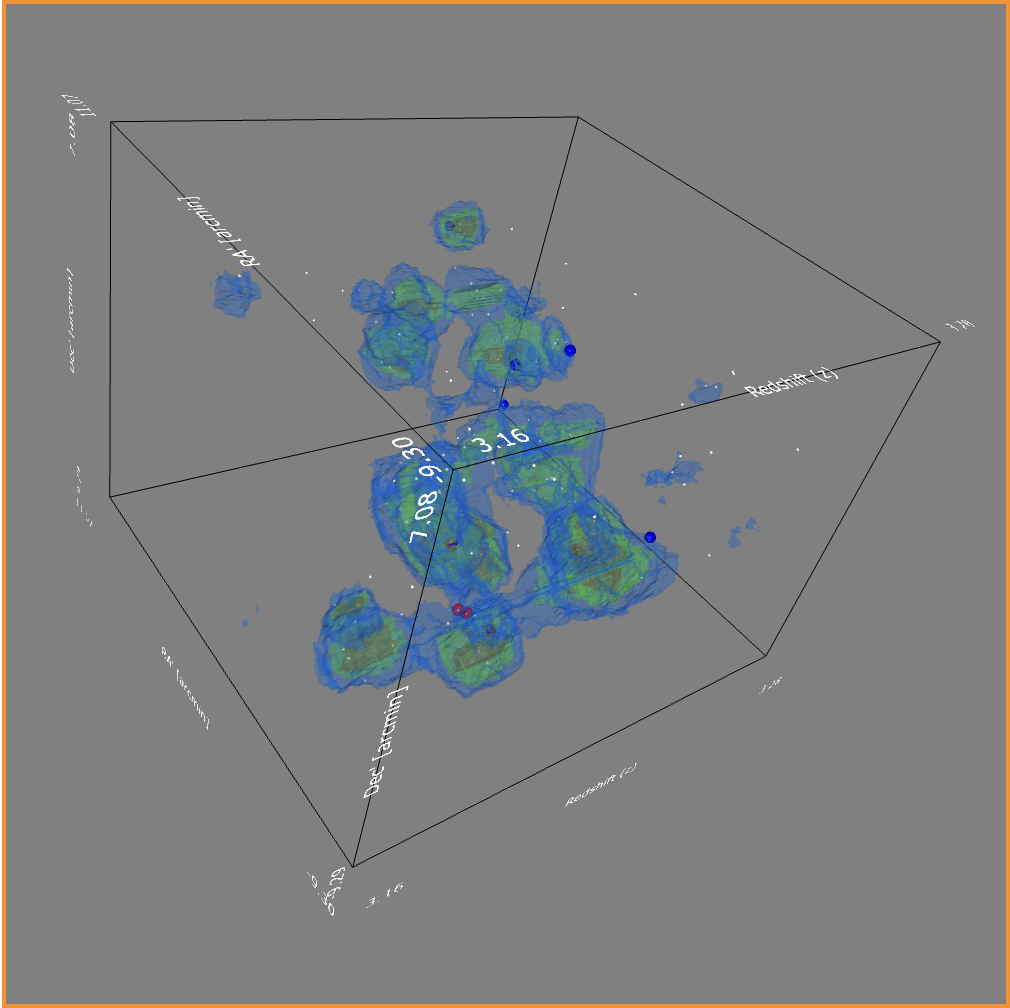}
\end{center}
\caption{An isometric view of UDS-PCl-z3.2 from the MCVT mapping. We show $3/5/8\sigma$ contours as blue/green/red surfaces. MQGs are shown as red spheres, MSFGs as blue spheres, and all other spectroscopically confirmed galaxies as small white spheres. An interactive version of this figure, including all protoclusters in Table~\ref{tab:struct} plus UDS-Peak-z4.6 and EGS-Peak-z4.9, will be available in the online journal via the X3D-pathway \citep{Vogt2016}.}
\label{fig:noninteractive}
\end{figure*}

Simulations show that the galaxies which are found in $z=0$ clusters and other cosmic structures are found spread across tens of comoving megaparsecs in the high redshift universe \citep{Chiang2013, Muldrew2015, Remus2023}. To identify potential extended large-scale structure around the overdense peaks in \S\ref{sssec:peaks}, we lower the density threshold to $3\sigma_{\ldg}$, taking all voxels above this threshold to be ``structure voxels.'' As we did for the peaks, we identify contiguous groups of structure voxels and calculate masses and barycenters of structures. Because these extended structures are extended and lumpy, we do not attempt to correct for elongation along the line of sight. Structures containing a robust peak detection and with estimated mass in excess of that of the Fornax Cluster \citep[$M_\mathrm{struct} > 10^{13.85}$;][]{Drinkwater2001} are labeled ``protoclusters.'' Identifiers for new protoclusters are assigned as PCl-[field]-z[redshift], otherwise the name from the literature is given. We show the properties of these protoclusters in Table~\ref{tab:struct} and a 3D plot of PCl-UDS-z3.2 in Figure~\ref{fig:noninteractive} (an interactive version with all protoclusters is available in the online journal).

\begin{deluxetable*}{cccccccccc}
\centerwidetable
\tablewidth{0pt}
\tablecaption{Protoclusters \label{tab:struct}}
\tablehead{\colhead{ID} & \colhead{$z_\mathrm{Bary}$}& \colhead{RA$_\mathrm{Bary}$} & \colhead{Dec$_\mathrm{Bary}$} & \colhead{$\logmpeak$} & 
\colhead{Volume} & $n_\mathrm{massive}$ & $n_\mathrm{NIRSpec}$ \\ & & \colhead{[Degrees]} & \colhead{[Degrees]} & & \colhead{[cMpc${^3}$]}\\}
\startdata
UDS-PCl-z3.2 & 3.216 & 34.296989 & -5.235807 & 14.4 & 3720 & 3 & 60\\
UDS-PCl-z3.7 & 3.701 & 34.274170 & -5.240151 & 14.0 & 1795 & 1 & 15\\
UDS-PCl-z3.8 & 3.804 & 34.290020 & -5.258747 & 14.4 & 3726 & 1 & 24\\
UDS-PCl-z4.0 & 3.980 & 34.358517 & -5.211499 & 14.3 & 3308 & 3 & 46\\
EGS-PCl-z3.2 & 3.229 & 214.896000 & 52.879972 & 14.1 & 1886 & 3 & 55\\
Cosmic Vine$^1$ & 3.437 & 214.875225 & 52.867714 & 14.0 & 1342 & 3 & 47\\
\hline
\hline
& & & & & & \\
\enddata
\tablecomments{$^{1}$ \citet{Jin2024a}}
\end{deluxetable*}

\textbf{UDS-PCl-z3.2}: Previous studies identified an overdensity around the quiescent galaxy ZF-UDS-7329 \citep{Glazebrook2024, Turner2025, Jespersen2025}. In our density mapping, we identify several overdense peaks (two robust detections and two moderate confidence detections; Tables~\ref{tab:peaks} and \ref{tab:cpeaks}) contained within an extended protocluster. ZF-UDS-7329 lies just outside the $5\sigma$ contour of the  UDS-Peak-z3.2-1 but within UDS-PCl-z3.2. Another $z=3.2$ MQG, RUBIES-47714, also lies close by, just barely outside the $5\sigma$ peak and protocluster bounds. 

\textbf{UDS-PCl-z3.7 and UDS-PCl-z3.8}: These two newly discovered protoclusters lie in the UDS, each containing a massive quiescent galaxy and with other neighboring massive galaxies. Their $\sim75$ cMpc separation along the line of sight, while significant, is not unheard of for massive and extended high-redshift structures \citep[e.g.,][]{Cucciati2018}. In Figure~\ref{fig:crosssec} and the interactive version of Figure~\ref{fig:noninteractive}, there is a hint of intervening filamentary structure between them at $z\sim3.75$. If so, UDS-PCl-z3.7 and UDS-PCl-z3.8 could be part of a single large-scale proto-supercluster, and are therefore prime candidates for further spectroscopic followup.

\textbf{UDS-PCl-z4.0}: This structure was first identified as an overdensity of VANDELS spectroscopic detections in \citet{Tanaka2024}. A $z=4$ overdensity of quiescent galaxies lying just south of the PRIMER UDS footprint \citep{Tanaka2024}, indicating that this structure extends well outside of the volume observed here. We identify five $\logmstar\geq10.5$ and one $10.0\leq\logmstar<10.5$ quiescent galaxies in and around this structure. This structure was independently identified by \citet{Sun2025b} as ``the Bigfoot''.

\textbf{EGS-PCl-z3.2}: We find four massive galaxies (two of which are quiescent) in and around EGS-PCl-z3.2, a newly discovered protocluster. Unlike most other $3<z<4$ protoclusters in this work which are typically spread out with many sub-peaks within and nearby the structures, EGS-PCl-z3.2 appears relatively compact with only a few minor foreground and background sub-peaks. This could be an indication that this structure may be more evolved than the others presented here and gravitationally collapsing, or there may be extended overdense structure outside of the relatively small EGS footprint.

\textbf{Cosmic Vine}: First identified in \citet{Jin2024a}, the $z=3.44$ Cosmic Vine extends across the EGS field. The Cosmic Vine is also notable for a pair of merging quiescent galaxies in its core \citet{Ito2025}. One of these galaxies was presented in \citet{Nanayakkara2024}, featured heavily in the discussion relating morphology and environment in \citet{Kawinwanichakij2025}, and is included in our analysis (ID 2565-31322 in Figure~\ref{fig:prisms1} and Table~\ref{tab:prospector_output}). The second galaxy's spectrum was taken in G235M by the DeepDive program and, while included in the density mapping, was not modeled in this work. Another galaxy with a prominent Balmer break in its spectrum, WIDE-3098 at $z=3.4407$, lies outside the EGS footprint (and is therefore not included in this work's analyses), to the northeast on the sky, due east in the RA$^\prime$ Dec$^\prime$ frame, between the ``tongs.'' This could suggest that the Cosmic Vine structure extends outside of the EGS footprint and is larger than the extent inferred in this work or \citet{Jin2024a}.

\textbf{UDS-Peak-z4.6 and EGS-Peak-z4.9}: While not detected as protoclusters, these two peaks are notable in that they host particularly MQGs at $z>4.5$ and were previously identified in \citet{Carnall2024} and \citet{DeGraaff2024}, respectively. While UDS-Peak-z4.6 is fairly isolated in the UDS field, it contains a pair of $\logmstar\approx11$ quiescent galaxies, PRIMER-EXCELS-117560 and PRIMER-EXCELS-109760. EGS-Peak-z4.9 was identified as an overdensity in \citet{DeGraaff2024} and a protocluster in \citet{UrbanoStawinski2024}, though due to its location on the edge of the EGS we lack the coverage to determine whether it extends beyond the field. Within the bounds of the EGS, we estimate the mass contained within the $3\sigma$ contour to be $\logmstruct=13.7$, just barely below the protocluster classification cutoff. We speculate that if it extends beyond the field, then it would easily meet the standard for a bona fide ``protocluster.'' The small clump of star-forming galaxies just south of EGS-Peak-z4.9 identified in \citet{DeGraaff2024} is also recovered in this work (Figure~\ref{fig:MCVT}), though we do not detect structure between the two overdensities.

\section{Results}
\label{sec:results}

\subsection{The Effects of Environment on $3<z<5$ Massive Quiescent Galaxies}

\begin{figure*}[!htb]
\centering
\includegraphics[width=\linewidth]{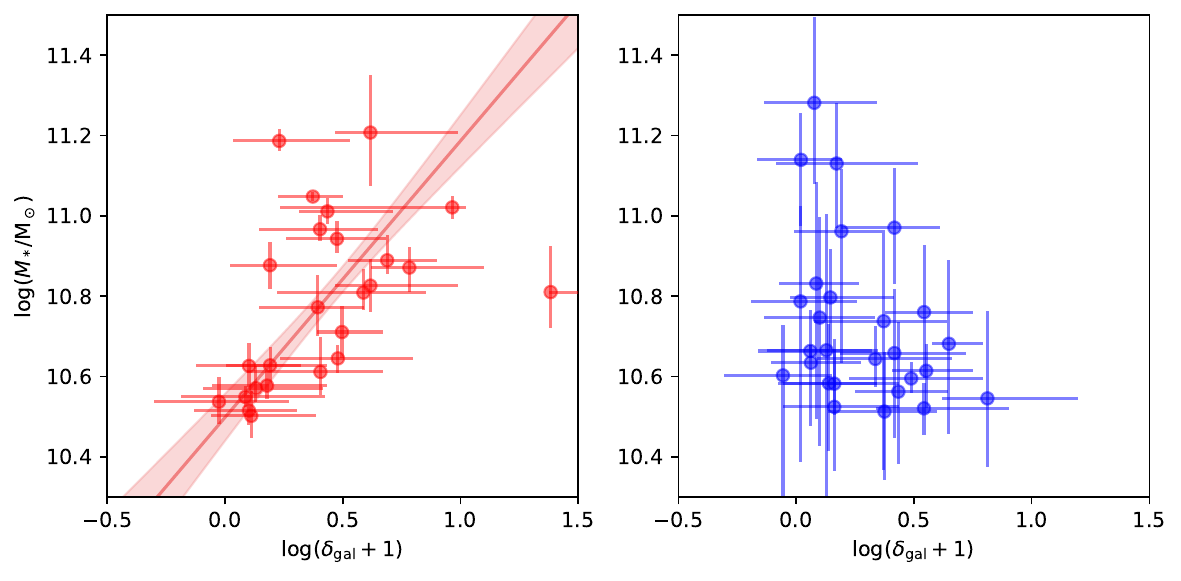}
\caption{\textbf{Left:} Stellar mass as a function of $\ldg$ for MQGs. The red line shows the best-fit lines to the MQG population using orthogonal distance regression and the semi-transparent bands show the $1\sigma$ confidence limits. 
\textbf{Right:} As the left panel, but for the MSFG population. We do not fit a line to the MSFG population due to its higher uncertainties on $\logmstar$, though we do note that there seems to be a visual flat or negative relation between stellar mass and local overdensity however.}
\label{fig:masssfrdensity}
\end{figure*}

With SED models of bright galaxies and density maps in hand and structures identified, we match our spectroscopy and SED models to the density maps. Sources with spectroscopic redshifts (including bright galaxies for which we have SED models) are assigned local $\delta_\mathrm{gal}$ based on the nearest voxel in the density grid. In Figure~\ref{fig:masssfrdensity} we show $\ldg$ versus stellar mass for massive galaxies (red for MQGs, blue for MSFGs). For MQGs, we fit a line to these quantities using an orthogonal distance regression (ODR) fit to account for the uncertainties on $\ldg$ and the inferred stellar population properties (light red, with the confidence band showing the $1\sigma$ uncertainties). The positive trend from the ODR fit visually agrees with the data.

Due to generally higher uncertainties on the inferred stellar mass in the MSFG population, we do not fit a trendline. A simple visual comparison between the two clearly shows that the ODR fit to the MQG sample would be a poor fit to the MSFGs. Instead, there seems to be a flat or negative relation between stellar mass and local overdensity for the star-forming galaxies.

While the locations of the MQGs shown in Figure~\ref{fig:crosssec} indicates that they generally lie in or near overdense regions, this is not entirely unexpected. Massive galaxies are biased tracers of cosmic structure and more overdense regions have more massive halos, and therefore, more massive galaxies. We see this trend for MQGs, where higher masses coincide with higher density values. For massive star-forming galaxies however, the trend appears to be reversed or nonexistent, which \emph{is} unexpected. This could indicate an increased efficiency for quenching of massive galaxies in overdense regions. Future analyses studying differences in the stellar mass function between overdense structure and the field will determine if this is the case \citep[e.g.,][]{Vanderburg2020, Forrest2024, Edward2024}.

\subsection{Where Massive Galaxies Live}

With the identified structures in EGS and UDS and our sample of massive galaxies, we aim to investigate how these populations of massive quiescent and star-forming galaxies are distributed within these structures. To do so, we match each galaxy with its nearest $5\sigma$ overdensity down to $\logmpeak=12$. For a given galaxy, we measure the comoving distance from the galaxy to the contour shell enclosing the $5\sigma$ peak, $D_\mathrm{5\sigma}$ (galaxies which lie within the peak have $D_\mathrm{5\sigma}=0$). We identify galaxies either within the volumes enclosed by the $5\sigma$ peaks or located nearby, within $0<D_\mathrm{5\sigma}\leq2.5$ of the peak's border. 

Next, we consider the massive galaxy sample. Of the massive galaxies for which the posterior median $\log M_*/M_\odot\geq10.5$, we find, as expected, a higher fraction of these sources are located within $5\sigma$ overdensities than the overall $m_\mathrm{f444w}<27.5$ spectroscopic population. However the uncertainties for quantities inferred from SED modeling like sSFR, SFR, and mass are non-negligible (especially for dusty star-forming galaxies, see Figure~\ref{fig:massdustssfr}) and these quantities are often degenerate. In Figure~\ref{fig:massdustssfr}, one can see that the mass uncertainties of several ``massive'' star-forming galaxies extend far into the lower mass region and sSFRs from quiescent and star-forming galaxies cross the $\log (\mathrm{sSFR/yr^{-1}})=-10$ line. 

Therefore, instead of universally considering a galaxy as ``quiescent'' or ``star-forming,'' we instead incorporate the uncertainties from their inferred posteriors. 2D histograms of the stacked posteriors in sSFR, stellar mass, and dust optical depth in  Figure~\ref{fig:massdustssfr} for all 200 galaxies fit with \texttt{Prospector}. For each galaxy, we perform 500 random draws from its posterior, reclassifying it for each draw as massive quiescent, massive star forming, or rejecting it (low mass) depending on that draw's stellar mass and SFR.

\begin{figure*}[!htb]
\centering
\includegraphics[width=\linewidth]{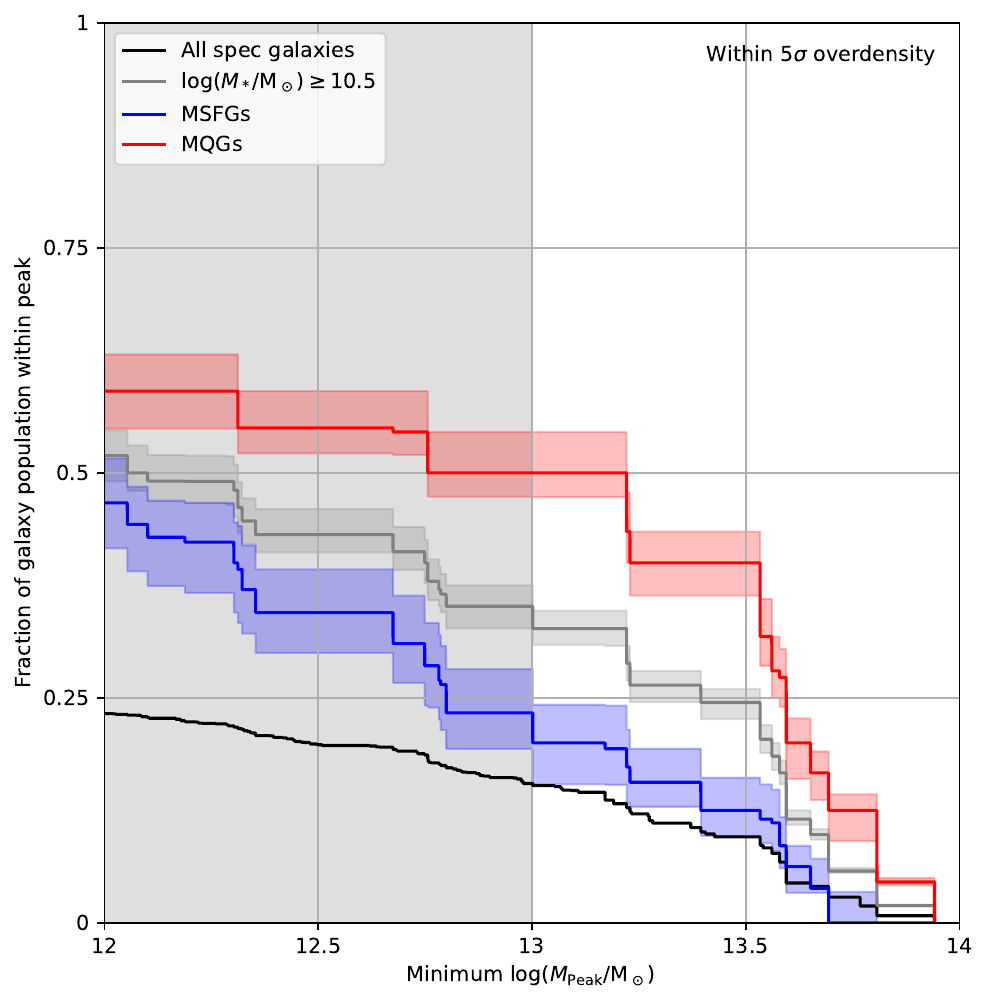}
\caption{The fractions of spectroscopically-confirmed galaxy subpopulations identified within overdense peaks above a minimum $M_\mathrm{peak}$. As one moves right along the x-axis, the number of overdense peaks above the minimum $\logmpeak$ declines, and therefore the fractions of galaxies within these peaks also declines. We show MQGs (red), MSFGs (blue), all massive galaxies (gray), and the overall spectroscopic sample (black). Not only are a higher fraction of MQGs found within overdense peaks than any other subpopulation, $\sim50\%$ also are found primarily in robustly detected $\logmpeak\geq13$ overdense peaks. We show all peaks down to $\logmpeak=12$, though the peaks in the shaded region $12\leq\logmpeak<13$ tend to be physically smaller, with fewer spectroscopically confirmed galaxies, and typically have lower median and maximum density.}%
\label{fig:sfqstruct}
\end{figure*}

In Figure~\ref{fig:sfqstruct} we show how these populations depend on the local structure. The trend line show the fractions of galaxies of a given population which reside within an overdense peak above a given minimum peak mass. As $M_\mathrm{peak}$ is restricted to higher masses, the fractions of all populations in close proximity to these peaks fall. This is expected because the number of peaks selected (and the number of galaxies around them) naturally decreases as the minimum $\logmpeak$ increases.

However we see that the relative declines in these populations are remarkably different. At first, the fraction of MSFGs and MQGs within such structures are fairly close for all $\logmpeak\geq12$ and lie above the overall spectroscopic sample. As the $\logmpeak$ minimum increases, the fraction of massive star-forming galaxies decreases more rapidly than MQGs, with $25\%/50\%$ massive star-forming/quiescent galaxies found within robustly detected $5\sigma$ peaks with $\logmpeak\geq13$. Not only are massive galaxies (gray line) found preferentially within overdense peaks, MQGs appear to cluster especially strongly.

\begin{figure*}[!htb]
\begin{center}
\includegraphics[width=\linewidth]{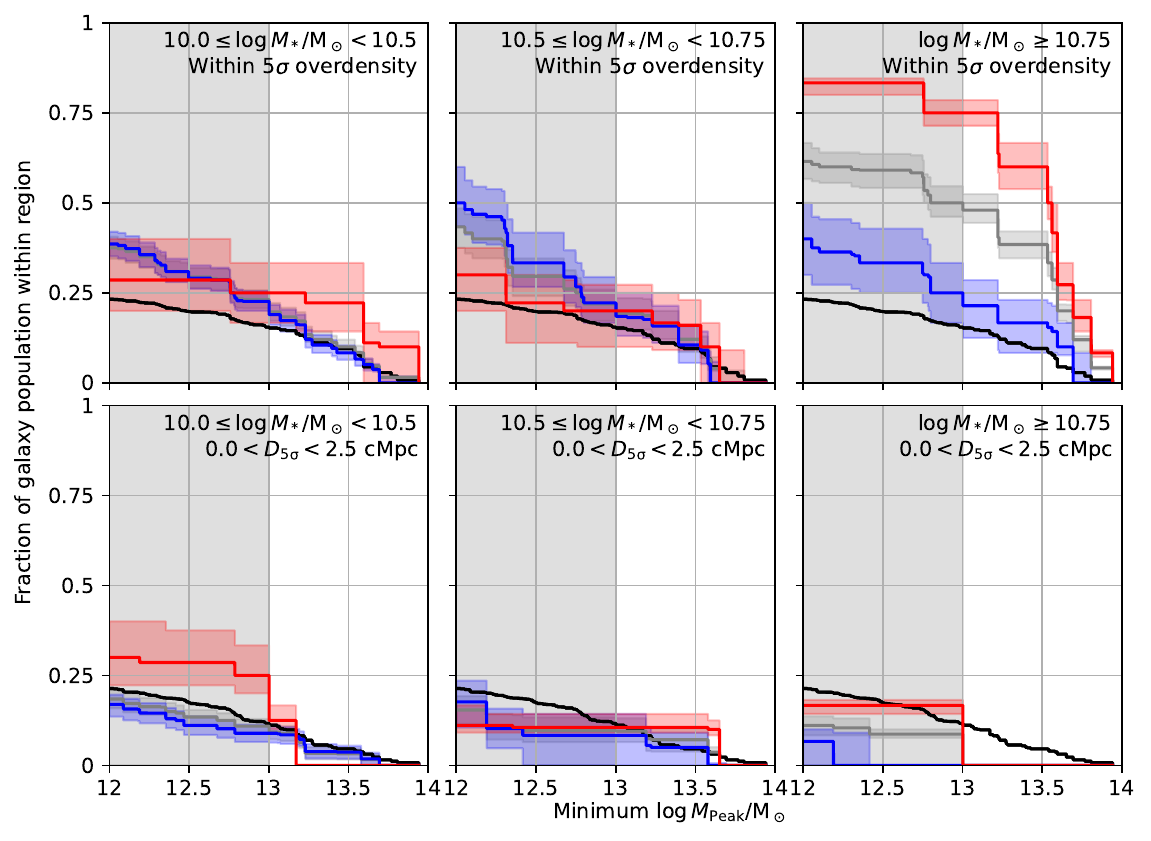}
\end{center}
\caption{As Figure~\ref{fig:sfqstruct}, but splitting MQGs and MSFGs into different stellar mass bins (low to high, left-to-right) and different regions (top to bottom, inside overdense peak versus immediately nearby overdense peak). The black line representing all spectroscopically confirmed galaxies is uniform across the mass bins. At lower stellar masses, the clustering of MQGs is relatively consistent with MSFGs and the general spectroscopic sample; as stellar mass increases however, MQGs appear to cluster much more strongly in overdense peaks.}%
\label{fig:sfqstruct_2}
\end{figure*}

To investigate trends with stellar mass, we consider galaxies with $10\leq\logmstar<10.5$ (which may suffer from mass incompleteness) and break down the massive $\logmstar\geq10.5$ sample into two mass bins of roughly equal posterior mass: $10.5\leq\logmstar<10.75$ and $\logmstar\geq10.75$. We show how the star-forming and quiescent subpopulations of these sets depend on $\logmpeak$ in the top row of Figure~\ref{fig:sfqstruct_2}. We see little change between the mass bins for the star-forming population, but significant change for the high-mass quiescent bin. In the bottom row, we show the region ``nearby'' the peak (within 2.5 cMpc of the border).
In most instances, massive galaxies found within this region behave similar to the overall spectroscopic sample. There appears to be some preference for $10.0\leq\logmstar<10.5$ quiescent galaxies to either be found within massive peaks or in regions around lower-mass peaks, though the number of sources is low ($\lesssim10$), uncertainties on stellar population properties are large, and mass completeness in this bin is a concern. On the high mass end, we see that \emph{all} $\logmstar\geq10.75$ quiescent galaxies are found within 2.5 cMpc of an overdense peak with $\logmpeak\geq12$ (the two quiescent galaxies in the bottom right panel are ZF-UDS-7329 and RUBIES-47714, both of which lie within 0.5 cMpc of the boundary of the same peak, see Figure~\ref{fig:noninteractive}), and there is a strong trend for these galaxies to live within overdensities with $\logmpeak\geq13$.

\section{Discussion}
\label{sec:disc}

All of these trends point toward more efficient and rapid mass assembly and quenching in overdense regions than the field at $3<z<5$. That these high redshift MQGs also appear to be correlated with massive proto-structures also implies that even at high-redshift cosmic structure plays a significant role in the formation and evolution of massive galaxies. 

In an observational analysis across $\sim20$ square degrees, \citet{Shibuya2025} found that galaxy merger fraction increased almost linearly with $\delta_{gal}$ at $z\sim2-5$, indicating that merger rates are enhanced in overdense regions at high redshift. High merger rates are also seen in simulations, with \citet{Huko2022} showing that the ex-situ rapidly increases with stellar mass in $\logmstar>10.5$ galaxies up to $z\sim4$, where major mergers provided $\sim50\%$ and minor mergers and accretion $\sim25\%$ each of ex-situ mass. That MQGs are found predominantly in overdense regions, with the most massive MQGs uniformly distributed in or around early overdense cosmic structure, therefore suggests that galaxy interactions and mergers in their environments could be driving the trends seen with stellar mass (e.g., Figure~\ref{fig:masssfrdensity}).

To investigate the effects of environment in MQG formation, we use the inferred SFHs from \texttt{Prospector} to estimate the stellar mass of MQG protenigors at $z>5$ and draw comparisons with observations and simulations. 

\begin{figure}[!htb]
\begin{center}
\includegraphics[width=\linewidth]{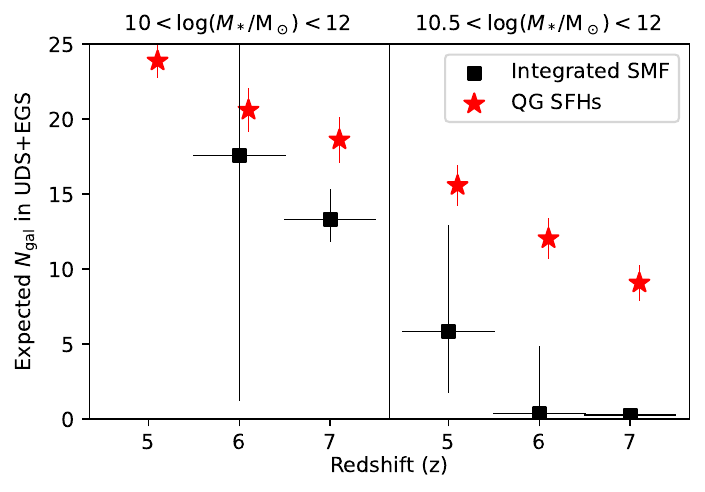}
\end{center}
\caption{Expected numbers of $10<\logmstar<12$ (left) and $10.5<\logmstar<12$ (right) galaxies at $z=5, 6,$ and $7$ from the integrated galaxy SMF in \citet{Weibel2024} (black squares) and the SFHs of the MQGs in this work (red stars; slightly offset along the x-axis for visibility). We find relatively good agreement between the integrated SMF between $10<\logmstar<12$ and the number of observed MQGs with $\logmstar>10$ at $z=5,6,7$. However, above the knee of the stellar mass function ($\logmstar>10.5$), there appears to be an excess of MQGs with high mass relative to the SMF. The uncertainties on SMF points were calculated by bootstrapping, which likely overestimates uncertainties in the $z\sim5,6$ bins due to anticorrelation between SMF variables and underestimates uncertainties in the $z\sim7$ bin as \citet{Weibel2024} fixed $\log M^*=10$.}
\label{fig:numbers}
\end{figure}

First, we integrate the Schechter function fit to the galaxy stellar mass function (SMF) in \citet{Weibel2024} between $10.0<\logmstar<12$ and $10.5<\logmstar<12$ in an equivalent comoving volume to that spanned by our spectroscopic sample $3<z<5$ in UDS and EGS ($\approx333$ square arcminutes total). We perform this for the $z\sim5$, $z\sim6$, and $z\sim7$ stellar mass functions (measured in redshift bins $4.5<z<5.5$, $5.5<z<6.5$, $7.5<z<8.5$). Because the integral is strongly dependent on $\log(M^*)$ which is anticorrelated with $\log(\Phi^*)$ and $\alpha$, the uncertainties we obtain from bootstrapping are likely \emph{over}estimated in the $z\sim5,6$ bins; in the $z\sim7$ bin \citet{Weibel2024} fixed $\log M^*=10$, so uncertainties here could be \emph{under}estimated. The number of galaxies $N_\mathrm{gal}$ expected in this volume are shown as black squares in Figure~\ref{fig:numbers} (at $z\sim5$ the $10.0\leq\logmstar\leq12$ bin has $N_\mathrm{gal}\approx90$ and lies far above the regions of the panel). 

Then for the 30 quiescent galaxies (red models in Figures~\ref{fig:prisms1}, \ref{fig:prisms2}, and \ref{fig:lilprisms}), we count the number of $\logmstar>10.0$ and $\logmstar>10.5$ progenitors at $z=5$, $z=6$, and $z=7$ and obtain uncertainties by bootstrapping draws from the posteriors (red stars in Figure~\ref{fig:numbers}). While the number of MQGs with $\logmstar>10.0$ at high redshift agrees with the number obtained from the observed SMF, at $\logmstar>10.5$ $N_\mathrm{gal}$ from the SFHs is markedly higher than $N_\mathrm{gal}$ from the SMF. In other words, single progenitors formed according to the inferred SFHs would be significantly more massive and numerous at $z>5$ than actual, observed high-redshift galaxies.

\begin{figure}[!htb]
\begin{center}
\includegraphics[width=\linewidth]{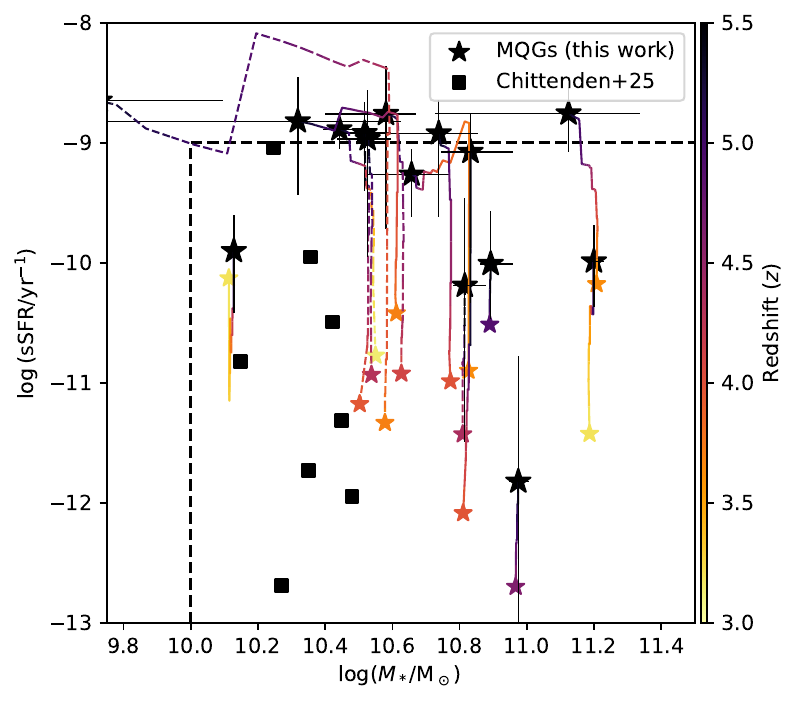}
\end{center}
\caption{A comparison of the 14 MQG progenitorx with $\log (M_{*,16}/\mathrm{M_\odot})>10$ and $\log\ \mathrm{(sSFR_{16}/yr^{-1})<-9}$ at $z=5.5$ (black stars) with the simulated quiescent galaxies from the THESAN simulation \citep{Chittenden2025} (black squares). Inferred median SFHs are shown as colored dashed/solid lines for MQGs outside/inside overdensities and the solid colored stars show the median observed surviving stellar mass and sSFR. MQG progenitors of comparable stellar masses to the simulated quiescent galaxies tend to have higher sSFRs and MQG progenitors with comparable sSFRs tend to have higher stellar masses. Finally, of the seven MQG progenitors with median stellar mass and sSFR in the quiescent region (black dashed lines), all but one are found in an overdensity and most are significantly more massive than the simulated quiescent galaxies.}
\label{fig:history}
\end{figure}

Next, we turn to high-redshift simulations of quiescent galaxies. \citet{Chittenden2025} identified nine $\logmstar>10$ galaxies in the THESAN-1 simulation box (side length 95.5 cMpc) with sSFR $<1\ \mathrm{Gyr^{-1}}$ at $z=5.5$ (Black squares in Figure~\ref{fig:history}). To compare with these findings, we trace back the SFHs of all 30 $\logmstar\geq10$ quiescent galaxies and examine their total stellar mass and sSFR at $z=5.5$. We show the stellar mass and sSFR from the MQG SFHs at $z=5.5$ as black stars in Figure~\ref{fig:history} (SFHs up to $z_\mathrm{obs}$ are shown as colored lines terminating in colored stars). MQGs outside of overdense structure (dashed lines) tend to have comparable or slightly higher stellar mass and are more star-forming at $z=5.5$ than the THESAN quiescent galaxies. However, of the eight MQGs with the most mass at $z=5.5$, seven are classified as overdensity MQGs (solid lines). Of the seven MQGs with median sSFR and stellar mass in the quiescent region (black dashed line), six are overdensity MQGs, and most are significantly higher mass than the simulated quiescent galaxies in the $z=5.5$ snapshot.

The lack of simulated or observed galaxies at similarly high masses might suggest that, while the stellar mass in overdensity MQGs at $3<z<5$ did form early, at $z=5.5$ a significant fraction of these stars were still located in other, neighboring halos and would not assemble onto the main progenitor halo until later. Both simulations \citep{Chittenden2025} and observations \citep{Weibel2024} demonstrate that galaxies are able to assemble $\logmstar\sim10$ by $z\approx5$, but the relative dearth of $\logmstar\geq10.5$ high-redshift galaxies compared to the number inferred from SFHs of MQGs would suggest that the total mass in the MQG SFH includes stars which had already formed by $z=5.5$ but had not yet been accreted onto the main progenitor halo.

However, this comparison depend on the ability to accurately infer SFHs from PRISM spectroscopy and broadband photometry. Our \texttt{Prospector} modeling to low-resolution ($R\sim100$) PRISM spectra utilizes stellar libraries with solar-scaled stellar abundances, which cannot reproduce the $\alpha$-enhancement expected in such early-forming quiescent galaxies. If these galaxies' stellar populations are $\alpha$-enhanced \citep{Park2024a}, that would drive the inferred formation time towards later cosmic times and lower stellar masses. The young age of the universe at $3<z<5$ may play an additional confounding role here. At low redshifts, post-starburst galaxies are sometimes referred to as E+A or K+A due to their spectral signatures of recently-formed (e.g., A-stars from a recent starburst event; hence the A) on top of an old stellar population (e.g., K-stars or a typical low-$z$ elliptical galaxy spectrum; hence the E or K). However quiescent galaxies at high-redshift are almost all post-starbursts \citep[e.g.,][]{Wild2016}; if an overdensity MQG at $z=3$ underwent multiple mergers (either gas-rich and driving starbursts or dry and depositing stars with a different stellar age) since $z=8$, is the age resolution of the stellar population synthesis models sufficient to recover these different subpopulations all with ages $\lesssim1$ Gyr? High resolution grating spectroscopy, the implementation of $\alpha$-enhanced stellar templates, and further testing of stellar population models will be necessary to better constrain the SFHs of these galaxies and either lessen or confirm the tension with high-redshift observations presented here.

\section{Conclusions}
\label{sec:conc}

In this paper, we presented an analysis of the environments of massive galaxies in the EGS and UDS fields using RUBIES spectroscopy and archival data from the DJA. We summarize our findings below.

\begin{itemize}
    \item Using a Monte Carlo Voronoi Tesselation technique to combine photometry and over 2,000 spectroscopic redshifts in the UDS and EGS fields, we produce galaxy overdensity maps identify 12 overdense peaks containing $\logmstar\geq10.5$ galaxies at $3<z<5$ and six protoclusters.
    \item We investigate how stellar mass trends with environment $\ldg$ for MQGs and MSFGs. As might be expected from hierarchical structure formation, we observe a positive trend between MQG stellar mass and overdensity; however this trend is curiously absent from the MSFG population. This could suggest an increased efficiency of quenching mechanisms in overdense environments. 
    \item We find that massive galaxies cluster strongly (again, as would be expected from hierarchical structure formation): a high fraction ($\approx30\%$) live within massive ($\logmpeak\geq13$) overdense peaks, with an even stronger preference for quiescent galaxies to lie in such overdensities $\approx50\%$. We explore how this trend depends on stellar mass and proximity to the overdensity. We find that over $75\%$ of $\logmstar\geq10.75$ quiescent galaxies are located within massive overdense peaks, with the only two counterexamples (ZF-UDS-7329 and RUBIES-47714) situated 0.5 cMpc from their shared peak's border region.

    \item Using the inferred SFHs from the quiescent galaxy sample, we compare the total mass at $z>5$ of the observed MQGs with high-redshift observations and simulations. While $\logmstar>10$ galaxies are found in both observations and simulations in good agreement with the SFHs, at $\logmstar>10.5$ the agreement breaks down. Under the assumption that MQGs formed all of their stars in-situ (i.e., in the main progenitor halo), the number of $z>5$ massive progenitors would exceed the number inferred from the galaxy stellar mass function fit to high-redshift observations. The inferred SFHs also predict stellar masses exceeding those of simulated quiescent galaxies at $z=5.5$, with the descendant MQGs of the most significant outliers living in overdense regions.

    \item We argue that these trends point toward ex-situ star formation playing an important role in quiescent galaxy mass assembly. Increased merger rates in overdense environments would naturally lead to greater numbers of higher mass MQGs. If the stars found in overdensity MQGs formed across multiple halos at high redshift before assembling onto the main halo at later cosmic times, the tension with observations and simulations from the ``single-halo progenitor'' assumption would be lessened.

\end{itemize}

\begin{acknowledgments}

Some of the data products presented herein were retrieved from the Dawn JWST Archive (DJA). DJA is an initiative of the Cosmic Dawn Center (DAWN), which is funded by the Danish National Research Foundation under grant DNRF140.

This work is based in part on observations made with the NASA/ESA/CSA James Webb Space Telescope. Support for programs GO-4233 and GO-3659 was provided by NASA through a grant from the Space Telescope Science Institute, which is operated by the Association of Universities for Research in Astronomy, Inc., under NASA contract NAS 5-03127.

The authors acknowledge the CEERS and PRIMER teams for developing their observing program with a zero-exclusive access period.
\end{acknowledgments}

%
\facilities{HST, JWST}

\software{Matplotlib \citep{2007CSE.....9...90H}, 
        NumPy \citep{2020Natur.585..357H},
        astropy \citep{2013A&A...558A..33A,2018AJ....156..123A,2022ApJ...935..167A},
        Prospector \citep{Johnson2017, Johnson2021}, 
        dynesty \citep{Speagle2020}, joblib  \citep{joblib}
        mayavi \citep{mayavi}
          }


\appendix
\restartappendixnumbering

\section{Dusty Star-Forming SEDs}
\label{app:dusty}

\begin{figure*}[!htb]
\begin{center}
\includegraphics[width=\linewidth]{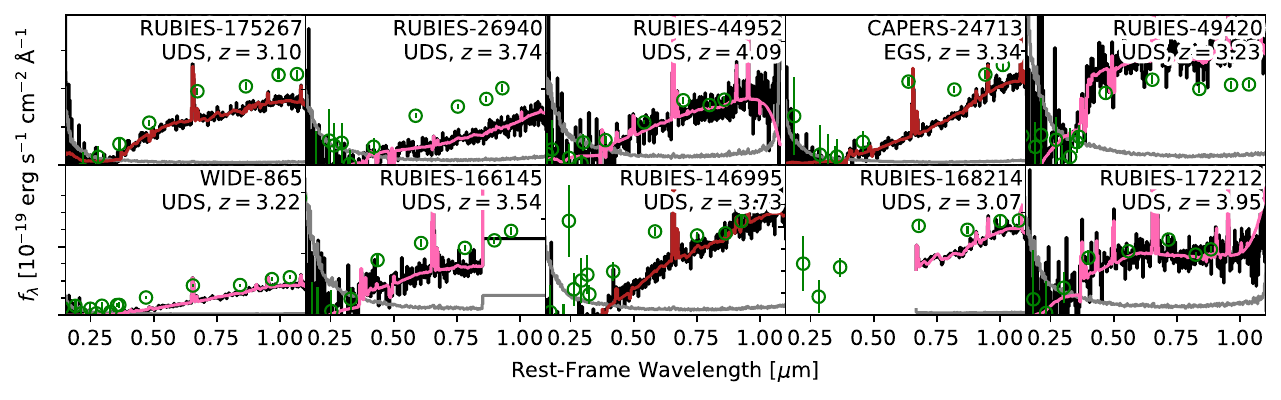}
\end{center}
\caption{PRISM spectra (black) and best-fit \texttt{Prospector} models for galaxies with $\logmstar\geq10.0$ and $\log (\mathrm{sSFR/yr^{-1}})<-10$ (red) or $\log (\mathrm{sSFR_{16}/yr^{-1}})<-10$ (pink). Unlike Figures~\ref{fig:prisms1}, \ref{fig:prisms2}, and \ref{fig:lilprisms}, here we leave the PRISM spectrum uncalibrated and show the model spectrum scaled to the PRISM instead. Based on their high dust optical depths we classify them as ``dusty star-forming galaxies'' by default, though classifying them as quiescent leaves our conclusions unchanged.}
\label{fig:dusties}
\end{figure*}

\section{Monte Carlo Voronoi Tesselation}
\label{app:mcvt}

\subsection{Ingredients}

This subsection describes the three input data products. The two dimensional field grid is the plane which we will partition into cells centered on galaxies (i.e., Voronoi tesselation), the coordinates of which are taken from the source coordinate catalog if their entry in the redshift catalog indicates that they are members of the redshift slice.

\subsubsection{Field Grid}
\label{sssec:grid}

First, we construct a grid of pixels upon which we will perform the Voronoi tesselation. Before we begin our density estimation method, we adjust our coordinate system for geometric convenience. Because the UDS is an equatorial field, the angular distance between two sources within it will be fairly close to a cartesian approximation with RA and Dec as $x$ and $y$ coordinates (a difference of $<0.01\%$). This approximation, however, cannot be made in the EGS, which lies at Dec $\sim53$ where separation is very non-cartesian due to spherical geometry.

To ameliorate the effects of spherical geometry, we reorient our coordinate axes. We take the median RA and Dec in each of the photometric catalogs to be the ``center'' of each field and apply an invertible transformation to our coordinate frame which places the center at the origin while preserving angular separations. We refer to these new axes as \rap\ and \decp. Finally, for computational efficiency, in the case of the EGS we rotate our axes by 49 degrees counter-clockwise such that the field lies flat along the \rap\ axis (in the UDS, which is left un-rotated, \rap\ and \decp\ are equivalent $\Delta \mathrm{RA}$ and $\Delta \mathrm{Dec}$). Because our fields are equatorial in \rap\ and \decp, we therefore perform our analyses using cartesian approximation in these coordinates, which is a good approximation of the angular separation in RA and Dec. 

On top of the field in \rap\ and \decp\, we overlay a cartesian grid of $2.5^{\prime\prime}\times2.5^{\prime\prime}$ pixels which extends $5^\prime$ beyond the field edges in each direction (each pixel $\mrap_i,\ \mdecp_j$ can easily be translated back to traditional RA and Dec coordinates using the inverse transformation). Using the F444W mosaics from the DJA and the locations of sources in the photometric catalogs, we make a pixel mask of the grid identifying which pixels do and do not have F444W coverage (hereafter referred to as the ``edge mask''). 

\subsubsection{Source Coordinate and Redshift Catalogs}
\label{sssec:sccat}

Second, we produce a catalog of points to drop onto the grid. As the name suggests, this ``source coordinate catalog'' is just the \rap\ and \decp\ coordinates of a subset of sources in the photometric catalog. Both UDS and EGS F444W reach $5\sigma$ depths at $\sim28$ magnitudes, so first we remove all sources fainter than 27.5 magnitudes in F444W. We require each source to have detections in at least three other bands, a fit photometric redshift ($z_\mathrm{phot}>-1$), and a photometry quality flag (USE$\_$PHOT $=1$). We find 68,944 sources meet this criteria in the UDS and 29,052 in EGS. 

Next, we construct a redshift array to pair with the catalog of \rap\ and \decp\ coordinates. For each source in the source coordinate catalog, we randomly sample its associated \texttt{eazy} $p(z)$ distribution 100 times to draw 100 probabilistic redshifts. This produces an $N\times100$ grid, where N is the number of selected sources. The majority of our sources observed in imaging alone (i.e., they have only photometric redshifts) which is sufficient to identify particularly massive proto-structures and inform a measure of the average galaxy density in the field \citep[e.g., Figure 13 in][]{Chiang2013}, however we wish to also detect less massive, smaller proto-structures and map associated substructure. 

The wealth of available spectroscopy in the UDS and EGS fields will help by increasing the precision of our density measurements. Many sources have entries in multiple spectroscopic catalogs, so we assign redshifts based on the following ranking: $z_\mathrm{spec,\ MSA}$, $z_\mathrm{spec,\ ground}$, $z_\mathrm{best,\ 3DHST}$, $z_\mathrm{phot}$. The treatment of galaxies with $z_\mathrm{phot}$ is described above; below we detail how redshifts for the various $z_\mathrm{spec}$ categories are resampled.

$\mathbf{z_\mathrm{\bf{spec,\ MSA}}}$: For galaxies with a NIRSpec MSA spectrum for which we measured a redshift, we instead replace the redshifts sampled from the photometric $p(z)$ with a set sampled from the spectroscopic $p(z)$ instead (see \S~\ref{ssec:spec}).

$\mathbf{z_\mathrm{\bf{spec,\ ground}}}$: Uncertainties on galaxies from ground-based catalogs are not provided, so we resample redshifts from a gaussian centered on $z_\mathrm{spec,\ ground}$ with a conservative width $\sigma_z = 0.001(1+z_\mathrm{spec,\ ground})$. The UDS VANDELS catalog contains quality flags indicating the probability of the given spectroscopic redshift being correct, so for each VANDELS source we perform 100 weighted coin flips (where the weight is set by the source's flag probability); we use the resampled $z_\mathrm{spec}$ when heads is thrown and the redshifts sampled from the photometric $p(z)$ distribution when the result is tails (i.e., we leave the redshift entry in the grid as-is). Redshifts from Deeper than DEEP \citep{Stawinski2024} also have a 1-4 flag system but no associated probabilities. We retain sources with flags 3 and 4 and assume the same weighting for these as the VANDELS catalog. DEEP2/3 redshifts are sourced from the 3DHST catalog which only retained sourced with the highest quality flag (flag 4). UDS ground-based redshifts in the 3DHST come from various small spectroscopic programs and are ungraded, but treated as ``robust.''

$\mathbf{z_\mathrm{\bf{spec,\ 3dHST}}}$: Because of the relatively low resolution of the HST grism, sources with grism redshifts in the 3DHST catalog (referred to as $z_\mathrm{grism}$ therein) have larger uncertainties than typical spectroscopic redshifts. Fortunately, the catalog provides upper and lower $1\sigma$ and $2\sigma$ uncertainties so we are able to statistically account for this intrinsic limitation. First we perform a quality cut: for any source with upper or lower $\sigma_{z_\mathrm{spec}} > 0.0375(1+z_\mathrm{spec})$ (roughly the 90th percentile of $\sigma_{z_\mathrm{spec}}$ values), we ignore its entry in the 3DHST catalog and default to redshifts sampled from the photometric $p(z)$. Sources which fail this typically have uncertainties consistent with confusion of a single line as either [OIII] 5008 or H$\alpha$ and are typically lower redshift than the range we sample here, so this choice has no effects on our results. Then, from the remaining sources with sufficiently precise $z_\mathrm{spec}$, we resample redshifts from an asymmetric gaussian cumulative distribution function where the gaussian CDF above/below the median ($z_\mathrm{spec}$) has dispersion equal to the upper/lower $1\sigma$ uncertainties from the 3DHST catalog. Incorporating the given $2\sigma$ uncertainties into our statistical resampling of 3DHST redshifts has negligible results on our results, so we just use the $1\sigma$ uncertainties for simplicity.

\subsection{Overdensity Mapping with Monte Carlo Voronoi Tesselation}
\label{ssec:mcvt}

In this section, we explain the method of quantifying environment at a given redshift using our input data (the field grid, source coordinate catalog, and redshift catalogs). For a target redshift $z_\mathrm{slice}$, we identify members of the redshift slice by iterating over the redshift catalog. For each source, it is identified as a member of the redshift slice if the line-of-sight comoving separation between it and the slice's redshift is less than 5 cMpc (i.e., $|D_\mathrm{C,\ l.o.s}(z_\mathrm{slice}) - D_\mathrm{C,\ l.o.s}(z_{\mathrm{source},h})| < 5$ cMpc) in a given iteration (index $h$). This means that galaxies with broad $p(z)$ distributions, such as those with only a few bands of photometry, are unlikely to fall into the same redshift slice multiple times; galaxies with narrower $p(z)$ distributions, such as those with features falling in medium bands, will fall into the slice somewhat more frequently; and galaxies with extremely narrow $p(z)$s (e.g., sources with MSA spectra) will almost always fall in the slice.

Having selected the galaxies in a given 10 cMpc-wide slice of the field, we next correct for edge effects by filling in the region outside the field F444W footprint with mock galaxies. Using the field grid edgemask, we can estimate an average source density $\overline{\Sigma}_{\mathrm{MCVT}}^{h}(z_\mathrm{slice})$ within the field for the iteration. We then multiply this surface density by the area of the region outside the footprint to obtain the expected number of sources that would lie in this region $m$ (not necessarily an integer). To account for the uncertainty of this estimate, we randomly sample an integer $M$ from a Poisson distribution with expectation value $m$. We randomly distribute $M$ galaxies across the field grid pixels that lie outside the F444W footprint.

While it is possible to infer a single expectation value of e.g. a Poisson distribution from all 100 $\overline{\Sigma}_{\mathrm{MCVT}}^{h}(z_\mathrm{slice})$ measurements, doing so would result in less variance in the number of sources outside the footprint, which would result in less variance in density measurements for sources near the footprint edges. We opt for this approach to inflate uncertainties for these sources and better reflect the poor constraints on density measurements at the footprint edges.

Having assigned the locations of real galaxies within the footprint and fake galaxies without, we proceed with the Voronoi tesselation. We partition the field grid by assigning each pixel at $\mrap_i,\ \mdecp_j$ to the galaxy, real or fake, nearest to it. All pixels associated with a given source are grouped together into a cell and are assigned a value equal to the inverse of the area of the cell. Thus, the values of all pixels $i,\ j$ of the cell are simply the surface density, $\Sigma_{\mathrm{MCVT}}^{h,i,j}(z_\mathrm{slice})$.

Having successfully produced a surface density map for iteration $h$, we save the grid of $\Sigma_{\mathrm{MCVT}}^{h,i,j}(z_\mathrm{slice})$ pixels and repeat the process for iteration $h+1$. After 100 iterations, for each pixel $i,j$ we take the median across the index $h$ of all 100 $\Sigma_{\mathrm{MCVT}}^{h,i,j}(z_\mathrm{slice})$ to be $\Sigma_{\mathrm{MCVT}}^{i,j}(z_\mathrm{slice})$. We also take the 84th and 16th percentile values of $\Sigma_{\mathrm{MCVT}}^{h,i,j}(z_\mathrm{slice})$ across index $h$ to be the uncertainties on the surface density for pixel $i,j$. Typically uncertainties from the stacked MCVT maps are around $\sim0.2$ dex and we do not find evidence for any strong dependence on local structure (e.g., uncertainties do not vary significantly near overdense peaks versus out near the median density).

We refer to this method as Monte Carlo Voronoi Tesselation, or MCVT for short; similar to VMC \citep[e.g.,][]{Lemaux2018} but distinct in setup and execution. In Figure~\ref{fig:MCVT} we show a single iteration on the left and the stacked median on the right (in the online journal an animation shows all iterations and the development of the median image). Red points denote the location of the redshift slice members for a given iteration and the field grid pixels are colored (with arbitrary scaling) by their $\log \Sigma_{\mathrm{MCVT}}^{h,i,j}(z_\mathrm{slice})$ on the left (the individual iteration maps) and $\log \Sigma_{\mathrm{MCVT}}^{i,j}(z_\mathrm{slice})$ on the right (the stacked median map). White contours indicate the boundaries of the F444W footprint. As $k$ progresses, structure which are not detectable in individual iteration maps begin to appear in the median image.

We then convert the MCVT surface density to a galaxy overdensity measure (sometimes called density contrast), $\delta_\mathrm{gal}^{i,j} = (\Sigma_\mathrm{MCVT}^{i,j} - \widetilde{\Sigma}_\mathrm{MCVT}^\mathrm{footprint}) / \widetilde{\Sigma}_\mathrm{MCVT}^\mathrm{footprint}$, where $\widetilde{\Sigma}_\mathrm{MCVT}^\mathrm{footprint}$ is the median of $\Sigma_\mathrm{MCVT}^{i,j}$ for all $i,j$ such that $\mrap_i,\ \mdecp_j$ lies inside the F444W footprint.

Figure~\ref{fig:MCVT} shows a single iteration on the left and the stacked median on the right (in the online journal an animation shows all iterations and the development of the median image). Red points denote the location of the redshift slice members for a given iteration and the field grid pixels are colored (with arbitrary scaling) by their $\log \Sigma_{\mathrm{MCVT}}^{h,i,j}(z_\mathrm{slice})$ on the left (the individual iteration maps) and $\log \Sigma_{\mathrm{MCVT}}^{i,j}(z_\mathrm{slice})$ on the right (the stacked median map). White contours indicate the boundaries of the F444W footprint. As $k$ progresses, structure which are not detectable in individual iteration maps begin to appear in the median image.

\section{Overdensity Candidates}
\label{app:candidates}

\begin{deluxetable*}{ccccccc}
\centerwidetable
\tablewidth{0pt}
\tablecaption{Moderate Confidence Overdensity Peaks/Protogroup Candidates \label{tab:cpeaks}}
\tablehead{\colhead{ID} & \colhead{$z_\mathrm{Bary}$}& \colhead{RA$_\mathrm{Bary}$} & \colhead{Dec$_\mathrm{Bary}$} & \colhead{$\logmpeak$} & \colhead{Volume} & \colhead{$n_\mathrm{NIRSpec}$}\\
& & \colhead{[Degrees]} & \colhead{[Degrees]} & & \colhead{[cMpc$^3$]}}
\startdata
UDS-cPeak-z3.2-1 & 3.198 & 34.313715 & -5.221222 & 13.2 & 228 	& 4 \\
UDS-cPeak-z3.2-2 & 3.226 & 34.413566 & -5.154879 & 13.2 & 217 	& 9 \\
UDS-cPeak-z3.2-3 & 3.238 & 34.356984 & -5.246196 & 13.1 & 182 	& 4 \\
UDS-cPeak-z3.6 & 3.599 & 34.491399 & -5.148159 & 13.4 & 389 	& 8 \\
UDS-cPeak-z3.7 & 3.705 & 34.453155 & -5.153699 & 13.3 & 278 	& 9 \\
UDS-cPeak-z3.8 & 3.839 & 34.255607 & -5.244232 & 13.0 & 167 	& 0 \\
UDS-cPeak-z5.0 & 4.954 & 34.375104 & -5.260380 & 13.4 & 406 	& 3 \\
EGS-cPeak-z3.3 & 3.350 & 214.859354 & 52.850926 & 13.2 & 169	& 10\\
EGS-cPeak-z3.5 & 3.483 & 214.879893 & 52.882881 & 13.1 & 178	& 2 \\
EGS-cPeak-z3.6-1 & 3.620 & 214.882649 & 52.857065 & 13.1 & 157	& 4 \\
EGS-cPeak-z3.6-2 & 3.630 & 214.807821 & 52.779133 & 13.3 & 258	& 5 \\
EGS-cPeak-z3.8 & 3.771 & 214.887038 & 52.923152 & 13.1 & 152	& 3 \\
EGS-cPeak-z3.9 & 3.852 & 214.913536 & 52.879846 & 13.5 & 458	& 6 \\
EGS-cPeak-z4.4 & 4.352 & 214.857767 & 52.876192 & 13.2 & 230	& 6 \\
EGS-cPeak-z4.5 & 4.479 & 214.938785 & 52.940256 & 13.3 & 260	& 9 \\
EGS-cPeak-z4.6-1 & 4.598 & 214.922473 & 52.913510 & 13.4 & 376	& 8 \\
EGS-cPeak-z4.6-2 & 4.633 & 215.012347 & 52.993217 & 13.1 & 186	& 5 \\
EGS-cPeak-z4.8 & 4.809 & 214.868356 & 52.878316 & 13.8 & 790	& 20\\
\enddata
\tablecomments{As Table~\ref{tab:peaks}, but for the moderate confidence detections ($n_\mathrm{Massive}=0$). We label these as ``candidate'' overdensities.}
\end{deluxetable*}

\bibliography{sample701}{}

\begin{thebibliography}{}
\expandafter\ifx\csname natexlab\endcsname\relax\def\natexlab#1{#1}\fi
\providecommand{\url}[1]{\href{#1}{#1}}
\providecommand{\dodoi}[1]{doi:~\href{http://doi.org/#1}{\nolinkurl{#1}}}
\providecommand{\doeprint}[1]{\href{http://ascl.net/#1}{\nolinkurl{http://ascl.net/#1}}}
\providecommand{\doarXiv}[1]{\href{https://arxiv.org/abs/#1}{\nolinkurl{https://arxiv.org/abs/#1}}}

\bibitem[{M.~G. Abadi {et~al.}(1999)Abadi, Moore, \& Bower}]{Abadi1999}
Abadi, M.~G., Moore, B., \& Bower, R.~G. 1999, \bibinfo{title}{Ram pressure
  stripping of spiral galaxies in clusters,} \mnras, 308, 947,
  \dodoi{10.1046/j.1365-8711.1999.02715.x}

\bibitem[{S. Alonso {et~al.}(2012)Alonso, Mesa, Padilla, \&
  Lambas}]{Alonso2012}
Alonso, S., Mesa, V., Padilla, N., \& Lambas, D.~G. 2012,
  \bibinfo{title}{Galaxy interactions,} \aap, 539, A46,
  \dodoi{10.1051/0004-6361/201117901}

\bibitem[{ {Astropy Collaboration} {et~al.}(2013){Astropy Collaboration},
  {Robitaille}, {Tollerud}, {Greenfield}, {Droettboom}, {Bray}, {Aldcroft},
  {Davis}, {Ginsburg}, {Price-Whelan}, {Kerzendorf}, {Conley}, {Crighton},
  {Barbary}, {Muna}, {Ferguson}, {Grollier}, {Parikh}, {Nair}, {Unther},
  {Deil}, {Woillez}, {Conseil}, {Kramer}, {Turner}, {Singer}, {Fox}, {Weaver},
  {Zabalza}, {Edwards}, {Azalee Bostroem}, {Burke}, {Casey}, {Crawford},
  {Dencheva}, {Ely}, {Jenness}, {Labrie}, {Lim}, {Pierfederici}, {Pontzen},
  {Ptak}, {Refsdal}, {Servillat}, \& {Streicher}}]{2013A&A...558A..33A}
{Astropy Collaboration}, {Robitaille}, T.~P., {Tollerud}, E.~J., {et~al.} 2013,
  \bibinfo{title}{{Astropy: A community Python package for astronomy},} \aap,
  558, A33, \dodoi{10.1051/0004-6361/201322068}

\bibitem[{ {Astropy Collaboration} {et~al.}(2018){Astropy Collaboration},
  {Price-Whelan}, {Sip{\H{o}}cz}, {G{\"u}nther}, {Lim}, {Crawford}, {Conseil},
  {Shupe}, {Craig}, {Dencheva}, {Ginsburg}, {VanderPlas}, {Bradley},
  {P{\'e}rez-Su{\'a}rez}, {de Val-Borro}, {Aldcroft}, {Cruz}, {Robitaille},
  {Tollerud}, {Ardelean}, {Babej}, {Bach}, {Bachetti}, {Bakanov}, {Bamford},
  {Barentsen}, {Barmby}, {Baumbach}, {Berry}, {Biscani}, {Boquien}, {Bostroem},
  {Bouma}, {Brammer}, {Bray}, {Breytenbach}, {Buddelmeijer}, {Burke},
  {Calderone}, {Cano Rodr{\'\i}guez}, {Cara}, {Cardoso}, {Cheedella}, {Copin},
  {Corrales}, {Crichton}, {D'Avella}, {Deil}, {Depagne}, {Dietrich}, {Donath},
  {Droettboom}, {Earl}, {Erben}, {Fabbro}, {Ferreira}, {Finethy}, {Fox},
  {Garrison}, {Gibbons}, {Goldstein}, {Gommers}, {Greco}, {Greenfield},
  {Groener}, {Grollier}, {Hagen}, {Hirst}, {Homeier}, {Horton}, {Hosseinzadeh},
  {Hu}, {Hunkeler}, {Ivezi{\'c}}, {Jain}, {Jenness}, {Kanarek}, {Kendrew},
  {Kern}, {Kerzendorf}, {Khvalko}, {King}, {Kirkby}, {Kulkarni}, {Kumar},
  {Lee}, {Lenz}, {Littlefair}, {Ma}, {Macleod}, {Mastropietro}, {McCully},
  {Montagnac}, {Morris}, {Mueller}, {Mumford}, {Muna}, {Murphy}, {Nelson},
  {Nguyen}, {Ninan}, {N{\"o}the}, {Ogaz}, {Oh}, {Parejko}, {Parley}, {Pascual},
  {Patil}, {Patil}, {Plunkett}, {Prochaska}, {Rastogi}, {Reddy Janga},
  {Sabater}, {Sakurikar}, {Seifert}, {Sherbert}, {Sherwood-Taylor}, {Shih},
  {Sick}, {Silbiger}, {Singanamalla}, {Singer}, {Sladen}, {Sooley},
  {Sornarajah}, {Streicher}, {Teuben}, {Thomas}, {Tremblay}, {Turner},
  {Terr{\'o}n}, {van Kerkwijk}, {de la Vega}, {Watkins}, {Weaver}, {Whitmore},
  {Woillez}, {Zabalza}, \& {Astropy Contributors}}]{2018AJ....156..123A}
{Astropy Collaboration}, {Price-Whelan}, A.~M., {Sip{\H{o}}cz}, B.~M., {et~al.}
  2018, \bibinfo{title}{{The Astropy Project: Building an Open-science Project
  and Status of the v2.0 Core Package},} \aj, 156, 123,
  \dodoi{10.3847/1538-3881/aabc4f}

\bibitem[{ {Astropy Collaboration} {et~al.}(2022){Astropy Collaboration},
  {Price-Whelan}, {Lim}, {Earl}, {Starkman}, {Bradley}, {Shupe}, {Patil},
  {Corrales}, {Brasseur}, {N{\"o}the}, {Donath}, {Tollerud}, {Morris},
  {Ginsburg}, {Vaher}, {Weaver}, {Tocknell}, {Jamieson}, {van Kerkwijk},
  {Robitaille}, {Merry}, {Bachetti}, {G{\"u}nther}, {Aldcroft},
  {Alvarado-Montes}, {Archibald}, {B{\'o}di}, {Bapat}, {Barentsen},
  {Baz{\'a}n}, {Biswas}, {Boquien}, {Burke}, {Cara}, {Cara}, {Conroy},
  {Conseil}, {Craig}, {Cross}, {Cruz}, {D'Eugenio}, {Dencheva}, {Devillepoix},
  {Dietrich}, {Eigenbrot}, {Erben}, {Ferreira}, {Foreman-Mackey}, {Fox},
  {Freij}, {Garg}, {Geda}, {Glattly}, {Gondhalekar}, {Gordon}, {Grant},
  {Greenfield}, {Groener}, {Guest}, {Gurovich}, {Handberg}, {Hart},
  {Hatfield-Dodds}, {Homeier}, {Hosseinzadeh}, {Jenness}, {Jones}, {Joseph},
  {Kalmbach}, {Karamehmetoglu}, {Ka{\l}uszy{\'n}ski}, {Kelley}, {Kern},
  {Kerzendorf}, {Koch}, {Kulumani}, {Lee}, {Ly}, {Ma}, {MacBride}, {Maljaars},
  {Muna}, {Murphy}, {Norman}, {O'Steen}, {Oman}, {Pacifici}, {Pascual},
  {Pascual-Granado}, {Patil}, {Perren}, {Pickering}, {Rastogi}, {Roulston},
  {Ryan}, {Rykoff}, {Sabater}, {Sakurikar}, {Salgado}, {Sanghi}, {Saunders},
  {Savchenko}, {Schwardt}, {Seifert-Eckert}, {Shih}, {Jain}, {Shukla}, {Sick},
  {Simpson}, {Singanamalla}, {Singer}, {Singhal}, {Sinha}, {Sip{\H{o}}cz},
  {Spitler}, {Stansby}, {Streicher}, {{\v{S}}umak}, {Swinbank}, {Taranu},
  {Tewary}, {Tremblay}, {de Val-Borro}, {Van Kooten}, {Vasovi{\'c}}, {Verma},
  {de Miranda Cardoso}, {Williams}, {Wilson}, {Winkel}, {Wood-Vasey}, {Xue},
  {Yoachim}, {Zhang}, {Zonca}, \& {Astropy Project
  Contributors}}]{2022ApJ...935..167A}
{Astropy Collaboration}, {Price-Whelan}, A.~M., {Lim}, P.~L., {et~al.} 2022,
  \bibinfo{title}{{The Astropy Project: Sustaining and Growing a
  Community-oriented Open-source Project and the Latest Major Release (v5.0) of
  the Core Package},} \apj, 935, 167, \dodoi{10.3847/1538-4357/ac7c74}

\bibitem[{W.~M. Baker {et~al.}(2025{\natexlab{a}})Baker, Lim, D’Eugenio,
  Maiolino, Ji, Arribas, Bunker, Carniani, Charlot, de Graaff, Hainline,
  Looser, Lyu, Rinaldi, Robertson, Schaller, Schaye, Scholtz, Übler, Williams,
  Willmer, Willott, \& Zhu}]{Baker2025}
Baker, W.~M., Lim, S., D’Eugenio, F., {et~al.} 2025{\natexlab{a}},
  \bibinfo{title}{The abundance and nature of high-redshift quiescent galaxies
  from JADES spectroscopy and the FLAMINGO simulations,} \mnras, 539, 557,
  \dodoi{10.1093/mnras/staf475}

\bibitem[{W.~M. Baker {et~al.}(2025{\natexlab{b}})Baker, Valentino, Lagos, Ito,
  Jespersen, Gottumukkala, Hjorth, Langeroodi, \& Sedgewick}]{Baker2025b}
Baker, W.~M., Valentino, F., Lagos, C. d.~P., {et~al.} 2025{\natexlab{b}},
  \bibinfo{title}{{Exploring over 700 massive quiescent galaxies at z = 2-7:
  Demographics and stellar mass functions},} \doarXiv{2506.04119}

\bibitem[{I.~K. Baldry {et~al.}(2004)Baldry, Glazebrook, Brinkmann, Željko
  Ivezić, Lupton, Nichol, \& Szalay}]{Baldry2004}
Baldry, I.~K., Glazebrook, K., Brinkmann, J., {et~al.} 2004,
  \bibinfo{title}{Quantifying the Bimodal Color‐Magnitude Distribution of
  Galaxies,} \apj, 600, 681, \dodoi{10.1086/380092}

\bibitem[{M.~L. Balogh \& S.~L. Morris(2000)Balogh \& Morris}]{Balogh2000b}
Balogh, M.~L., \& Morris, S.~L. 2000, \bibinfo{title}{H <i>α</i> photometry of
  Abell 2390,} \mnras, 318, 703, \dodoi{10.1046/j.1365-8711.2000.03826.x}

\bibitem[{M.~L. Balogh {et~al.}(2000)Balogh, Navarro, \& Morris}]{Balogh2000a}
Balogh, M.~L., Navarro, J.~F., \& Morris, S.~L. 2000, \bibinfo{title}{The
  Origin of Star Formation Gradients in Rich Galaxy Clusters,} \apj, 540, 113,
  \dodoi{10.1086/309323}

\bibitem[{R.~L. Barone-Nugent {et~al.}(2014)Barone-Nugent, Trenti, Wyithe,
  Bouwens, Oesch, Illingworth, Carollo, Su, Stiavelli, Labbe, \& van
  Dokkum}]{BaroneNugent2014}
Barone-Nugent, R.~L., Trenti, M., Wyithe, J. S.~B., {et~al.} 2014,
  \bibinfo{title}{{MEASUREMENT OF GALAXY CLUSTERING AT z ∼ 7.2 AND THE
  EVOLUTION OF GALAXY BIAS FROM 3.8 {\textless} z {\textless} 8 IN THE XDF,
  GOODS-S, AND GOODS-N},} \apj, 793, 17, \dodoi{10.1088/0004-637X/793/1/17}

\bibitem[{P. Behroozi \& J. Silk(2018)Behroozi \& Silk}]{Behroozi2018}
Behroozi, P., \& Silk, J. 2018, \bibinfo{title}{The most massive galaxies and
  black holes allowed by ΛCDM,} \mnras, 477, 5382,
  \dodoi{10.1093/mnras/sty945}

\bibitem[{M.~R. Blanton {et~al.}(2003)Blanton, Hogg, Bahcall, Baldry,
  Brinkmann, Csabai, Eisenstein, Fukugita, Gunn, Željko Ivezić, Lamb, Lupton,
  Loveday, Munn, Nichol, Okamura, Schlegel, Shimasaku, Strauss, Vogeley, \&
  Weinberg}]{Blanton2003}
Blanton, M.~R., Hogg, D.~W., Bahcall, N.~A., {et~al.} 2003, \bibinfo{title}{The
  Broadband Optical Properties of Galaxies with Redshifts 0.02 \&lt; 
  <i>z</i>  \&lt; 0.22,} \apj, 594, 186, \dodoi{10.1086/375528}

\bibitem[{M. Boylan-Kolchin(2023)Boylan-Kolchin}]{Boylan-Kolchin2023}
Boylan-Kolchin, M. 2023, \bibinfo{title}{Stress testing ΛCDM with
  high-redshift galaxy candidates,} Nature Astronomy, 7, 731,
  \dodoi{10.1038/s41550-023-01937-7}

\bibitem[{G. Brammer(2023{\natexlab{a}})Brammer}]{grizli}
Brammer, G. 2023{\natexlab{a}}, grizli, 1.9.11 Zenodo,
  \dodoi{10.5281/zenodo.8370018}

\bibitem[{G. Brammer(2023{\natexlab{b}})Brammer}]{msaexp}
Brammer, G. 2023{\natexlab{b}}, msaexp: NIRSpec analyis tools, 0.6.17 Zenodo,
  \dodoi{10.5281/zenodo.8319596}

\bibitem[{G. Brammer \& F. Valentino(2025)Brammer \& Valentino}]{brammer2025}
Brammer, G., \& Valentino, F. 2025, The DAWN JWST Archive: Compilation of
  Public NIRSpec Spectra, 4.4 Zenodo, \dodoi{10.5281/zenodo.15472354}

\bibitem[{G.~B. Brammer {et~al.}(2008)Brammer, van Dokkum, \&
  Coppi}]{Brammer2008}
Brammer, G.~B., van Dokkum, P.~G., \& Coppi, P. 2008, \bibinfo{title}{EAZY: A
  Fast, Public Photometric Redshift Code,} \apj, 686, 1503,
  \dodoi{10.1086/591786}

\bibitem[{G.~B. Brammer {et~al.}(2009)Brammer, Whitaker, van Dokkum,
  Marchesini, Labbé, Franx, Kriek, Quadri, Illingworth, Lee, Muzzin, \&
  Rudnick}]{Brammer2009}
Brammer, G.~B., Whitaker, K.~E., van Dokkum, P.~G., {et~al.} 2009,
  \bibinfo{title}{THE DEAD SEQUENCE: A CLEAR BIMODALITY IN GALAXY COLORS FROM z
  = 0 to z = 2.5,} \apjl, 706, L173, \dodoi{10.1088/0004-637X/706/1/L173}

\bibitem[{D. Calzetti {et~al.}(2000)Calzetti, Armus, Bohlin, Kinney, Koornneef,
  \& Storchi‐Bergmann}]{Calzetti2000a}
Calzetti, D., Armus, L., Bohlin, R.~C., {et~al.} 2000, \bibinfo{title}{The Dust
  Content and Opacity of Actively Star‐forming Galaxies,} \apj, 533, 682,
  \dodoi{10.1086/308692}

\bibitem[{A.~C. Carnall {et~al.}(2023)Carnall, McLure, Dunlop, McLeod, Wild,
  Cullen, Magee, Begley, Cimatti, Donnan, Hamadouche, Jewell, \&
  Walker}]{Carnall2023c}
Carnall, A.~C., McLure, R.~J., Dunlop, J.~S., {et~al.} 2023, \bibinfo{title}{{A
  massive quiescent galaxy at redshift 4.658},} \nat, 619, 716,
  \dodoi{10.1038/s41586-023-06158-6}

\bibitem[{A.~C. Carnall {et~al.}(2024)Carnall, Cullen, McLure, McLeod, Begley,
  Donnan, Dunlop, Shapley, Rowlands, Almaini, Arellano-Córdova, Barrufet,
  Cimatti, Ellis, Grogin, Hamadouche, Illingworth, Koekemoer, Leung, Lovell,
  Pérez-González, Santini, Stanton, \& Wild}]{Carnall2024}
Carnall, A.~C., Cullen, F., McLure, R.~J., {et~al.} 2024, \bibinfo{title}{The
  JWST EXCELS survey: too much, too young, too fast? Ultra-massive quiescent
  galaxies at 3 < z < 5,} \mnras, 534, 325, \dodoi{10.1093/mnras/stae2092}

\bibitem[{G. Chabrier(2003)Chabrier}]{Chabrier2003a}
Chabrier, G. 2003, \bibinfo{title}{{Galactic Stellar and Substellar Initial
  Mass Function},} \pasp, 115, 763, \dodoi{10.1086/376392}

\bibitem[{J.~B. Champagne {et~al.}(2025{\natexlab{a}})Champagne, Wang, Zhang,
  Yang, Fan, Hennawi, Sun, Bañados, Bosman, Costa, Eilers, Endsley, Jin, Jun,
  Li, Lin, Liu, Loiacono, Lupi, Mazzucchelli, Pudoka, Protušovà, Rojas-Ruiz,
  Tee, Trebitsch, Venemans, Zhuang, \& Zou}]{Champagne2025}
Champagne, J.~B., Wang, F., Zhang, H., {et~al.} 2025{\natexlab{a}},
  \bibinfo{title}{A Quasar-anchored Protocluster at z = 6.6 in the ASPIRE
  Survey. I. Properties of [O iii ] Emitters in a 10 Mpc Overdensity
  Structure,} \apj, 981, 113, \dodoi{10.3847/1538-4357/adb1bd}

\bibitem[{J.~B. Champagne {et~al.}(2025{\natexlab{b}})Champagne, Wang, Yang,
  Fan, Hennawi, Sun, Bañados, Bosman, Costa, Habouzit, Jin, Jun, Li, Liu,
  Loiacono, Lupi, Mazzucchelli, Pudoka, Rojas-Ruiz, Tee, Trebitsch, Zhang,
  Zhuang, \& Zou}]{Champagne2025a}
Champagne, J.~B., Wang, F., Yang, J., {et~al.} 2025{\natexlab{b}},
  \bibinfo{title}{A Quasar-anchored Protocluster at z = 6.6 in the ASPIRE
  Survey. II. An Environmental Analysis of Galaxy Properties in an Overdense
  Structure,} \apj, 981, 114, \dodoi{10.3847/1538-4357/adb1bc}

\bibitem[{N. Chartab {et~al.}(2020)Chartab, Mobasher, Darvish, Finkelstein,
  Guo, Kodra, Lee, Newman, Pacifici, Papovich, Sattari, Shahidi, Dickinson,
  Faber, Ferguson, Giavalisco, \& Jafariyazani}]{Chartab2020}
Chartab, N., Mobasher, B., Darvish, B., {et~al.} 2020,
  \bibinfo{title}{Large-scale Structures in the CANDELS Fields: The Role of the
  Environment in Star Formation Activity,} \apj, 890, 7,
  \dodoi{10.3847/1538-4357/ab61fd}

\bibitem[{Y.-K. Chiang {et~al.}(2013)Chiang, Overzier, \&
  Gebhardt}]{Chiang2013}
Chiang, Y.-K., Overzier, R., \& Gebhardt, K. 2013, \bibinfo{title}{ANCIENT
  LIGHT FROM YOUNG COSMIC CITIES: PHYSICAL AND OBSERVATIONAL SIGNATURES OF
  GALAXY PROTO-CLUSTERS,} \apj, 779, 127, \dodoi{10.1088/0004-637X/779/2/127}

\bibitem[{H.~G. Chittenden {et~al.}(2025)Chittenden, Glazebrook, Nanayakkara,
  Kawinwanichakij, Lagos, Kimmig, \& Remus}]{Chittenden2025}
Chittenden, H.~G., Glazebrook, K., Nanayakkara, T., {et~al.} 2025,
  \bibinfo{title}{On the unique evolutionary mechanisms of massive quiescent
  galaxies in the epoch of reionisation,} \doarXiv{2504.19696}

\bibitem[{J. Choi {et~al.}(2016)Choi, Dotter, Conroy, Cantiello, Paxton, \&
  Johnson}]{Choi2016}
Choi, J., Dotter, A., Conroy, C., {et~al.} 2016, \bibinfo{title}{MESA
  ISOCHRONES AND STELLAR TRACKS (MIST). I. SOLAR-SCALED MODELS,} \apj, 823,
  102, \dodoi{10.3847/0004-637X/823/2/102}

\bibitem[{C. Conroy \& J.~E. Gunn(2010)Conroy \& Gunn}]{Conroy2010a}
Conroy, C., \& Gunn, J.~E. 2010, \bibinfo{title}{THE PROPAGATION OF
  UNCERTAINTIES IN STELLAR POPULATION SYNTHESIS MODELING. III. MODEL
  CALIBRATION, COMPARISON, AND EVALUATION,} \apj, 712, 833,
  \dodoi{10.1088/0004-637X/712/2/833}

\bibitem[{C. Conroy {et~al.}(2009)Conroy, Gunn, \& White}]{Conroy2009}
Conroy, C., Gunn, J.~E., \& White, M. 2009, \bibinfo{title}{THE PROPAGATION OF
  UNCERTAINTIES IN STELLAR POPULATION SYNTHESIS MODELING. I. THE RELEVANCE OF
  UNCERTAIN ASPECTS OF STELLAR EVOLUTION AND THE INITIAL MASS FUNCTION TO THE
  DERIVED PHYSICAL PROPERTIES OF GALAXIES,} \apj, 699, 486,
  \dodoi{10.1088/0004-637X/699/1/486}

\bibitem[{M.~C. Cooper {et~al.}(2007)Cooper, Newman, Weiner, Yan, Willmer,
  Bundy, Coil, Conselice, Davis, Faber, Gerke, Guhathakurta, Koo, \&
  Noeske}]{Cooper2007a}
Cooper, M.~C., Newman, J.~A., Weiner, B.~J., {et~al.} 2007,
  \bibinfo{title}{{The DEEP2 Galaxy Redshift Survey: the role of galaxy
  environment in the cosmic star formation history},} \mnras, 383, 1058,
  \dodoi{10.1111/j.1365-2966.2007.12613.x}

\bibitem[{O. Cucciati {et~al.}(2018)Cucciati, Lemaux, Zamorani, Fèvre, Tasca,
  Hathi, Lee, Bardelli, Cassata, Garilli, Brun, Maccagni, Pentericci, Thomas,
  Vanzella, Zucca, Lubin, Amorin, Cassarà, Cimatti, Talia, Vergani, Koekemoer,
  Pforr, \& Salvato}]{Cucciati2018}
Cucciati, O., Lemaux, B.~C., Zamorani, G., {et~al.} 2018, \bibinfo{title}{The
  progeny of a cosmic titan: a massive multi-component proto-supercluster in
  formation at z = 2.45 in VUDS,} \aap, 619, A49,
  \dodoi{10.1051/0004-6361/201833655}

\bibitem[{S.~E. Cutler {et~al.}(2024)Cutler, Whitaker, Weaver, 王, Pan,
  Bezanson, Furtak, Labbe, Leja, Price, Cheng, Clausen, Cullen, Dayal,
  de~Graaff, Dickinson, Dunlop, Feldmann, Franx, Giavalisco, Glazebrook,
  Greene, Grogin, Illingworth, Koekemoer, Kokorev, Marchesini, Maseda, Miller,
  Nanayakkara, Nelson, Setton, Shipley, \& Suess}]{Cutler2024}
Cutler, S.~E., Whitaker, K.~E., Weaver, J.~R., {et~al.} 2024,
  \bibinfo{title}{Two Distinct Classes of Quiescent Galaxies at Cosmic Noon
  Revealed by JWST PRIMER and UNCOVER,} \apjl, 967, L23,
  \dodoi{10.3847/2041-8213/ad464c}

\bibitem[{B. Darvish {et~al.}(2015)Darvish, Mobasher, Sobral, Scoville, \&
  Aragon-Calvo}]{Darvish2015}
Darvish, B., Mobasher, B., Sobral, D., Scoville, N., \& Aragon-Calvo, M. 2015,
  \bibinfo{title}{A COMPARATIVE STUDY OF DENSITY FIELD ESTIMATION FOR GALAXIES:
  NEW INSIGHTS INTO THE EVOLUTION OF GALAXIES WITH ENVIRONMENT IN COSMOS OUT TO
  z ∼ 3,} \apj, 805, 121, \dodoi{10.1088/0004-637X/805/2/121}

\bibitem[{A. de~Graaff {et~al.}(2024)de~Graaff, Rix, Carniani, Suess, Charlot,
  Curtis-Lake, Arribas, Baker, Boyett, Bunker, Cameron, Chevallard, Curti,
  Eisenstein, Franx, Hainline, Hausen, Ji, Johnson, Jones, Maiolino, Maseda,
  Nelson, Parlanti, Rawle, Robertson, Tacchella, {\"{U}}bler, Williams,
  Willmer, \& Willott}]{DeGraaff2023}
de~Graaff, A., Rix, H.~W., Carniani, S., {et~al.} 2024,
  \bibinfo{title}{{Ionised gas kinematics and dynamical masses of z {\&} 6
  galaxies from JADES/NIRSpec high-resolution spectroscopy},} \aap, 684,
  \dodoi{10.1051/0004-6361/202347755}

\bibitem[{A. de~Graaff {et~al.}(2025{\natexlab{a}})de~Graaff, Setton, Brammer,
  Cutler, Suess, Labbé, Leja, Weibel, Maseda, Whitaker, Bezanson, Boogaard,
  Cleri, Lucia, Franx, Greene, Hirschmann, Matthee, McConachie, Naidu, Oesch,
  Price, Rix, Valentino, Wang, \& Williams}]{DeGraaff2024}
de~Graaff, A., Setton, D.~J., Brammer, G., {et~al.} 2025{\natexlab{a}},
  \bibinfo{title}{Efficient formation of a massive quiescent galaxy at redshift
  4.9,} Nature Astronomy, 9, 280, \dodoi{10.1038/s41550-024-02424-3}

\bibitem[{A. de~Graaff {et~al.}(2025{\natexlab{b}})de~Graaff, Brammer, Weibel,
  Lewis, Maseda, Oesch, Bezanson, Boogaard, Cleri, Cooper, Gottumukkala,
  Greene, Hirschmann, Hviding, Katz, Labbé, Leja, Matthee, McConachie, Miller,
  Naidu, Price, Rix, Setton, Suess, Wang, Whitaker, \& Williams}]{DeGraaff2025}
de~Graaff, A., Brammer, G., Weibel, A., {et~al.} 2025{\natexlab{b}},
  \bibinfo{title}{RUBIES: A complete census of the bright and red distant
  Universe with JWST/NIRSpec,} \aap, 697, A189,
  \dodoi{10.1051/0004-6361/202452186}

\bibitem[{C.~T. Donnan {et~al.}(2024)Donnan, McLure, Dunlop, McLeod, Magee,
  Arellano-Córdova, Barrufet, Begley, Bowler, Carnall, Cullen, Ellis, Fontana,
  Illingworth, Grogin, Hamadouche, Koekemoer, Liu, Mason, Santini, \&
  Stanton}]{Donnan2024}
Donnan, C.~T., McLure, R.~J., Dunlop, J.~S., {et~al.} 2024,
  \bibinfo{title}{<i>JWST</i> PRIMER: a new multifield determination of the
  evolving galaxy UV luminosity function at redshifts <i>z</i> ≃ 9 – 15,}
  \mnras, 533, 3222, \dodoi{10.1093/mnras/stae2037}

\bibitem[{A. Dotter(2016)Dotter}]{Dotter2016}
Dotter, A. 2016, \bibinfo{title}{MESA ISOCHRONES AND STELLAR TRACKS (MIST) 0:
  METHODS FOR THE CONSTRUCTION OF STELLAR ISOCHRONES,} \apjs, 222, 8,
  \dodoi{10.3847/0067-0049/222/1/8}

\bibitem[{A. Dressler(1980)Dressler}]{Dressler1980}
Dressler, A. 1980, \bibinfo{title}{Galaxy morphology in rich clusters -
  Implications for the formation and evolution of galaxies,} \apj, 236, 351,
  \dodoi{10.1086/157753}

\bibitem[{M.~J. Drinkwater {et~al.}(2001)Drinkwater, Gregg, \&
  Colless}]{Drinkwater2001}
Drinkwater, M.~J., Gregg, M.~D., \& Colless, M. 2001,
  \bibinfo{title}{Substructure and Dynamics of the Fornax Cluster,} \apj, 548,
  L139, \dodoi{10.1086/319113}

\bibitem[{A.~H. Edward {et~al.}(2024)Edward, Balogh, Bahe, Cooper, Hatch,
  Marchioni, Muzzin, Noble, Rudnick, Vulcani, Wilson, Lucia, Demarco, Forrest,
  Hirschmann, Castignani, Cerulo, Finn, Hewitt, Jablonka, Kodama, Maurogordato,
  Nantais, \& Xie}]{Edward2024}
Edward, A.~H., Balogh, M.~L., Bahe, Y.~M., {et~al.} 2024, \bibinfo{title}{The
  stellar mass function of quiescent galaxies in 2 < z < 2.5 protoclusters,}
  \mnras, 527, 8598, \dodoi{10.1093/mnras/stad3751}

\bibitem[{J. Falcón-Barroso {et~al.}(2011)Falcón-Barroso, Sánchez-Blázquez,
  Vazdekis, Ricciardelli, Cardiel, Cenarro, Gorgas, \&
  Peletier}]{Falcn-Barroso2011}
Falcón-Barroso, J., Sánchez-Blázquez, P., Vazdekis, A., {et~al.} 2011,
  \bibinfo{title}{An updated MILES stellar library and stellar population
  models,} \aap, 532, A95, \dodoi{10.1051/0004-6361/201116842}

\bibitem[{S.~L. Finkelstein {et~al.}(2023)Finkelstein, Bagley, Ferguson,
  Wilkins, Kartaltepe, Papovich, Yung, Haro, Behroozi, Dickinson, Kocevski,
  Koekemoer, Larson, Bail, Morales, Pérez-González, Burgarella, Davé,
  Hirschmann, Somerville, Wuyts, Bromm, Casey, Fontana, Fujimoto, Gardner,
  Giavalisco, Grazian, Grogin, Hathi, Hutchison, Jha, Jogee, Kewley,
  Kirkpatrick, Long, Lotz, Pentericci, Pierel, Pirzkal, Ravindranath, Ryan,
  Trump, Yang, Bhatawdekar, Bisigello, Buat, Calabrò, Castellano, Cleri,
  Cooper, Croton, Daddi, Dekel, Elbaz, Franco, Gawiser, Holwerda,
  Huertas-Company, Jaskot, Leung, Lucas, Mobasher, Pandya, Tacchella, Weiner,
  \& Zavala}]{Finkelstein2023}
Finkelstein, S.~L., Bagley, M.~B., Ferguson, H.~C., {et~al.} 2023,
  \bibinfo{title}{CEERS Key Paper. I. An Early Look into the First 500 Myr of
  Galaxy Formation with JWST,} \apjl, 946, L13,
  \dodoi{10.3847/2041-8213/acade4}

\bibitem[{S.~L. Finkelstein {et~al.}(2024)Finkelstein, Leung, Bagley,
  Dickinson, Ferguson, Papovich, Akins, Haro, Davé, Dekel, Kartaltepe,
  Kocevski, Koekemoer, Pirzkal, Somerville, Yung, Amorín, Backhaus, Behroozi,
  Bisigello, Bromm, Casey, Óscar A.~Chávez~Ortiz, Cheng, Chworowsky, Cleri,
  Cooper, Davis, de~la Vega, Elbaz, Franco, Fontana, Fujimoto, Giavalisco,
  Grogin, Holwerda, Huertas-Company, Hirschmann, Iyer, Jogee, Jung, Larson,
  Lucas, Mobasher, Morales, Morley, Mukherjee, Pérez-González, Ravindranath,
  Rodighiero, Rowland, Tacchella, Taylor, Trump, \& Wilkins}]{Finkelstein2024}
Finkelstein, S.~L., Leung, G. C.~K., Bagley, M.~B., {et~al.} 2024,
  \bibinfo{title}{The Complete CEERS Early Universe Galaxy Sample: A
  Surprisingly Slow Evolution of the Space Density of Bright Galaxies at z ∼
  8.5–14.5,} \apjl, 969, L2, \dodoi{10.3847/2041-8213/ad4495}

\bibitem[{B. Forrest {et~al.}(2023)Forrest, Lemaux, Shah, Staab, McConachie,
  Cucciati, Gal, Hung, Lubin, Cassarà, Cassata, Chang, Cooper, Decarli, Gomez,
  Gururajan, Hathi, Kashino, Marchesini, Marsan, McDonald, Muzzin, Shen,
  Stawinski, Talia, Vergani, Wilson, \& Zamorani}]{Forrest2023}
Forrest, B., Lemaux, B.~C., Shah, E., {et~al.} 2023,
  \bibinfo{title}{Elentári:a massive proto-supercluster at <scp> <i>z</i> ∼
  3.3 </scp> in the <scp>cosmos</scp> field,} \mnras, 526, L56,
  \dodoi{10.1093/mnrasl/slad114}

\bibitem[{B. Forrest {et~al.}(2024)Forrest, Lemaux, Shah, Staab, Gal, Lubin,
  Cooper, Cucciati, Hung, McConachie, Muzzin, Wilson, Bardelli, Cassarà,
  Chang, Giddings, Golden-Marx, Hathi, Stawinski, \& Zucca}]{Forrest2024}
Forrest, B., Lemaux, B.~C., Shah, E.~A., {et~al.} 2024,
  \bibinfo{title}{Environmental Effects on the Stellar Mass Function in a z ∼
  3.3 Overdensity of Galaxies in the COSMOS Field*,} \apj, 971, 169,
  \dodoi{10.3847/1538-4357/ad5e78}

\bibitem[{Y. Fudamoto {et~al.}(2025)Fudamoto, Helton, Lin, Sun, Behroozi,
  Hsiao, Egami, Bunker, Harikane, Ouchi, Liu, Liu, Maiolino, Ji, Jin, Tee,
  Wang, Willmer, Xu, \& Zhu}]{Fudamoto2025}
Fudamoto, Y., Helton, J.~M., Lin, X., {et~al.} 2025, \bibinfo{title}{SAPPHIRES:
  A Galaxy Over-Density in the Heart of Cosmic Reionization at $z=8.47$,}
  \url{http://arxiv.org/abs/2503.15597}

\bibitem[{B. Garilli {et~al.}(2021)Garilli, McLure, Pentericci, Franzetti,
  Gargiulo, Carnall, Cucciati, Iovino, Amorin, Bolzonella, Bongiorno,
  Castellano, Cimatti, Cirasuolo, Cullen, Dunlop, Elbaz, Finkelstein, Fontana,
  Fontanot, Fumana, Guaita, Hartley, Jarvis, Juneau, Maccagni, McLeod, Nandra,
  Pompei, Pozzetti, Scodeggio, Talia, Calabr{\`{o}}, Cresci, Fynbo, Hathi,
  Hibon, Koekemoer, Magliocchetti, Salvato, Vietri, Zamorani, Almaini,
  Balestra, Bardelli, Begley, Brammer, Bell, Bowler, Brusa, Buitrago, Caputi,
  Cassata, Charlot, Citro, Cristiani, Curtis-Lake, Dickinson, Fazio, Ferguson,
  Fiore, Franco, Georgakakis, Giavalisco, Grazian, Hamadouche, Jung, Kim,
  Khusanova, {Le F{\`{e}}vre}, Longhetti, Lotz, Mannucci, Maltby, Matsuoka,
  Mendez-Hernandez, Mendez-Abreu, Mignoli, Moresco, Nonino, Pannella, Papovich,
  Popesso, Roberts-Borsani, Rosario, Saldana-Lopez, Santini, Saxena, Schaerer,
  Schreiber, Stark, Tasca, Thomas, Vanzella, Wild, Williams, \&
  Zucca}]{Garilli2021}
Garilli, B., McLure, R., Pentericci, L., {et~al.} 2021, \bibinfo{title}{{The
  VANDELS ESO public spectroscopic survey},} \aap, 647, A150,
  \dodoi{10.1051/0004-6361/202040059}

\bibitem[{K. Glazebrook {et~al.}(2024)Glazebrook, Nanayakkara, Schreiber,
  Lagos, Kawinwanichakij, Jacobs, Chittenden, Brammer, Kacprzak, Labbe,
  Marchesini, Marsan, Oesch, Papovich, Remus, Tran, Esdaile, \&
  Chandro-Gomez}]{Glazebrook2024}
Glazebrook, K., Nanayakkara, T., Schreiber, C., {et~al.} 2024,
  \bibinfo{title}{A massive galaxy that formed its stars at z ≈ 11,} \nat,
  628, 277, \dodoi{10.1038/s41586-024-07191-9}

\bibitem[{J.~E. Gunn \& I.~G. J.~Richard(1972)Gunn \& J.~Richard}]{Gunn1972}
Gunn, J.~E., \& J.~Richard, I.~G. 1972, \bibinfo{title}{On the Infall of Matter
  Into Clusters of Galaxies and Some Effects on Their Evolution,} \apj, 176, 1,
  \dodoi{10.1086/151605}

\bibitem[{C.~R. {Harris} {et~al.}(2020){Harris}, {Millman}, {van der Walt},
  {Gommers}, {Virtanen}, {Cournapeau}, {Wieser}, {Taylor}, {Berg}, {Smith},
  {Kern}, {Picus}, {Hoyer}, {van Kerkwijk}, {Brett}, {Haldane}, {del R{\'\i}o},
  {Wiebe}, {Peterson}, {G{\'e}rard-Marchant}, {Sheppard}, {Reddy}, {Weckesser},
  {Abbasi}, {Gohlke}, \& {Oliphant}}]{2020Natur.585..357H}
{Harris}, C.~R., {Millman}, K.~J., {van der Walt}, S.~J., {et~al.} 2020,
  \bibinfo{title}{{Array programming with NumPy},} \nat, 585, 357,
  \dodoi{10.1038/s41586-020-2649-2}

\bibitem[{K.~E. Heintz {et~al.}(2024)Heintz, Watson, Brammer, Vejlgaard,
  Hutter, Strait, Matthee, Oesch, Jakobsson, Tanvir, Laursen, Naidu, Mason,
  Killi, Jung, Hsiao, Abdurro’uf, Coe, Haro, Finkelstein, \&
  Toft}]{Heintz2024}
Heintz, K.~E., Watson, D., Brammer, G., {et~al.} 2024, \bibinfo{title}{Strong
  damped Lyman-a absorption in young star-forming galaxies at redshifts 9 to
  11,} Science, 384, 890, \dodoi{10.1126/science.adj0343}

\bibitem[{J.~M. Helton {et~al.}(2024{\natexlab{a}})Helton, Sun, Woodrum,
  Hainline, Willmer, Rieke, Rieke, Tacchella, Robertson, Johnson, Alberts,
  Eisenstein, Hausen, Bonaventura, Bunker, Charlot, Curti, Curtis-Lake, Looser,
  Maiolino, Willott, Witstok, Boyett, Chen, Egami, Endsley, Hviding, Jaffe, Ji,
  Lyu, \& Sandles}]{Helton2024}
Helton, J.~M., Sun, F., Woodrum, C., {et~al.} 2024{\natexlab{a}},
  \bibinfo{title}{The JWST Advanced Deep Extragalactic Survey: Discovery of an
  Extreme Galaxy Overdensity at z = 5.4 with JWST/NIRCam in GOODS-S,} \apj,
  962, 124, \dodoi{10.3847/1538-4357/ad0da7}

\bibitem[{J.~M. Helton {et~al.}(2024{\natexlab{b}})Helton, Sun, Woodrum,
  Hainline, Willmer, Rieke, Rieke, Alberts, Eisenstein, Tacchella, Robertson,
  Johnson, Baker, Bhatawdekar, Bunker, Chen, Egami, Ji, Maiolino, Willott, \&
  Witstok}]{Helton2024a}
Helton, J.~M., Sun, F., Woodrum, C., {et~al.} 2024{\natexlab{b}},
  \bibinfo{title}{Identification of High-redshift Galaxy Overdensities in
  GOODS-N and GOODS-S,} \apj, 974, 41, \dodoi{10.3847/1538-4357/ad6867}

\bibitem[{N.~K. Hine {et~al.}(2016)Hine, Geach, Alexander, Lehmer, Chapman, \&
  Matsuda}]{Hine2016}
Hine, N.~K., Geach, J.~E., Alexander, D.~M., {et~al.} 2016, \bibinfo{title}{An
  enhanced merger fraction within the galaxy population of the SSA22
  protocluster at z  = 3.1,} \mnras, 455, 2363, \dodoi{10.1093/mnras/stv2448}

\bibitem[{D. Hung {et~al.}(2020)Hung, Lemaux, Gal, Tomczak, Lubin, Cucciati,
  Pelliccia, Shen, Fèvre, Wu, Kocevski, Mei, \& Squires}]{Hung2020}
Hung, D., Lemaux, B.~C., Gal, R.~R., {et~al.} 2020,
  \bibinfo{title}{Establishing a new technique for discovering large-scale
  structure using the ORELSE survey,} \mnras, 491, 5524,
  \dodoi{10.1093/mnras/stz3164}

\bibitem[{D. Hung {et~al.}(2025)Hung, Lemaux, Cucciati, Forrest, Shah, Gal,
  Giddings, Sikorski, Golden-Marx, Lubin, Hathi, Zamorani, Shen, Bardelli,
  Cassarà, Lucia, Fontanot, Garilli, Guaita, Hirschmann, Lee, Newman,
  Ramakrishnan, Vergani, Xie, \& Zucca}]{Hung2024}
Hung, D., Lemaux, B.~C., Cucciati, O., {et~al.} 2025,
  \bibinfo{title}{Discovering Large-scale Structure at 2 < z < 5 in the
  C3VO Survey,} \apj, 980, 155, \dodoi{10.3847/1538-4357/ada616}

\bibitem[{J.~D. {Hunter}(2007){Hunter}}]{2007CSE.....9...90H}
{Hunter}, J.~D. 2007, \bibinfo{title}{{Matplotlib: A 2D Graphics Environment},}
  Computing in Science and Engineering, 9, 90, \dodoi{10.1109/MCSE.2007.55}

\bibitem[{F. Huško {et~al.}(2022)Huško, Lacey, \& Baugh}]{Huko2022}
Huško, F., Lacey, C.~G., \& Baugh, C.~M. 2022, \bibinfo{title}{The buildup of
  galaxies and their spheroids: The contributions of mergers, disc
  instabilities, and star formation,} \mnras, 518, 5323,
  \dodoi{10.1093/mnras/stac3152}

\bibitem[{R.~E. Hviding {et~al.}(2025)Hviding, de~Graaff, Miller, Setton,
  Greene, Labbé, Brammer, Bezanson, Boogaard, Cleri, Leja, Maseda, McConachie,
  Matthee, Naidu, Oesch, Wang, Whitaker, \& Williams}]{Hviding2025}
Hviding, R.~E., de~Graaff, A., Miller, T.~B., {et~al.} 2025,
  \bibinfo{title}{RUBIES: A Spectroscopic Census of Little Red Dots; All
  V-Shaped Point Sources Have Broad Lines,} \doarXiv{2506.05459}

\bibitem[{K. Ito {et~al.}(2023)Ito, Tanaka, Valentino, Toft, Brammer, Gould,
  Ilbert, Kashikawa, Kubo, Liang, McCracken, \& Weaver}]{Ito2023}
Ito, K., Tanaka, M., Valentino, F., {et~al.} 2023, \bibinfo{title}{COSMOS2020:
  Discovery of a Protocluster of Massive Quiescent Galaxies at z = 2.77,}
  \apjl, 945, L9, \dodoi{10.3847/2041-8213/acb49b}

\bibitem[{K. Ito {et~al.}(2025{\natexlab{a}})Ito, Valentino, Farcy, Lucia,
  Lagos, Hirschmann, Brammer, de~Graaff, Blánquez-Sesé, Ceverino, Faisst,
  Fontanot, Gillman, Hamadouche, Heintz, Jin, Jespersen, Kubo, Lee, Magdis,
  Man, Onodera, Rizzo, Shimakawa, Tanaka, Toft, Whitaker, Xie, \&
  Zhu}]{Ito2025}
Ito, K., Valentino, F., Farcy, M., {et~al.} 2025{\natexlab{a}},
  \bibinfo{title}{A merging pair of massive quiescent galaxies at <i>z</i> =
  3.44 in the Cosmic Vine,} \aap, 697, A111,
  \dodoi{10.1051/0004-6361/202453211}

\bibitem[{K. Ito {et~al.}(2025{\natexlab{b}})Ito, Valentino, Brammer,
  Hamadouche, Whitaker, Kokorev, Zhu, Kakimoto, Wu, Antwi-Danso, Baker,
  Ceverino, Faisst, Farcy, Fujimoto, Gallazzi, Gillman, Gottumukkala, Heintz,
  Hirschmann, Jespersen, Kubo, Lee, Magdis, Onodera, Shimakawa, Tanaka, Toft,
  \& Weaver}]{Ito2025b}
Ito, K., Valentino, F., Brammer, G., {et~al.} 2025{\natexlab{b}}
  \doarXiv{2506.22642}

\bibitem[{C.~K. Jespersen {et~al.}(2025)Jespersen, Carnall, \&
  Lovell}]{Jespersen2025}
Jespersen, C.~K., Carnall, A.~C., \& Lovell, C.~C. 2025,
  \bibinfo{title}{Explaining Ultramassive Quiescent Galaxies at 3 \&lt;
  <i>z</i> \&lt; 5 in the Context of Their Environments,} \apjl, 988, L19,
  \dodoi{10.3847/2041-8213/adeb7c}

\bibitem[{H.-Y. Jian {et~al.}(2012)Jian, Lin, \& Chiueh}]{Jian2012}
Jian, H.-Y., Lin, L., \& Chiueh, T. 2012, \bibinfo{title}{ENVIRONMENTAL
  DEPENDENCE OF THE GALAXY MERGER RATE IN A ΛCDM UNIVERSE,} \apj, 754, 26,
  \dodoi{10.1088/0004-637X/754/1/26}

\bibitem[{S. Jin {et~al.}(2024)Jin, Sillassen, Magdis, Brinch, Shuntov,
  Brammer, Gobat, Valentino, Carnall, Lee, Vijayan, Gillman, Kokorev, Bail,
  Greve, Gullberg, Gould, \& Toft}]{Jin2024a}
Jin, S., Sillassen, N.~B., Magdis, G.~E., {et~al.} 2024, \bibinfo{title}{Cosmic
  Vine: A z = 3.44 large-scale structure hosting massive quiescent galaxies,}
  /aap, 683, \dodoi{10.1051/0004-6361/202348540}

\bibitem[{T. joblib developers(2025)joblib developers}]{joblib}
joblib developers, T. 2025, joblib, 1.5.1 Zenodo,
  \dodoi{10.5281/zenodo.15496554}

\bibitem[{B. {Johnson} \& J. {Leja}(2017){Johnson} \& {Leja}}]{Johnson2017}
{Johnson}, B., \& {Leja}, J. 2017, {Bd-J/Prospector: Initial Release}, v0.1
  Zenodo, \dodoi{10.5281/zenodo.1116491}

\bibitem[{B.~D. {Johnson} {et~al.}(2021){Johnson}, {Leja}, {Conroy}, \&
  {Speagle}}]{Johnson2021}
{Johnson}, B.~D., {Leja}, J., {Conroy}, C., \& {Speagle}, J.~S. 2021,
  \bibinfo{title}{{Stellar Population Inference with Prospector},} \apjs, 254,
  22, \dodoi{10.3847/1538-4365/abef67}

\bibitem[{T. Kakimoto {et~al.}(2024)Kakimoto, Tanaka, Onodera, Shimakawa, Wu,
  Gould, Ito, Jin, Kubo, Suzuki, Toft, Valentino, \& Yabe}]{Kakimoto2024}
Kakimoto, T., Tanaka, M., Onodera, M., {et~al.} 2024, \bibinfo{title}{A Massive
  Quiescent Galaxy in a Group Environment at z = 4.53,} \apj, 963, 49,
  \dodoi{10.3847/1538-4357/ad1ff1}

\bibitem[{B.~S. Kalita {et~al.}(2021)Kalita, Daddi, D’Eugenio, Valentino,
  Rich, Gómez-Guijarro, Coogan, Delvecchio, Elbaz, Neill, Puglisi, \&
  Strazzullo}]{Kalita2021}
Kalita, B.~S., Daddi, E., D’Eugenio, C., {et~al.} 2021, \bibinfo{title}{An
  Ancient Massive Quiescent Galaxy Found in a Gas-rich z ∼ 3 Group,} \apjl,
  917, L17, \dodoi{10.3847/2041-8213/ac16dc}

\bibitem[{G. Kauffmann {et~al.}(2003)Kauffmann, Heckman, White, Charlot,
  Tremonti, Brinchmann, Bruzual, Peng, Seibert, Bernardi, Blanton, Brinkmann,
  Castander, Csábai, Fukugita, Ivezic, Munn, Nichol, Padmanabhan, Thakar,
  Weinberg, \& York}]{Kauffmann2003}
Kauffmann, G., Heckman, T.~M., White, D. M.~S., {et~al.} 2003,
  \bibinfo{title}{Stellar masses and star formation histories for 10
  <sup>5</sup> galaxies from the Sloan Digital Sky Survey,} \mnras, 341, 33,
  \dodoi{10.1046/j.1365-8711.2003.06291.x}

\bibitem[{L. Kawinwanichakij {et~al.}(2025)Kawinwanichakij, Glazebrook,
  Nanayakkara, Kacprzak, Chittenden, Jacobs, Ángel Chandro-Gómez, Lagos,
  Marchesini, Martínez-Marín, Oesch, \& Remus}]{Kawinwanichakij2025}
Kawinwanichakij, L., Glazebrook, K., Nanayakkara, T., {et~al.} 2025,
  \bibinfo{title}{Stellar Mass-Size Relation and Morphology of Massive
  Quiescent Galaxies at $3 < z < 4$ with JWST,} \dodoi{2505.03089}

\bibitem[{M. Killi {et~al.}(2024)Killi, Watson, Brammer, McPartland,
  Antwi-Danso, Newshore, Coe, Allen, Fynbo, Gould, Heintz, Rusakov, \&
  Vejlgaard}]{Killi2024}
Killi, M., Watson, D., Brammer, G., {et~al.} 2024, \bibinfo{title}{Deciphering
  the JWST spectrum of a ‘little red dot’ at <i>z</i> ∼ 4.53: An obscured
  AGN and its star-forming host,} \aap, 691, A52,
  \dodoi{10.1051/0004-6361/202348857}

\bibitem[{M. Kriek \& C. Conroy(2013)Kriek \& Conroy}]{Kriek2013}
Kriek, M., \& Conroy, C. 2013, \bibinfo{title}{THE DUST ATTENUATION LAW IN
  DISTANT GALAXIES: EVIDENCE FOR VARIATION WITH SPECTRAL TYPE,} \apj, 775, L16,
  \dodoi{10.1088/2041-8205/775/1/L16}

\bibitem[{M. Kubo {et~al.}(2021)Kubo, Umehata, Matsuda, Kajisawa, Steidel,
  Yamada, Tanaka, Hatsukade, Tamura, Nakanishi, Kohno, Lee, \&
  Matsuda}]{Kubo2021b}
Kubo, M., Umehata, H., Matsuda, Y., {et~al.} 2021, \bibinfo{title}{A Massive
  Quiescent Galaxy Confirmed in a Protocluster at z = 3.09,} \apj, 919, 6,
  \dodoi{10.3847/1538-4357/ac0cf8}

\bibitem[{C.~d.~P. Lagos {et~al.}(2024)Lagos, Valentino, Wright, de~Graaff,
  Glazebrook, {De Lucia}, Robotham, Nanayakkara, Chandro-Gomez, Bravo, Baugh,
  Harborne, Hirschmann, Fontanot, Xie, \& Chittenden}]{Lagos2024a}
Lagos, C. d.~P., Valentino, F., Wright, R.~J., {et~al.} 2024,
  \bibinfo{title}{{The diverse star formation histories of early massive,
  quenched galaxies in modern galaxy formation simulations},} \mnras, 536,
  2324, \dodoi{10.1093/mnras/stae2626}

\bibitem[{N. Laporte {et~al.}(2022)Laporte, Zitrin, Dole, Roberts-Borsani,
  Furtak, \& Witten}]{Laporte2022}
Laporte, N., Zitrin, A., Dole, H., {et~al.} 2022, \bibinfo{title}{A lensed
  protocluster candidate at z = 7.66 identified in JWST observations of the
  galaxy cluster SMACS0723−7327,} \aap, 667, L3,
  \dodoi{10.1051/0004-6361/202244719}

\bibitem[{R.~B. Larson {et~al.}(1980)Larson, Tinsley, \& Caldwell}]{Larson1980}
Larson, R.~B., Tinsley, B.~M., \& Caldwell, C.~N. 1980, \bibinfo{title}{The
  evolution of disk galaxies and the origin of S0 galaxies,} \apj, 237, 692,
  \dodoi{10.1086/157917}

\bibitem[{B.~D. Lehmer {et~al.}(2009)Lehmer, Alexander, Geach, Smail,
  Basu-Zych, Bauer, Chapman, Matsuda, Scharf, Volonteri, \&
  Yamada}]{Lehmer2009b}
Lehmer, B.~D., Alexander, D.~M., Geach, J.~E., {et~al.} 2009,
  \bibinfo{title}{{The chandra deep protocluster survey: Evidence for an
  enhancement of AGN activity in the SSA22 protocluster at z = 3.09},} \apj,
  691, 687, \dodoi{10.1088/0004-637X/691/1/687}

\bibitem[{J. Leja {et~al.}(2019)Leja, Carnall, Johnson, Conroy, \&
  Speagle}]{Leja2019}
Leja, J., Carnall, A.~C., Johnson, B.~D., Conroy, C., \& Speagle, J.~S. 2019,
  \bibinfo{title}{How to Measure Galaxy Star Formation Histories. II.
  Nonparametric Models,} \apj, 876, 3, \dodoi{10.3847/1538-4357/ab133c}

\bibitem[{J. Leja {et~al.}(2017)Leja, Johnson, Conroy, van Dokkum, \&
  Byler}]{Leja2017}
Leja, J., Johnson, B.~D., Conroy, C., van Dokkum, P.~G., \& Byler, N. 2017,
  \bibinfo{title}{Deriving Physical Properties from Broadband Photometry with
  Prospector: Description of the Model and a Demonstration of its Accuracy
  Using 129 Galaxies in the Local Universe,} \apj, 837, 170,
  \dodoi{10.3847/1538-4357/aa5ffe}

\bibitem[{B.~C. Lemaux {et~al.}(2018)Lemaux, Fèvre, Cucciati, Ribeiro, Tasca,
  Zamorani, Ilbert, Thomas, Bardelli, Cassata, Hathi, Pforr, Smolčić,
  Delvecchio, Novak, Berta, McCracken, Koekemoer, Amorín, Garilli, Maccagni,
  Schaerer, \& Zucca}]{Lemaux2018}
Lemaux, B.~C., Fèvre, O.~L., Cucciati, O., {et~al.} 2018, \bibinfo{title}{The
  VIMOS Ultra-Deep Survey: Emerging from the dark, a massive proto-cluster at z
  ~ 4.57,} \aap, 615, A77, \dodoi{10.1051/0004-6361/201730870}

\bibitem[{B.~C. Lemaux {et~al.}(2022)Lemaux, Cucciati, Fèvre, Zamorani, Lubin,
  Hathi, Ilbert, Pelliccia, Amorín, Bardelli, Cassata, Gal, Garilli, Guaita,
  Giavalisco, Hung, Koekemoer, Maccagni, Pentericci, Ribeiro, Schaerer, Shah,
  Shen, Staab, Talia, Thomas, Tomczak, Tresse, Vanzella, Vergani, \&
  Zucca}]{Lemaux2022a}
Lemaux, B.~C., Cucciati, O., Fèvre, O.~L., {et~al.} 2022, \bibinfo{title}{The
  VIMOS Ultra Deep Survey: The reversal of the star-formation rate − density
  relation at 2 < z < 5,} \aap, 662, A33, \dodoi{10.1051/0004-6361/202039346}

\bibitem[{M. Lepore {et~al.}(2024)Lepore, Mascolo, Tozzi, Churazov,
  Mroczkowski, Borgani, Carilli, Gaspari, Ginolfi, Liu, Pentericci, Rasia,
  Rosati, Röttgering, Anderson, Dannerbauer, Miley, \& Norman}]{Lepore2024}
Lepore, M., Mascolo, L.~D., Tozzi, P., {et~al.} 2024, \bibinfo{title}{Feeding
  and feedback processes in the Spiderweb proto-intracluster medium,} \aap,
  682, A186, \dodoi{10.1051/0004-6361/202347538}

\bibitem[{X. Lin {et~al.}(2025)Lin, Egami, Sun, Zhang, Fan, Helton, Wang,
  Bunker, Cai, Eisenstein, Jaffe, Ji, Jin, Pudoka, Tacchella, Tee, Rinaldi,
  Robertson, Sun, Willmer, Willott, Zhang, \& Zhu}]{Lin2025}
Lin, X., Egami, E., Sun, F., {et~al.} 2025, \bibinfo{title}{The Luminosity
  Function and Clustering of H$α$ Emitting Galaxies at $z\approx4-6$ from a
  Complete NIRCam Grism Redshift Survey,} \url{http://arxiv.org/abs/2504.08028}

\bibitem[{S. Liu {et~al.}(2023)Liu, Zheng, Shi, Cai, Fan, Wang, Yuan, Xu, Pan,
  Liu, Qin, Zhang, \& Wen}]{Liu2023}
Liu, S., Zheng, X.~Z., Shi, D.~D., {et~al.} 2023, \bibinfo{title}{What boost
  galaxy mergers in two massive galaxy protoclusters at z  = 2.24?} \mnras,
  523, 2422, \dodoi{10.1093/mnras/stad1543}

\bibitem[{J.~M. Lotz {et~al.}(2013)Lotz, Papovich, Faber, Ferguson, Grogin,
  Guo, Kocevski, Koekemoer, Lee, McIntosh, Momcheva, Rudnick, Saintonge, Tran,
  van~der Wel, \& Willmer}]{Lotz2013}
Lotz, J.~M., Papovich, C., Faber, S.~M., {et~al.} 2013, \bibinfo{title}{CAUGHT
  IN THE ACT: THE ASSEMBLY OF MASSIVE CLUSTER GALAXIES AT z = 1.62,} \apj, 773,
  154, \dodoi{10.1088/0004-637X/773/2/154}

\bibitem[{A. Man \& S. Belli(2018)Man \& Belli}]{Man2018}
Man, A., \& Belli, S. 2018, \bibinfo{title}{{Star formation quenching in
  massive galaxies},} Nat. Astron., 2, 695, \dodoi{10.1038/s41550-018-0558-1}

\bibitem[{L.~D. Mascolo {et~al.}(2023)Mascolo, Saro, Mroczkowski, Borgani,
  Churazov, Rasia, Tozzi, Dannerbauer, Basu, Carilli, Ginolfi, Miley, Nonino,
  Pannella, Pentericci, \& Rizzo}]{DiMascolo2023}
Mascolo, L.~D., Saro, A., Mroczkowski, T., {et~al.} 2023,
  \bibinfo{title}{Forming intracluster gas in a galaxy protocluster at a
  redshift of 2.16,} \nat, 615, 809, \dodoi{10.1038/s41586-023-05761-x}

\bibitem[{M.~V. Maseda {et~al.}(2024)Maseda, de~Graaff, Franx, Rix, Carniani,
  Laseter, Dudzevičiūtė, Rawle, Parlanti, Arribas, Bunker, Cameron, Charlot,
  Curti, D’Eugenio, Jones, Kumari, Maiolino, Übler, Saxena, Smit, Willott,
  \& Witstok}]{Maseda2024}
Maseda, M.~V., de~Graaff, A., Franx, M., {et~al.} 2024, \bibinfo{title}{The
  NIRSpec Wide GTO Survey,} \aap, 689, A73, \dodoi{10.1051/0004-6361/202449914}

\bibitem[{I. McConachie {et~al.}(2022)McConachie, Wilson, Forrest, Marsan,
  Muzzin, Cooper, Annunziatella, Marchesini, Chan, Gomez, Abdullah, Saracco, \&
  Nantais}]{McConachie2022}
McConachie, I., Wilson, G., Forrest, B., {et~al.} 2022,
  \bibinfo{title}{Spectroscopic Confirmation of a Protocluster at z = 3.37 with
  a High Fraction of Quiescent Galaxies,} \apj, 926, 37,
  \dodoi{10.3847/1538-4357/ac2b9f}

\bibitem[{I. McConachie {et~al.}(2025{\natexlab{a}})McConachie, Antwi-Danso,
  Chang, Cooper, Edward, Forrest, Gomez, Lei, Lewis, Marchesini, Maseda,
  Muzzin, Noble, Stawinski, Webb, Wilson, \& Wisz}]{McConachie2025b}
McConachie, I., Antwi-Danso, J., Chang, W., {et~al.} 2025{\natexlab{a}},
  \bibinfo{title}{Excavating The Ruins: an Ancient $z=2.675$ Galaxy Which
  Formed in the First 500 Myr,} \doarXiv{2508.05752}

\bibitem[{I. McConachie {et~al.}(2025{\natexlab{b}})McConachie, Wilson,
  Forrest, Marsan, Muzzin, Cooper, Annunziatella, Marchesini, Gomez, Chang,
  Stawinski, McDonald, Webb, Noble, Lemaux, Shah, Staab, Lubin, \&
  Gal}]{McConachie2025a}
McConachie, I., Wilson, G., Forrest, B., {et~al.} 2025{\natexlab{b}},
  \bibinfo{title}{MAGAZ3NE: Evidence for Galactic Conformity in z ≳ 3
  Protoclusters*,} \apj, 978, 17, \dodoi{10.3847/1538-4357/ad8f36}

\bibitem[{S.~L. McGee {et~al.}(2011)McGee, Balogh, Wilman, Bower, Mulchaey,
  Parker, \& Oemler}]{McGee2011}
McGee, S.~L., Balogh, M.~L., Wilman, D.~J., {et~al.} 2011, \bibinfo{title}{The
  Dawn of the Red: star formation histories of group galaxies over the past 5
  billion years,} \mnras, 413, 996, \dodoi{10.1111/j.1365-2966.2010.18189.x}

\bibitem[{S. Mei {et~al.}(2023)Mei, Hatch, Amodeo, Afanasiev, {De Breuck},
  Stern, Cooke, Gonzalez, Noirot, Rettura, Seymour, Stanford, Vernet, \&
  Wylezalek}]{Mei2023}
Mei, S., Hatch, N.~A., Amodeo, S., {et~al.} 2023,
  \bibinfo{title}{{Morphology-density relation, quenching, and mergers in CARLA
  clusters and protoclusters at 1.4 {\textless} z {\textless} 2.8},} \aap, 670,
  A58, \dodoi{10.1051/0004-6361/202243551}

\bibitem[{I.~G. Momcheva {et~al.}(2016)Momcheva, Brammer, van Dokkum, Skelton,
  Whitaker, Nelson, Fumagalli, Maseda, Leja, Franx, Rix, Bezanson, Cunha,
  Dickey, Schreiber, Illingworth, Kriek, Labb{\'{e}}, Lange, Lundgren, Magee,
  Marchesini, Oesch, Pacifici, Patel, Price, Tal, Wake, van~der Wel, \&
  Wuyts}]{Momcheva2016a}
Momcheva, I.~G., Brammer, G.~B., van Dokkum, P.~G., {et~al.} 2016,
  \bibinfo{title}{{THE 3D-HST SURVEY: HUBBLE SPACE TELESCOPE WFC3/G141 GRISM
  SPECTRA, REDSHIFTS, AND EMISSION LINE MEASUREMENTS FOR ∼100,000 GALAXIES},}
  \apjs, 225, 27, \dodoi{10.3847/0067-0049/225/2/27}

\bibitem[{T. Morishita {et~al.}(2023)Morishita, Roberts-Borsani, Treu, Brammer,
  Mason, Trenti, Vulcani, Wang, Acebron, Bahé, Bergamini, Boyett, Bradac,
  Calabrò, Castellano, Chen, Lucia, Filippenko, Fontana, Glazebrook, Grillo,
  Henry, Jones, Kelly, Koekemoer, Leethochawalit, Lu, Marchesini, Mascia,
  Mercurio, Merlin, Metha, Nanayakkara, Nonino, Paris, Pentericci, Rosati,
  Santini, Strait, Vanzella, Windhorst, \& Xie}]{Morishita2023b}
Morishita, T., Roberts-Borsani, G., Treu, T., {et~al.} 2023,
  \bibinfo{title}{Early Results from GLASS-JWST. XIV. A Spectroscopically
  Confirmed Protocluster 650 Million Years after the Big Bang,} \apjl, 947,
  L24, \dodoi{10.3847/2041-8213/acb99e}

\bibitem[{S.~I. Muldrew {et~al.}(2015)Muldrew, Hatch, \& Cooke}]{Muldrew2015}
Muldrew, S.~I., Hatch, N.~A., \& Cooke, E.~A. 2015, \bibinfo{title}{What are
  protoclusters? – Defining high-redshift galaxy clusters and protoclusters,}
  \mnras, 452, 2528, \dodoi{10.1093/mnras/stv1449}

\bibitem[{A. Muzzin {et~al.}(2013)Muzzin, Marchesini, Stefanon, Franx,
  McCracken, Milvang-Jensen, Dunlop, Fynbo, Brammer, Labb{\'{e}}, \& van
  Dokkum}]{Muzzin2013}
Muzzin, A., Marchesini, D., Stefanon, M., {et~al.} 2013, \bibinfo{title}{{THE
  EVOLUTION OF THE STELLAR MASS FUNCTIONS OF STAR-FORMING AND QUIESCENT
  GALAXIES TO z = 4 FROM THE COSMOS/UltraVISTA SURVEY},} \apj, 777, 18,
  \dodoi{10.1088/0004-637X/777/1/18}

\bibitem[{T. Nanayakkara {et~al.}(2024)Nanayakkara, Glazebrook, Jacobs,
  Kawinwanichakij, Schreiber, Brammer, Esdaile, Kacprzak, Labbe, Lagos,
  Marchesini, Marsan, Oesch, Papovich, Remus, \& Tran}]{Nanayakkara2024}
Nanayakkara, T., Glazebrook, K., Jacobs, C., {et~al.} 2024, \bibinfo{title}{{A
  population of faint, old, and massive quiescent galaxies at
  {\$}{\$}3{\textless}z{\textless}4{\$}{\$} revealed by JWST NIRSpec
  Spectroscopy},} Sci. Rep., 14, 3724, \dodoi{10.1038/s41598-024-52585-4}

\bibitem[{T. Nanayakkara {et~al.}(2025)Nanayakkara, Glazebrook, Schreiber,
  Chittenden, Brammer, Esdaile, Jacobs, Kacprzak, Kawinwanichakij, Kimmig,
  Labbe, Lagos, Marchesini, Martìnez-Marìn, Marsan, Oesch, Papovich, Remus,
  \& Tran}]{Nanayakkara2025}
Nanayakkara, T., Glazebrook, K., Schreiber, C., {et~al.} 2025,
  \bibinfo{title}{The Formation Histories of Massive and Quiescent Galaxies in
  the 3 \&lt; <i>z</i> \&lt; 4.5 Universe,} \apj, 981, 78,
  \dodoi{10.3847/1538-4357/ada6ac}

\bibitem[{L. Napolitano {et~al.}(2025)Napolitano, Castellano, Pentericci, Haro,
  Fontana, Treu, Bergamini, Calabrò, Mascia, Morishita, Roberts-Borsani,
  Santini, Vanzella, Vulcani, Zakharova, Bakx, Dickinson, Grillo,
  Leethochawalit, Llerena, Merlin, Paris, Rojas-Ruiz, Rosati, Wang, Yoon, \&
  Zavala}]{Napolitano2025}
Napolitano, L., Castellano, M., Pentericci, L., {et~al.} 2025,
  \bibinfo{title}{Seven wonders of Cosmic Dawn: JWST confirms a high abundance
  of galaxies and AGN at z ≃ 9–11 in the GLASS field,} \aap, 693, A50,
  \dodoi{10.1051/0004-6361/202452090}

\bibitem[{J.~A. Newman {et~al.}(2013)Newman, Cooper, Davis, Faber, Coil,
  Guhathakurta, Koo, Phillips, Conroy, Dutton, Finkbeiner, Gerke, Rosario,
  Weiner, Willmer, Yan, Harker, Kassin, Konidaris, Lai, Madgwick, Noeske,
  Wirth, Connolly, Kaiser, Kirby, Lemaux, Lin, Lotz, Luppino, Marinoni,
  Matthews, Metevier, \& Schiavon}]{Newman2013}
Newman, J.~A., Cooper, M.~C., Davis, M., {et~al.} 2013, \bibinfo{title}{{THE
  DEEP2 GALAXY REDSHIFT SURVEY: DESIGN, OBSERVATIONS, DATA REDUCTION, AND
  REDSHIFTS},} \apjs, 208, 5, \dodoi{10.1088/0067-0049/208/1/5}

\bibitem[{J.~B. Oke \& J.~E. Gunn(1983)Oke \& Gunn}]{Oke1983}
Oke, J.~B., \& Gunn, J.~E. 1983, \bibinfo{title}{Secondary standard stars for
  absolute spectrophotometry,} The Astrophysical Journal, 266, 713,
  \dodoi{10.1086/160817}

\bibitem[{K.~C. Omori {et~al.}(2023)Omori, Bottrell, Walmsley, Yesuf, Goulding,
  Ding, Popping, Silverman, Takeuchi, \& Toba}]{Omori2023}
Omori, K.~C., Bottrell, C., Walmsley, M., {et~al.} 2023, \bibinfo{title}{Galaxy
  mergers in Subaru HSC-SSP: A deep representation learning approach for
  identification, and the role of environment on merger incidence,} \aap, 679,
  A142, \dodoi{10.1051/0004-6361/202346743}

\bibitem[{R. Pan {et~al.}(2025)Pan, Suess, Marchesini, Wang, Leja, Cutler,
  Whitaker, Bezanson, Price, Furtak, Weaver, Labbé, Brammer, Zhang, Dayal,
  Feldmann, Greene, Miller, Mitsuhashi, Nanayakkara, Nelson, Setton, \&
  Zitrin}]{Pan2025}
Pan, R., Suess, K.~A., Marchesini, D., {et~al.} 2025,
  \bibinfo{title}{UNCOVER/MegaScience: No Evidence of Environmental Quenching
  in a z$\sim$2.6 Proto-cluster,} \url{http://arxiv.org/abs/2504.06334}

\bibitem[{M. Park {et~al.}(2024)Park, Conroy, Johnson, Leja, Dotter, \&
  Cargile}]{Park2024a}
Park, M., Conroy, C., Johnson, B.~D., {et~al.} 2024,
  \bibinfo{title}{{{\$}\backslashalpha{\$}-MC: Self-consistent
  {\$}\backslashalpha{\$}-enhanced stellar population models covering a wide
  range of age, metallicity, and wavelength},} \doarXiv{2410.21375}

\bibitem[{P. Ramachandran \& G. Varoquaux(2011)Ramachandran \&
  Varoquaux}]{mayavi}
Ramachandran, P., \& Varoquaux, G. 2011, \bibinfo{title}{{Mayavi: 3D
  Visualization of Scientific Data},} Computing in Science \& Engineering, 13,
  40

\bibitem[{R.-S. Remus {et~al.}(2023)Remus, Dolag, \& Dannerbauer}]{Remus2023}
Remus, R.-S., Dolag, K., \& Dannerbauer, H. 2023, \bibinfo{title}{The Young and
  the Wild: What Happens to Protoclusters Forming at Redshift z ≈ 4?} \apj,
  950, 191, \dodoi{10.3847/1538-4357/accb91}

\bibitem[{P. Sanchez-Blazquez {et~al.}(2006)Sanchez-Blazquez, Peletier,
  Jimenez-Vicente, Cardiel, Cenarro, Falcon-Barroso, Gorgas, Selam, \&
  Vazdekis}]{Sanchez-Blazquez2006}
Sanchez-Blazquez, P., Peletier, R.~F., Jimenez-Vicente, J., {et~al.} 2006,
  \bibinfo{title}{Medium-resolution Isaac Newton Telescope library of empirical
  spectra,} \mnras, 371, 703, \dodoi{10.1111/j.1365-2966.2006.10699.x}

\bibitem[{C. Schreiber {et~al.}(2018)Schreiber, Labbé, Glazebrook, Bekiaris,
  Papovich, Costa, Elbaz, Kacprzak, Nanayakkara, Oesch, Pannella, Spitler,
  Straatman, Tran, \& Wang}]{Schreiber2018c}
Schreiber, C., Labbé, I., Glazebrook, K., {et~al.} 2018,
  \bibinfo{title}{Jekyll \&amp; Hyde: quiescence and extreme obscuration in a
  pair of massive galaxies 1.5 Gyr after the Big Bang,} \aap, 611, A22,
  \dodoi{10.1051/0004-6361/201731917}

\bibitem[{D.~J. Setton {et~al.}(2024)Setton, Khullar, Miller, Bezanson, Greene,
  Suess, Whitaker, Antwi-Danso, Atek, Brammer, Cutler, Dayal, Feldmann,
  Fujimoto, Furtak, Glazebrook, Goulding, Kokorev, Labbe, Leja, 马,
  Marchesini, Nanayakkara, Pan, Price, Siegel, Shipley, Weaver, van Dokkum,
  王, \& Williams}]{Setton2024}
Setton, D.~J., Khullar, G., Miller, T.~B., {et~al.} 2024,
  \bibinfo{title}{UNCOVER NIRSpec/PRISM Spectroscopy Unveils Evidence of Early
  Core Formation in a Massive, Centrally Dusty Quiescent Galaxy at z spec =
  3.97,} \apj, 974, 145, \dodoi{10.3847/1538-4357/ad6a18}

\bibitem[{E.~A. Shah {et~al.}(2024)Shah, Lemaux, Forrest, Cucciati, Hung,
  Staab, Hathi, Lubin, Gal, Shen, Zamorani, Giddings, Bardelli, Cassara,
  Cassata, Contini, Golden-Marx, Guaita, Gururajan, Koekemoer, McLeod, Tasca,
  Tresse, Vergani, \& Zucca}]{Shah2024}
Shah, E.~A., Lemaux, B., Forrest, B., {et~al.} 2024,
  \bibinfo{title}{Identification and characterization of six spectroscopically
  confirmed massive protostructures at 2.5 < z < 4.5,} \mnras, 529, 873,
  \dodoi{10.1093/mnras/stae519}

\bibitem[{L. Shen {et~al.}(2021)Shen, Lemaux, Lubin, Cucciati, Fèvre, Liu,
  Fang, Pelliccia, Tomczak, McKean, Miller, Fassnacht, Gal, Hung, Hathi,
  Bardelli, Vergani, \& Zucca}]{Shen2021}
Shen, L., Lemaux, B.~C., Lubin, L.~M., {et~al.} 2021,
  \bibinfo{title}{Implications of the Environments of Radio-detected Active
  Galactic Nuclei in a Complex Protostructure at z ∼ 3.3,} \apj, 912, 60,
  \dodoi{10.3847/1538-4357/abee75}

\bibitem[{T. Shibuya {et~al.}(2025)Shibuya, Ito, Asai, Kirihara, Fujimoto,
  Toba, Miura, Umayahara, Iwadate, Ali, \& Kodama}]{Shibuya2025}
Shibuya, T., Ito, Y., Asai, K., {et~al.} 2025, \bibinfo{title}{Galaxy
  morphologies revealed with Subaru HSC and super-resolution techniques. II.
  Environmental dependence of galaxy mergers at z ∼ 2–5,} \pasj, 77, 21,
  \dodoi{10.1093/pasj/psae096}

\bibitem[{J.~C. Siegel {et~al.}(2025)Siegel, Setton, Greene, Suess, Whitaker,
  Bezanson, Leja, Furtak, Cutler, de~Graaff, Feldmann, Khullar, Labbe,
  Marchesini, Miller, Nanayakkara, Pan, Price, Treiber, van Dokkum, 王, \&
  Weaver}]{Siegel2025}
Siegel, J.~C., Setton, D.~J., Greene, J.~E., {et~al.} 2025,
  \bibinfo{title}{UNCOVER: Significant Reddening in Cosmic Noon Quiescent
  Galaxies,} \apj, 985, 125, \dodoi{10.3847/1538-4357/adc7b7}

\bibitem[{R.~E. Skelton {et~al.}(2014)Skelton, Whitaker, Momcheva, Brammer, van
  Dokkum, Labb{\'{e}}, Franx, van~der Wel, Bezanson, {Da Cunha}, Fumagalli,
  {F{\"{o}}rster Schreiber}, Kriek, Leja, Lundgren, Magee, Marchesini, Maseda,
  Nelson, Oesch, Pacifici, Patel, Price, Rix, Tal, Wake, \&
  Wuyts}]{Skelton2014}
Skelton, R.~E., Whitaker, K.~E., Momcheva, I.~G., {et~al.} 2014,
  \bibinfo{title}{{3D-HST WFC3-SELECTED PHOTOMETRIC CATALOGS IN THE FIVE
  CANDELS/3D-HST FIELDS: PHOTOMETRY, PHOTOMETRIC REDSHIFTS, AND STELLAR
  MASSES},} \apjs, 214, 24, \dodoi{10.1088/0067-0049/214/2/24}

\bibitem[{M. Slob {et~al.}(2024)Slob, Kriek, Beverage, Suess, Barro, Bezanson,
  Brammer, Cheng, Conroy, de~Graaff, {F{\"{o}}rster Schreiber}, Franx, Lorenz,
  {Mancera Pi{\~{n}}a}, Marchesini, Muzzin, Newman, Price, Shapley, Stefanon,
  van Dokkum, \& Weisz}]{Slob2024}
Slob, M., Kriek, M., Beverage, A.~G., {et~al.} 2024, \bibinfo{title}{{The
  JWST-SUSPENSE Ultradeep Spectroscopic Program: Survey Overview and Star
  Formation Histories of Quiescent Galaxies at 1 {\textless} z {\textless} 3},}
  \apj, 973, 131, \dodoi{10.3847/1538-4357/ad65ff}

\bibitem[{J.~S. Speagle(2020)Speagle}]{Speagle2020}
Speagle, J.~S. 2020, \bibinfo{title}{dynesty: a dynamic nested sampling package
  for estimating Bayesian posteriors and evidences,} \mnras, 493, 3132,
  \dodoi{10.1093/mnras/staa278}

\bibitem[{P. Staab {et~al.}(2024)Staab, Lemaux, Forrest, Shah, Cucciati, Lubin,
  Gal, Hung, Shen, Giddings, Khusanova, Zamorani, Bardelli, Cassara, Cassata,
  Chiang, Fudamoto, Fukushima, Garilli, Giavalisco, Gruppioni, Guaita,
  Gururajan, Hathi, Kashino, Scoville, Talia, Vergani, \& Zucca}]{Staab2024}
Staab, P., Lemaux, B.~C., Forrest, B., {et~al.} 2024,
  \bibinfo{title}{Protoclusters as drivers of stellar mass growth in the early
  Universe, a case study: Taralay – a massive protocluster at z ∼ 4.57,}
  \mnras, 528, 6934, \dodoi{10.1093/mnras/stae301}

\bibitem[{S.~M.~U. Stawinski {et~al.}(2024{\natexlab{a}})Stawinski, Cooper,
  Forrest, Muzzin, Marchesini, Wilson, Gomez, McConachie, Marsan, Annuziatella,
  \& Chang}]{UrbanoStawinski2024}
Stawinski, S. M.~U., Cooper, M.~C., Forrest, B., {et~al.} 2024{\natexlab{a}},
  \bibinfo{title}{Spectroscopic Confirmation of an Ultra-Massive Galaxy in a
  Protocluster at <math> <mi>z</mi> <mo>∼</mo> <mn>4.9</mn> </math>,} OJA, 7,
  \dodoi{10.33232/001c.120087}

\bibitem[{S.~M.~U. Stawinski {et~al.}(2024{\natexlab{b}})Stawinski, Cooper,
  Finkelstein, Jung, P{\'{e}}rez-Gonz{\'{a}}lez, Bagley, Casey, Cooper, Hathi,
  Holwerda, Koekemoer, Kartaltepe, Fern{\'{a}}ndez, Larson, Lucas, \&
  Yung}]{Stawinski2024}
Stawinski, S. M.~U., Cooper, M.~C., Finkelstein, S.~L., {et~al.}
  2024{\natexlab{b}}, \bibinfo{title}{{Deeper than DEEP: a spectroscopic survey
  of z {\textgreater} 3 Ly $\alpha$ emitters in the Extended Groth Strip},}
  /mnras, 528, 5624, \dodoi{10.1093/mnras/stae361}

\bibitem[{I. Strateva {et~al.}(2001)Strateva, Željko Ivezić, Knapp,
  Narayanan, Strauss, Gunn, Lupton, Schlegel, Bahcall, Brinkmann, Brunner,
  Budavári, Csabai, Castander, Doi, Fukugita, Győry, Hamabe, Hennessy,
  Ichikawa, Kunszt, Lamb, McKay, Okamura, Racusin, Sekiguchi, Schneider,
  Shimasaku, \& York}]{Strateva2001}
Strateva, I., Željko Ivezić, Knapp, G.~R., {et~al.} 2001,
  \bibinfo{title}{Color Separation of Galaxy Types in the Sloan Digital Sky
  Survey Imaging Data,} \aj, 122, 1861, \dodoi{10.1086/323301}

\bibitem[{H. Sun {et~al.}(2025)Sun, Wang, Daddi, Hao, Xu, Elbaz, Zhou, Mo,
  Wang, Chen, Chen, Jin, Lyu, Sillassen, Wang, \& Yang}]{Sun2025b}
Sun, H., Wang, T., Daddi, E., {et~al.} 2025, \bibinfo{title}{The Bigfoot: A
  footprint of a Coma cluster progenitor at z=3.98,} \doarXiv{2508.21356}

\bibitem[{M. Tanaka {et~al.}(2024)Tanaka, Onodera, Shimakawa, Ito, Kakimoto,
  Kubo, Morishita, Toft, Valentino, \& Wu}]{Tanaka2024}
Tanaka, M., Onodera, M., Shimakawa, R., {et~al.} 2024, \bibinfo{title}{A
  Protocluster of Massive Quiescent Galaxies at z = 4,} \apj, 970, 59,
  \dodoi{10.3847/1538-4357/ad5316}

\bibitem[{M. Tang {et~al.}(2025)Tang, Stark, Plat, Feltre, Katz, Senchyna,
  Mason, Whitler, Chen, \& Topping}]{Tang2025}
Tang, M., Stark, D.~P., Plat, A., {et~al.} 2025, \bibinfo{title}{JWST/NIRSpec
  Observations of High Ionization Emission Lines in Galaxies at High Redshift,}
  \doarXiv{2505.06359}

\bibitem[{A.~R. Tomczak {et~al.}(2017)Tomczak, Lemaux, Lubin, Gal, Wu, Holden,
  Kocevski, Mei, Pelliccia, Rumbaugh, \& Shen}]{Tomczak2017a}
Tomczak, A.~R., Lemaux, B.~C., Lubin, L.~M., {et~al.} 2017,
  \bibinfo{title}{Glimpsing the imprint of local environment on the galaxy
  stellar mass function,} \mnras, 472, 3512, \dodoi{10.1093/mnras/stx2245}

\bibitem[{C. Turner {et~al.}(2025)Turner, Tacchella, D’Eugenio, Carniani,
  Curti, Glazebrook, Johnson, Lim, Looser, Maiolino, Nanayakkara, \&
  Wan}]{Turner2025}
Turner, C., Tacchella, S., D’Eugenio, F., {et~al.} 2025,
  \bibinfo{title}{Age-dating early quiescent galaxies: high star formation
  efficiency, but consistent with direct, higher-redshift observations,}
  \mnras, 537, 1826, \dodoi{10.1093/mnras/staf128}

\bibitem[{F. Valentino {et~al.}(2023)Valentino, Brammer, Gould, Kokorev,
  Fujimoto, Jespersen, Vijayan, Weaver, Ito, Tanaka, Ilbert, Magdis, Whitaker,
  Faisst, Gallazzi, Gillman, Giménez-Arteaga, Gómez-Guijarro, Kubo, Heintz,
  Hirschmann, Oesch, Onodera, Rizzo, Lee, Strait, \& Toft}]{Valentino2023}
Valentino, F., Brammer, G., Gould, K. M.~L., {et~al.} 2023, \bibinfo{title}{An
  Atlas of Color-selected Quiescent Galaxies at z \&gt; 3 in Public JWST
  Fields,} \apj, 947, 20, \dodoi{10.3847/1538-4357/acbefa}

\bibitem[{R.~F.~J. van~der Burg {et~al.}(2020)van~der Burg, Rudnick, Balogh,
  Muzzin, Lidman, Old, Shipley, Gilbank, McGee, Biviano, Cerulo, Chan, Cooper,
  Lucia, Demarco, Forrest, Gwyn, Jablonka, Kukstas, Marchesini, Nantais, Noble,
  Pintos-Castro, Poggianti, Reeves, Stefanon, Vulcani, Webb, Wilson, Yee, \&
  Zaritsky}]{Vanderburg2020}
van~der Burg, R. F.~J., Rudnick, G., Balogh, M.~L., {et~al.} 2020,
  \bibinfo{title}{The GOGREEN Survey: A deep stellar mass function of cluster
  galaxies at 1.0 < z < 1.4 and the complex nature of satellite quenching,}
  \aap, 638, A112, \dodoi{10.1051/0004-6361/202037754}

\bibitem[{A. van~der Wel {et~al.}(2014)van~der Wel, Franx, van Dokkum, Skelton,
  Momcheva, Whitaker, Brammer, Bell, Rix, Wuyts, Ferguson, Holden, Barro,
  Koekemoer, Chang, McGrath, Häussler, Dekel, Behroozi, Fumagalli, Leja,
  Lundgren, Maseda, Nelson, Wake, Patel, Labbé, Faber, Grogin, \&
  Kocevski}]{VanderWel2014}
van~der Wel, A., Franx, M., van Dokkum, P.~G., {et~al.} 2014,
  \bibinfo{title}{3D-HST+CANDELS: THE EVOLUTION OF THE GALAXY SIZE-MASS
  DISTRIBUTION SINCE <i>z</i> = 3,} \apj, 788, 28,
  \dodoi{10.1088/0004-637X/788/1/28}

\bibitem[{F.~P.~A. Vogt {et~al.}(2016)Vogt, Owen, Verdes-Montenegro, \&
  Borthakur}]{Vogt2016}
Vogt, F. P.~A., Owen, C.~I., Verdes-Montenegro, L., \& Borthakur, S. 2016,
  \bibinfo{title}{{ADVANCED DATA VISUALIZATION IN ASTROPHYSICS: THE X3D
  PATHWAY},} \apj, 818, 115, \dodoi{10.3847/0004-637X/818/2/115}

\bibitem[{C. Watson {et~al.}(2019)Watson, Tran, Tomczak, Alcorn, Salazar,
  Gupta, Momcheva, Papovich, van Dokkum, Brammer, Lotz, \&
  Willmer}]{Watson2019}
Watson, C., Tran, K.-V., Tomczak, A., {et~al.} 2019, \bibinfo{title}{Galaxy
  Merger Fractions in Two Clusters at Using the Hubble Space Telescope,} \apj,
  874, 63, \dodoi{10.3847/1538-4357/ab06ef}

\bibitem[{J.~R. Weaver {et~al.}(2023)Weaver, Davidzon, Toft, Ilbert, McCracken,
  Gould, Jespersen, Steinhardt, Lagos, Capak, Casey, Chartab, Faisst, Hayward,
  Kartaltepe, Kauffmann, Koekemoer, Kokorev, Laigle, Liu, Long, Magdis,
  McPartland, Milvang-Jensen, Mobasher, Moneti, Peng, Sanders, Shuntov,
  Sneppen, Valentino, Zalesky, \& Zamorani}]{Weaver2023a}
Weaver, J.~R., Davidzon, I., Toft, S., {et~al.} 2023,
  \bibinfo{title}{{COSMOS2020: The galaxy stellar mass function},} \aap, 677,
  A184, \dodoi{10.1051/0004-6361/202245581}

\bibitem[{J.~R. Weaver {et~al.}(2024)Weaver, Cutler, Pan, Whitaker, Labbé,
  Price, Bezanson, Brammer, Marchesini, Leja, 王, Furtak, Zitrin, Atek,
  Chemerynska, Coe, Dayal, van Dokkum, Feldmann, Schreiber, Franx, Fujimoto,
  Fudamoto, Glazebrook, de~Graaff, Greene, Juneau, Kassin, Kriek, Khullar,
  Maseda, Mowla, Muzzin, Nanayakkara, Nelson, Oesch, Pacifici, Papovich,
  Setton, Shapley, Shipley, Smit, Stefanon, Taylor, Weibel, \&
  Williams}]{Weaver2024}
Weaver, J.~R., Cutler, S.~E., Pan, R., {et~al.} 2024, \bibinfo{title}{The
  UNCOVER Survey: A First-look HST + JWST Catalog of 60,000 Galaxies near A2744
  and beyond,} \apjs, 270, 7, \dodoi{10.3847/1538-4365/ad07e0}

\bibitem[{A. Weibel {et~al.}(2024)Weibel, Oesch, Barrufet, Gottumukkala, Ellis,
  Santini, Weaver, Allen, Bouwens, Bowler, Brammer, Carnall, Cullen, Dayal,
  Dickinson, Donnan, Dunlop, Giavalisco, Grogin, Illingworth, Koekemoer, Labbe,
  Marchesini, McLeod, McLure, Naidu, Pérez-González, Shuntov, Stefanon, Toft,
  \& Xiao}]{Weibel2024}
Weibel, A., Oesch, P.~A., Barrufet, L., {et~al.} 2024, \bibinfo{title}{Galaxy
  build-up in the first 1.5 Gyr of cosmic history: insights from the stellar
  mass function at z ~ 4–9 from JWST NIRCam observations,} \mnras, 533, 1808,
  \dodoi{10.1093/mnras/stae1891}

\bibitem[{A.~R. Wetzel {et~al.}(2012)Wetzel, Tinker, \& Conroy}]{Wetzel2012}
Wetzel, A.~R., Tinker, J.~L., \& Conroy, C. 2012, \bibinfo{title}{Galaxy
  evolution in groups and clusters: star formation rates, red sequence
  fractions and the persistent bimodality,} \mnras, 424, 232,
  \dodoi{10.1111/j.1365-2966.2012.21188.x}

\bibitem[{V. Wild {et~al.}(2016)Wild, Almaini, Dunlop, Simpson, Rowlands,
  Bowler, Maltby, \& McLure}]{Wild2016}
Wild, V., Almaini, O., Dunlop, J., {et~al.} 2016, \bibinfo{title}{{The
  evolution of post-starburst galaxies from z=2 to 0.5},} \mnras, 463, 832,
  \dodoi{10.1093/mnras/stw1996}

\bibitem[{L. Wright {et~al.}(2024)Wright, Whitaker, Weaver, Cutler, 王,
  Carnall, Suess, Bezanson, Nelson, Miller, Ito, \& Valentino}]{Wright2024}
Wright, L., Whitaker, K.~E., Weaver, J.~R., {et~al.} 2024,
  \bibinfo{title}{Remarkably Compact Quiescent Candidates at 3 < z < 5 in
  JWST-CEERS,} \apjl, 964, L10, \dodoi{10.3847/2041-8213/ad2b6d}

\bibitem[{Y. Zhang {et~al.}(2025)Zhang, de~Graaff, Setton, Price, Bezanson, del
  P.~Lagos, Cutler, McConachie, Cleri, Cooper, Gottumukkala, Greene,
  Hirschmann, Khullar, Labbe, Leja, Maseda, Matthee, Miller, Nanayakkara,
  Suess, Wang, Whitaker, \& Williams}]{Zhang2025}
Zhang, Y., de~Graaff, A., Setton, D.~J., {et~al.} 2025, \bibinfo{title}{RUBIES
  spectroscopically confirms the high number density of quiescent galaxies from
  $\mathbf\{2<z<5\}$,} \doarXiv{2508.08577}

\end{thebibliography}
\bibliographystyle{aasjournalv7}

\end{document}